\documentclass[twocolumn,tighten]{aastex631}
\usepackage{float,graphicx,amsmath,multirow}
\usepackage{color}
\usepackage[version=4]{mhchem}
\usepackage{booktabs}
\usepackage{gensymb}

\newcommand{\cnp}{cyanonaphthalene}

\begin{document}

\title{Detection of Two Interstellar Polycyclic Aromatic Hydrocarbons via Spectral Matched Filtering}
\author{Brett A. McGuire}
\affiliation{Department of Chemistry, Massachusetts Institute of Technology, Cambridge, MA 02139, USA}
\affiliation{National Radio Astronomy Observatory, Charlottesville, VA 22903, USA}
\affiliation{Center for Astrophysics $\mid$ Harvard~\&~Smithsonian, Cambridge, MA 02138, USA}
\author{Ryan A. Loomis}
\altaffiliation{These authors contributed equally to this work.}
\affiliation{National Radio Astronomy Observatory, Charlottesville, VA 22903, USA}
\author{Andrew M. Burkhardt}
\altaffiliation{These authors contributed equally to this work.}
\affiliation{Center for Astrophysics $\mid$ Harvard~\&~Smithsonian, Cambridge, MA 02138, USA}
\author{Kin Long Kelvin Lee}
\affiliation{Center for Astrophysics $\mid$ Harvard~\&~Smithsonian, Cambridge, MA 02138, USA}
\author{Christopher N. Shingledecker}
\affiliation{Department of Physics and Astronomy, Benedictine College, Atchison, KS 66002, USA}
\affiliation{Center for Astrochemical Studies, Max Planck Intitute for Extraterrestrial Physics, Garching, Germany}
\affiliation{Institute for Theoretical Chemistry, University of Stuttgart, Stuttgart, Germany}
\author{Steven B. Charnley}
\affiliation{Astrochemistry Laboratory and the Goddard Center for Astrobiology, NASA Goddard Space Flight Center, Greenbelt, MD 20771, USA}
\author{Ilsa R. Cooke}
\affiliation{Univ Rennes, Centre National de la Recherche Scientifique, Institut de Physique de Rennes, Unité Mixte de Recherche 6251, F-35000 Rennes, France}
\author{Martin A. Cordiner}
\affiliation{Astrochemistry Laboratory and the Goddard Center for Astrobiology, NASA Goddard Space Flight Center, Greenbelt, MD 20771, USA}
\affiliation{Institute for Astrophysics and Computational Sciences, The Catholic University of America, Washington, DC 20064, USA}
\author{Eric Herbst}
\affiliation{Department of Chemistry, University of Virginia, Charlottesville, VA 22904, USA}
\affiliation{Department of Astronomy, University of Virginia, Charlottesville, VA 22904, USA}
\author{Sergei Kalenskii}
\affiliation{Astro Space Center, Lebedev Physical Institute, Russian Academy of Sciences, Moscow, Russia}
\author{Mark A. Siebert}
\affiliation{Department of Astronomy, University of Virginia, Charlottesville, VA 22904, USA}
\author{Eric R. Willis}
\affiliation{Department of Chemistry, University of Virginia, Charlottesville, VA 22904, USA}
\author{Ci Xue}
\affiliation{Department of Chemistry, University of Virginia, Charlottesville, VA 22904, USA}
\author{Anthony J. Remijan}
\affiliation{National Radio Astronomy Observatory, Charlottesville, VA 22903, USA}
\author{Michael C. McCarthy}
\affiliation{Center for Astrophysics $\mid$ Harvard~\&~Smithsonian, Cambridge, MA 02138, USA}

\correspondingauthor{Brett A. McGuire}
\email{brettmc@mit.edu}

\begin{abstract}
Ubiquitous unidentified infrared emission bands are seen in many astronomical sources.  Although these bands are widely, if not unanimously, attributed to the collective emission from polycyclic aromatic hydrocarbons, no single species from this class has been detected in space.  We present the discovery of two -CN functionalized polycyclic aromatic hydrocarbons, 1- and 2-\cnp, in the interstellar medium aided by spectral matched filtering.  Using radio observations with the Green Bank Telescope, we observe both bi-cyclic ring molecules in the molecular cloud TMC-1.  We discuss potential in situ gas-phase formation pathways from smaller organic precursor molecules.

\end{abstract}
%\keywords{Astrochemistry, ISM: molecules}

\section*{\hspace{1em}}
\vspace{-1.5em}

Aromatic molecules are a ubiquitous structural motif not only in the chemical make-up of life, but in the broader chemical evolution of the universe.  Indeed, as much as 10--25\% of all interstellar carbon is thought to be locked up in the form of polycyclic aromatic hydrocarbons (PAHs), thought to form primarily (if not exclusively) in the circumstellar envelopes of evolved stars \citep{Tielens:2008fx}. The presence of the aromatic molecule benzonitrile (BN; $c$-\ce{C6H5CN}) in the interstellar molecular cloud TMC-1 \citep{McGuire:2018it}, far removed from the envelope of an evolved star, was therefore unexpected.  Motivated by this discovery, a very high sensitivity spectral line survey of TMC-1 has been underway since Feb 2018 to critically ascertain its aromatic inventory in particular.

The vast majority ($>$80\%) of interstellar molecules have been discovered using pure rotational spectroscopy at radio frequencies  \citep{McGuire:2018mc}.  We therefore searched for PAHs using observations of the molecular cloud TMC-1 in the range of 8--33.5\,GHz where rotational emission lines of many aromatic molecules are predicted to be strongest at the 5-10\,K temperature of TMC-1 \citep{McGuire:2020bb}.  Using the 100-m Robert C. Byrd Green Bank Telescope (GBT), a large-scale observing project entitled GOTHAM (GBT Observations of TMC-1: Hunting Aromatic Molecules) is on-track to complete a high-sensitivity, high-resolution spectral line survey of TMC-1.  The first reduction of this data, comprising all observations obtained through May 2019 (hereafter `DR1'), had near-continuous frequency coverage between 8--11.6\,GHz and 18--29.5\,GHz, with a few small gaps.  Details of these observations are presented elsewhere \citep{McGuire:2020bb} and in the Appendix. Further observations were obtained through June 2020 (`DR2'), which extended the coverage to higher frequencies and improved the sensitivity in some regions already covered by DR1 (Fig.~\ref{gotham_coverage};  \citealt{McGuire:2020bb}). Both datasets have a spectral resolution of 1.4 kHz, which is equivalent to a velocity resolution of 0.05--0.02\,km\,s$^{-1}$ (varying with frequency).

\begin{figure}
    \centering
    \includegraphics[width=0.49\textwidth]{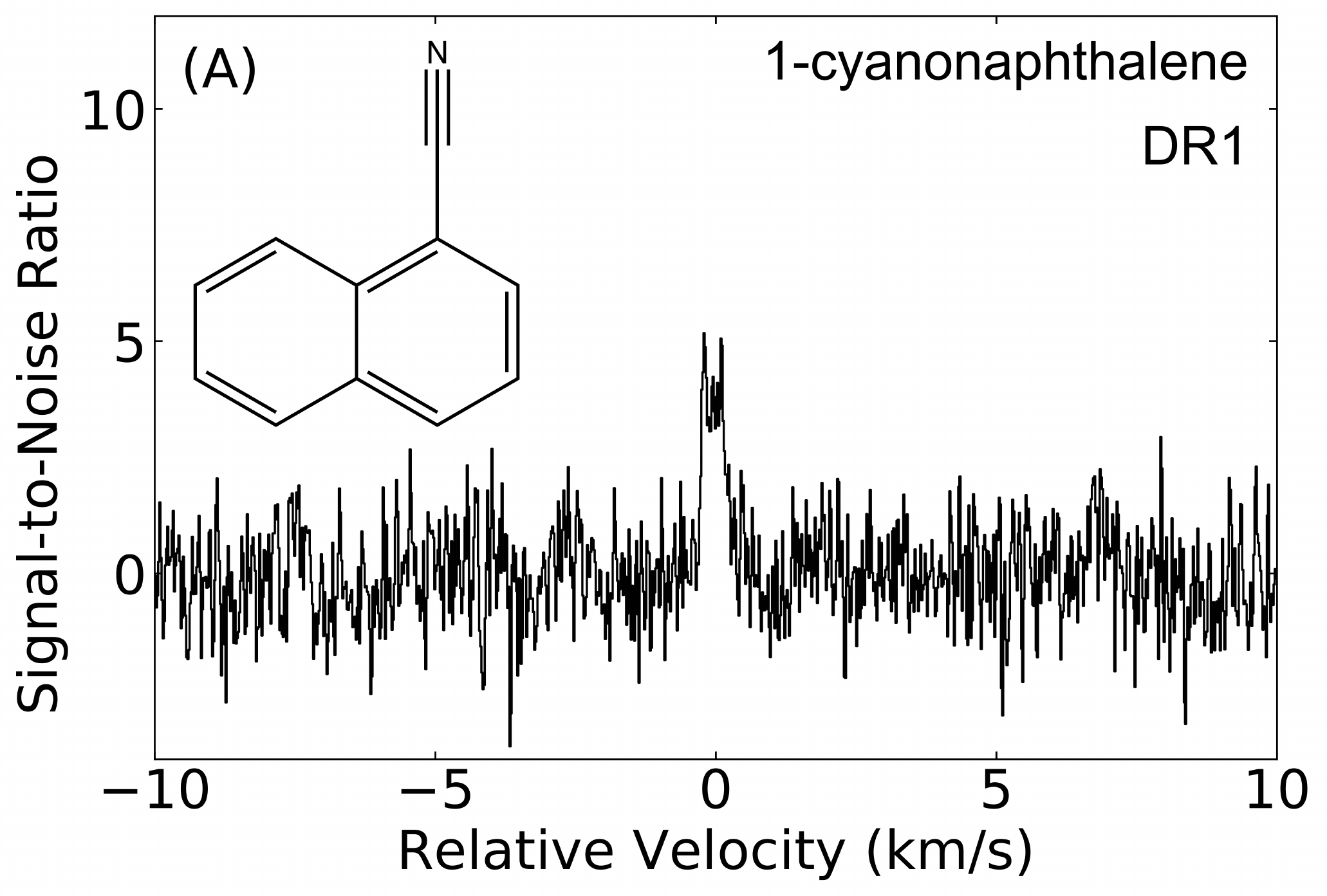}
    \includegraphics[width=0.49\textwidth]{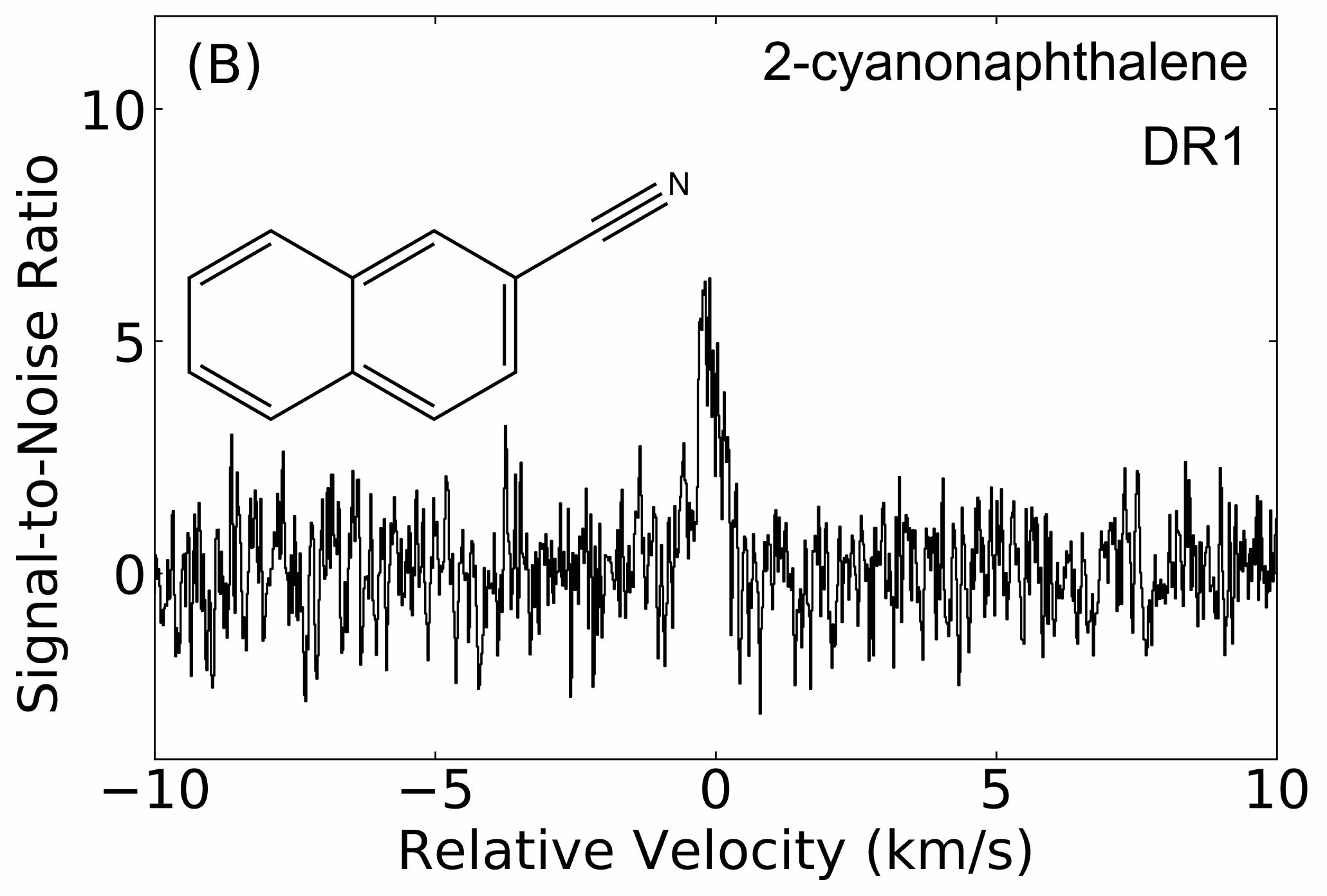}
    \caption{\textbf{Molecular structures and spectral stacks of 1- and 2-cyanonapthalene in the GOTHAM DR1 data.}  These molecules are derivatives of naphthalene, substituting a nitrile (\ce{-CN}) group for a hydrogen atom.  This produces two distinct isomers, both of which are highly polar, with dipole moments ($\mu$) along the $a$ and $b$ principal axes: (A) 1-CNN: $\mu_a$ = 3.6, $\mu_b$ = 3.0\,Debye; and (B) 2-CNN: $\mu_a$ = 5.1, $\mu_b$ = 1.0\,Debye \citep{McNaughton:2018op}. The stacks are shown relative to the TMC-1 systemic velocity of 5.8\,km\,s$^{-1}$.  The weighting process assumed an excitation temperature of $T_{ex}$~=~7\,K.  }
    \label{dr1_results}
\end{figure}

Given the detection of BN, a single benzene ring with an attached CN group, we searched for derivatives of naphthalene (two fused benzene rings; \ce{C10H8}).  Like benzene, naphthalene lacks a permanent dipole moment and thus possesses no pure rotational spectrum.  However, its --CN substituted derivatives, 1-cyanonaphthalene (1-CNN) and 2-cyanonaphthalene (2-CNN; collectively CNNs), have large permanent dipole moments (Fig.~\ref{dr1_results}) and laboratory measured rotational spectra \citep{McNaughton:2018op} that peak in the frequency region covered by our observations (Fig.~\ref{cnns_coverage}).

Similarly to BN \citep{McGuire:2018it}, initial searches for the CNNs in our DR1 data showed no individual rotational lines above the noise level of the observations. We therefore calculated a spectral stack by combining the positions of all the lines predicted in the dataset.  Because the average excitation temperature of molecules ($T_{\mathrm{ex}}\sim7$\,K) and the velocity (5.8\,km\,s$^{-1}$) of TMC-1 are known \citep{Gratier:2016fj,Kaifu:2004tk,McGuire:2020bb}, we average the signal from all rotational transitions covered by our observations weighted (up) by their predicted intensities and (down) by the local noise level of the observations (see Appendix). The two stacks for 1-CNN and 2-CNN derived from the DR1 data are shown in Fig.~\ref{dr1_results} and indicate the presence of both molecules. 

At this point, we had the higher-quality DR2 dataset in hand and we sought to determine the physical parameters ($T_{\mathrm{ex}}$, column density [$N_{\mathrm{T}}$], linewidth [$\Delta V$], source size [$\theta$], and velocity [$v_{\mathrm{lsr}}$]) that best reproduced the observations. A Markov-Chain Monte-Carlo (MCMC) analysis was used to derive the physical parameters that best reproduce the stacked emission, including radiative transfer corrections for optical depth \citep{Turner:1991um}, with more robust uncertainties than a least-squares fit.  For this more detailed analysis, we assume four partially overlapping Doppler velocity components with $v_{\mathrm{lsr}}$ between 5.5 and 6.1\,km\,s$^{-1}$ \citep{Dobashi:2018kd,Dobashi:2019ev} each with their own column density and (poorly constrained in our single-dish observations) source size.  A single excitation temperature ($T_{\mathrm{ex}}$) and linewidth ($\Delta V$; \citealt{McGuire:2020bb,Xue:2020aa,Loomis:2021aa}) is assumed for all four velocity components.  We use the physical parameters from the more strongly detected BN (Table~\ref{bn_results}; Fig.~\ref{bn_prior_triangle}) as Gaussian priors in this analysis.  

\begin{figure*}[tbh!]
    \centering
    \includegraphics[width=0.49\textwidth]{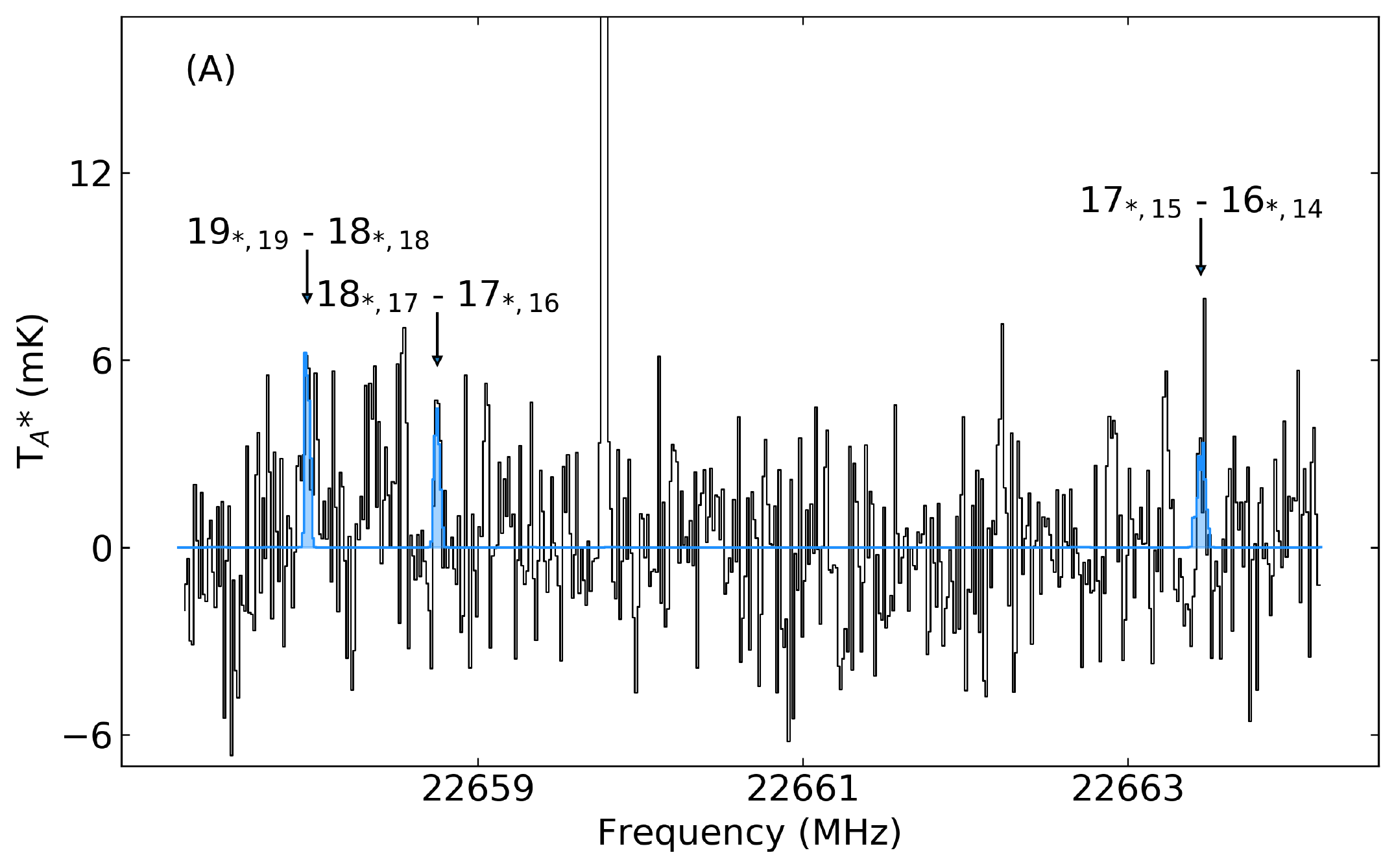}
    \includegraphics[width=0.49\textwidth]{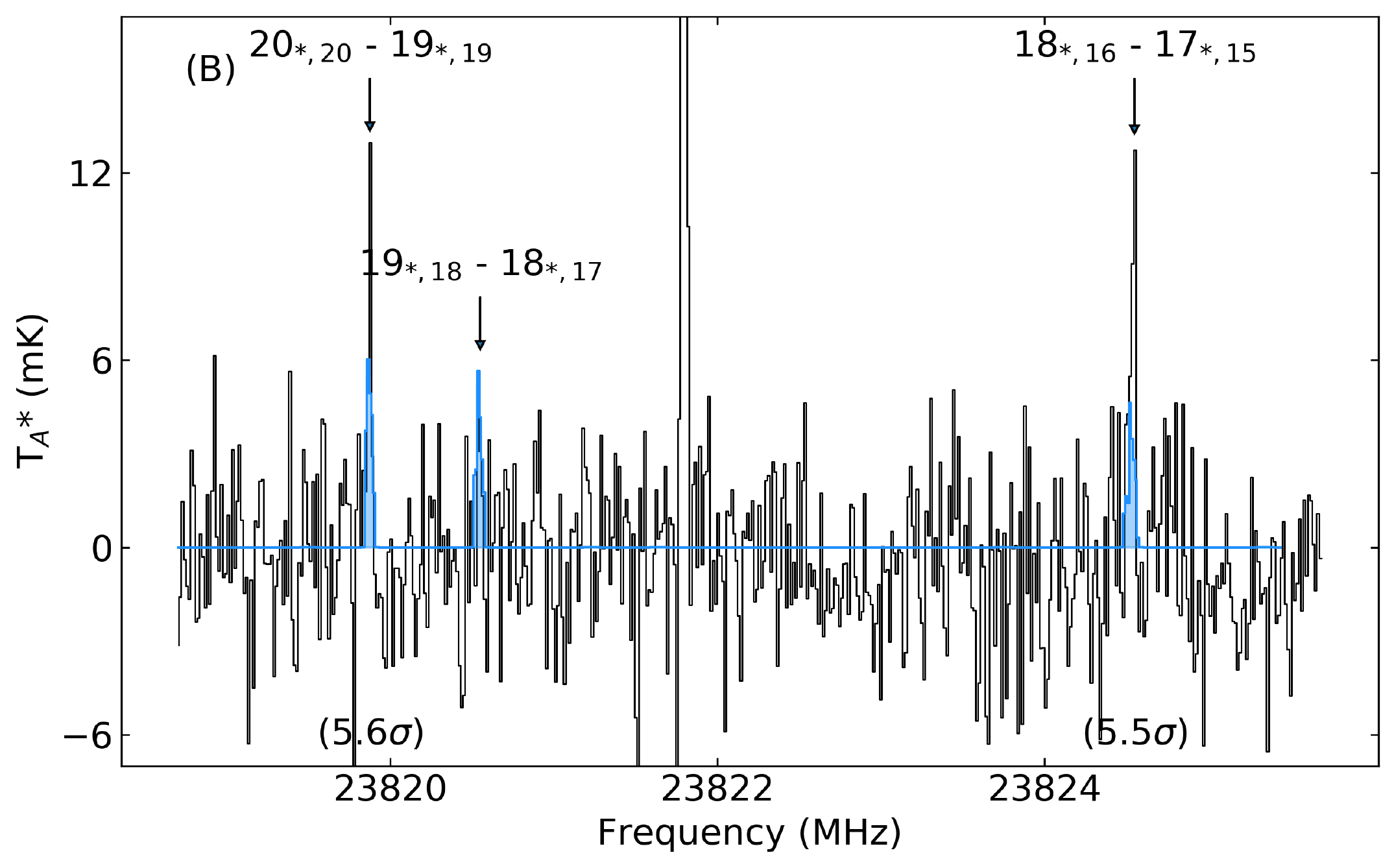}
    \includegraphics[width=0.49\textwidth]{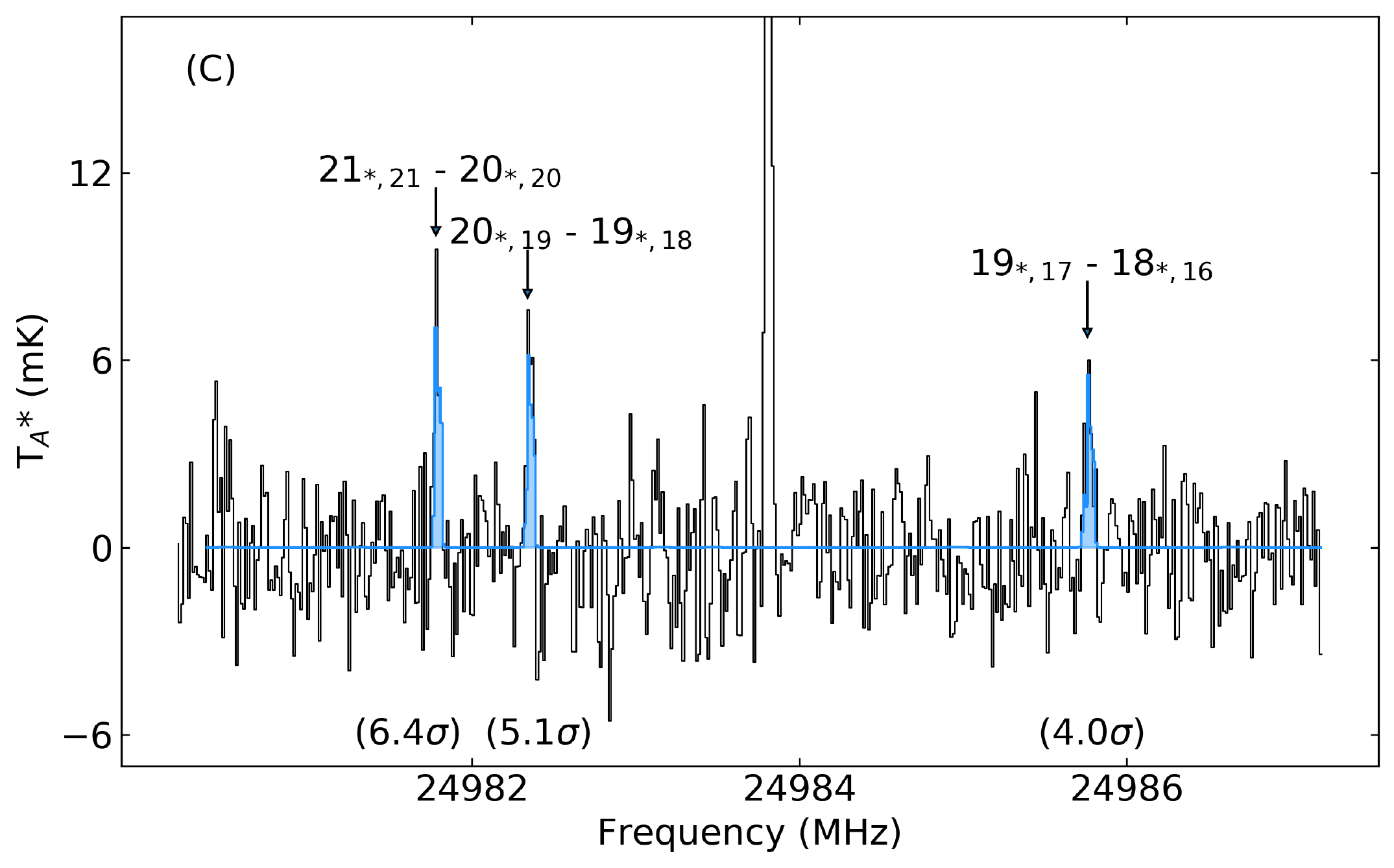}
    \includegraphics[width=0.49\textwidth]{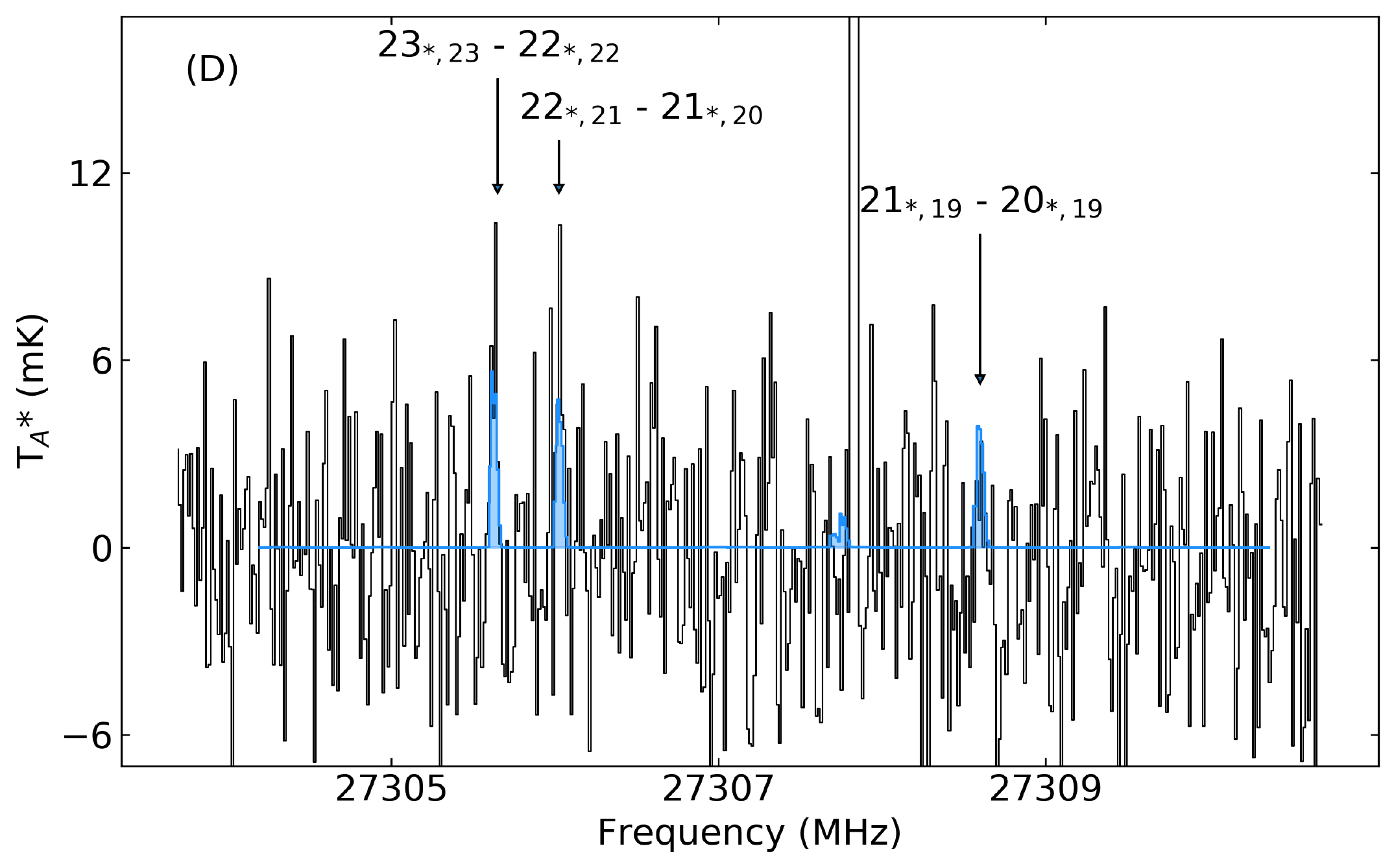}
    \caption{\textbf{GOTHAM DR2 spectra in the vicinity of predicted strong 1-CNN transitions.}  Observations are shown in black, with simulated spectra of 1-CNN using the parameters derived from our MCMC analysis overlaid in blue.  The observations have not been adjusted for the systemic velocity of TMC-1 (5.8\,km\,s$^{-1}$). The spectra have been smoothed with a Hanning window to a resolution of 14\,kHz for display.  The quantum numbers of the transitions, ignoring hyperfine structure, are labeled in each window.  Multiple closely spaced $K$-components of each transition contribute to each signal and are denoted by an asterisk.  Transitions with peak signal-to-noise ratios $>$4$\sigma$ are labeled beneath the spectra.  The y-axis is the atmosphere-corrected antenna temperature scale (T$_{\rm{A}}^*$).}
    \label{cnn_lines}
\end{figure*}

The results from the MCMC analysis for 1-CNN (Table~\ref{1-CNN_results}; Fig.~\ref{1-CNN_triangle}) and 2-CNN (Table~\ref{2-CNN_results}; Fig.~\ref{2-CNN_triangle}) give total column densities (the sum of all four velocity components) of 7.35$^{+3.33}_{-4.63}\times10^{11}$\,cm$^{-2}$ and 7.05$^{+3.23}_{-4.50}\times10^{11}$\,cm$^{-2}$, respectively.  At those column densities, we predict roughly a dozen features of 1-CNN, and none of 2-CNN, should be above the local noise level in parts of the DR2 data. Fig.~\ref{cnn_lines} shows the DR2 data along with simulated profiles; all other lines are predicted to be below the noise. There is evidence for at least five lines with peak signal-to-noise ratios (SNR) $>4\sigma$, and tentative evidence for several others.

Although these individual lines are weak, they provide us with additional confidence in the results of the stacking process. The information content of the entire spectrum can also be used to assess the presence or absence of each molecule, which has many weak transitions at or below the root mean square (RMS) noise level of the observations. We perform this test by using spectral stacking in concert with matched filtering (\citealt{Loomis:2021aa} and Appendix).   Although care must be taken with respect to interlopers and the noise characteristics of the data, this approach increases the SNR, with the averaged spectrum encapsulating the total information content of all observed lines, rather than examining each (lower-significance) line individually.

\begin{figure*}[hbt!]
    \centering
    \includegraphics[width=0.49\textwidth]{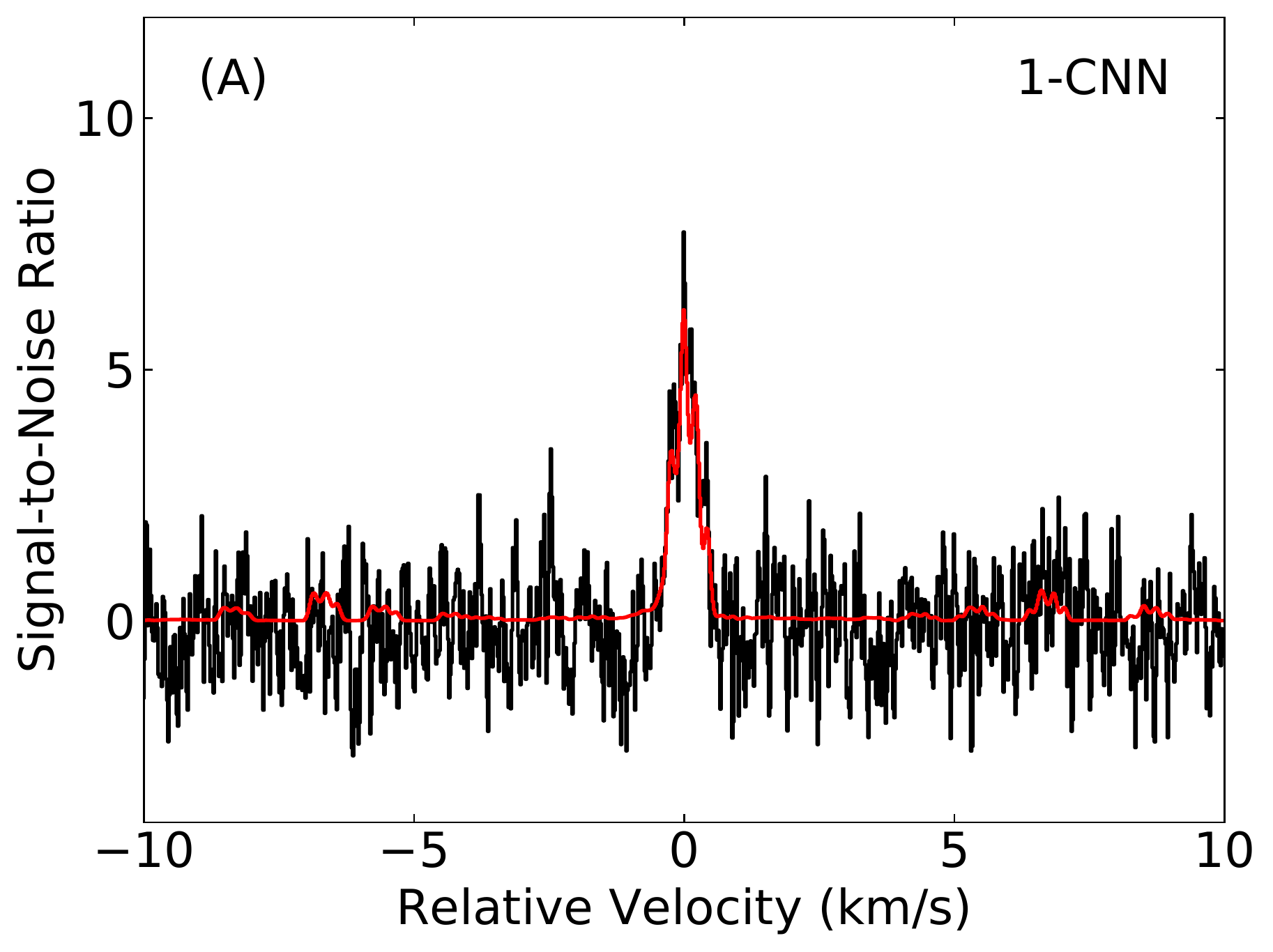}
    \includegraphics[width=0.49\textwidth]{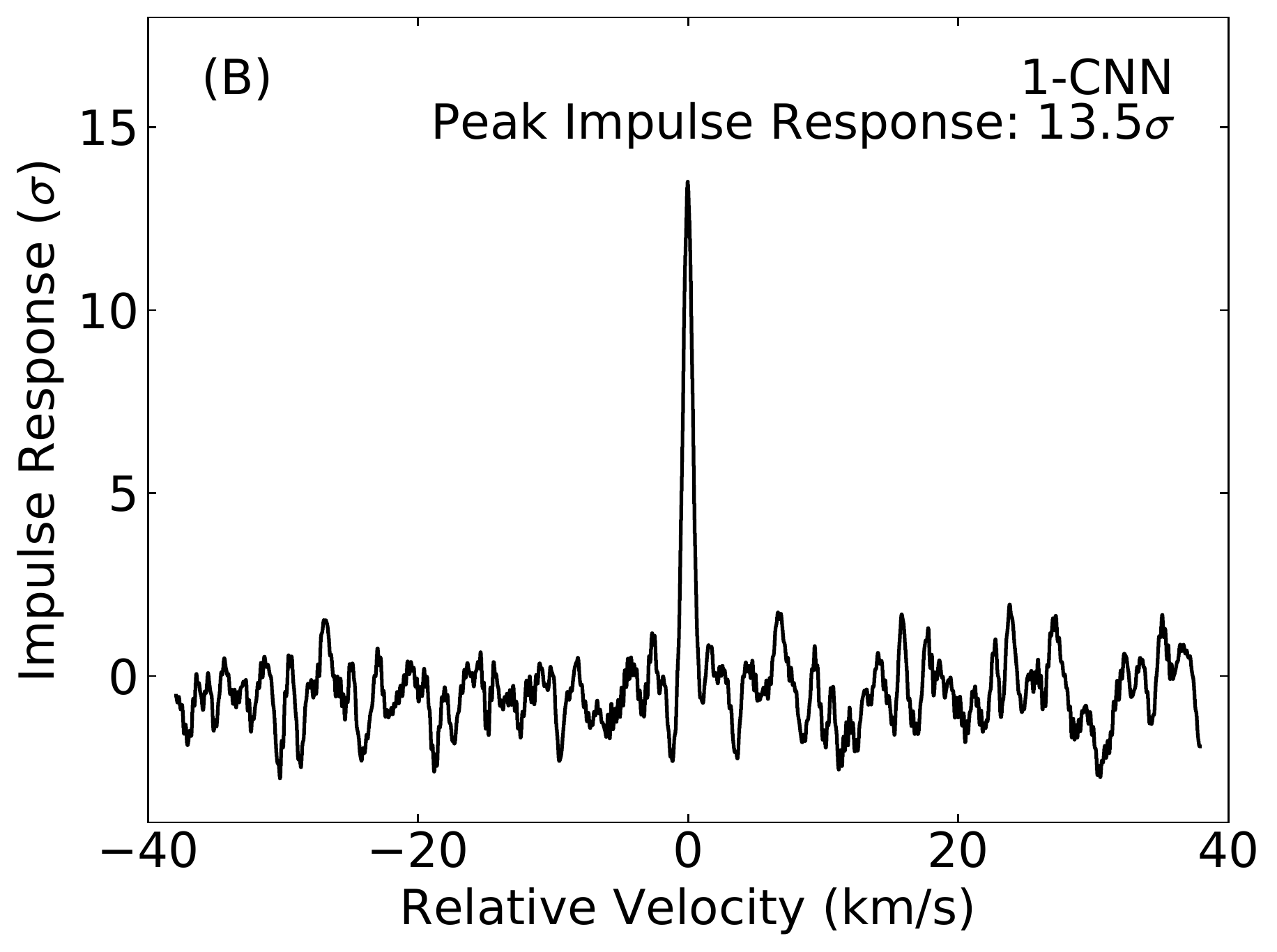}
    \includegraphics[width=0.49\textwidth]{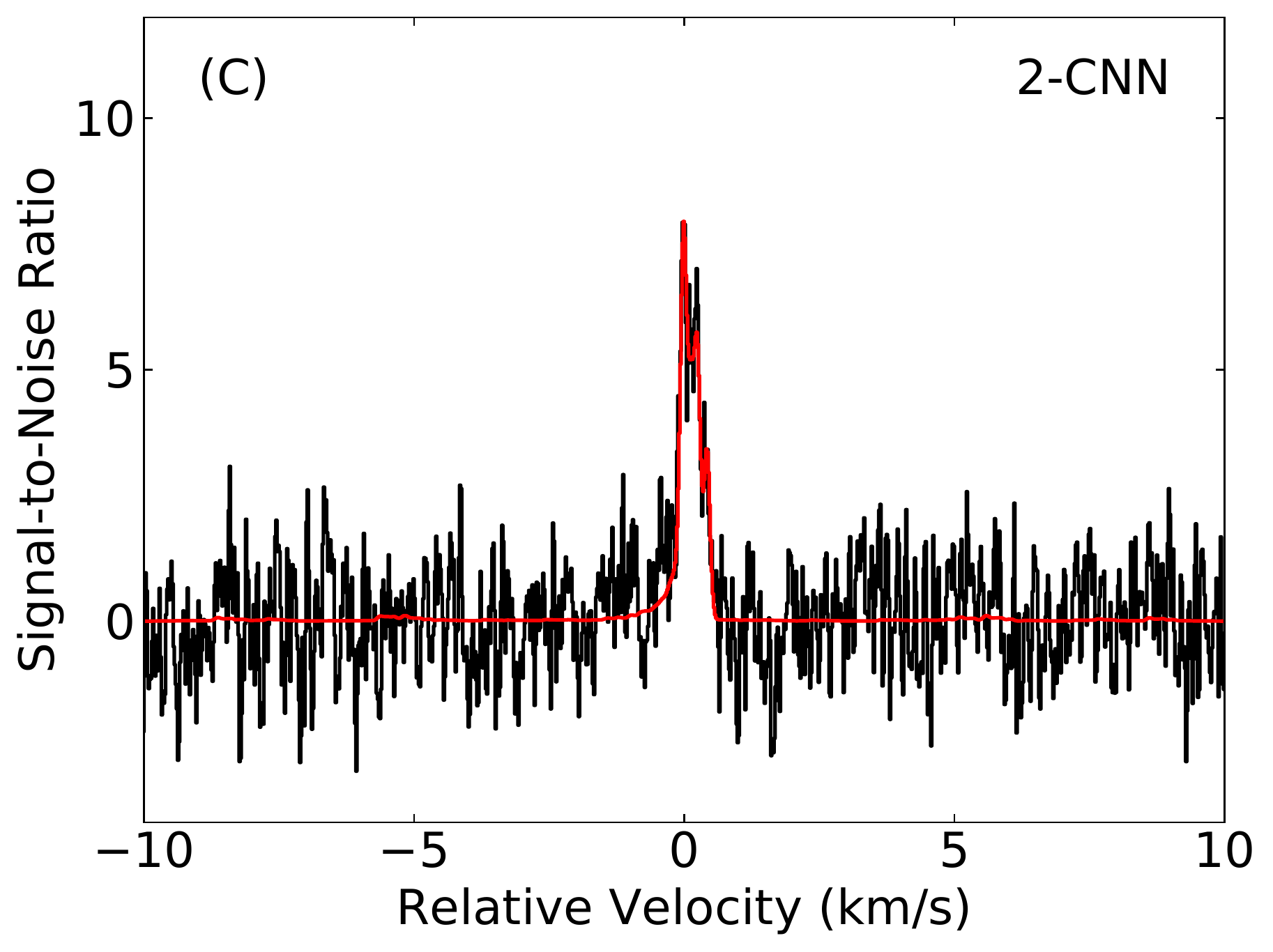}
    \includegraphics[width=0.49\textwidth]{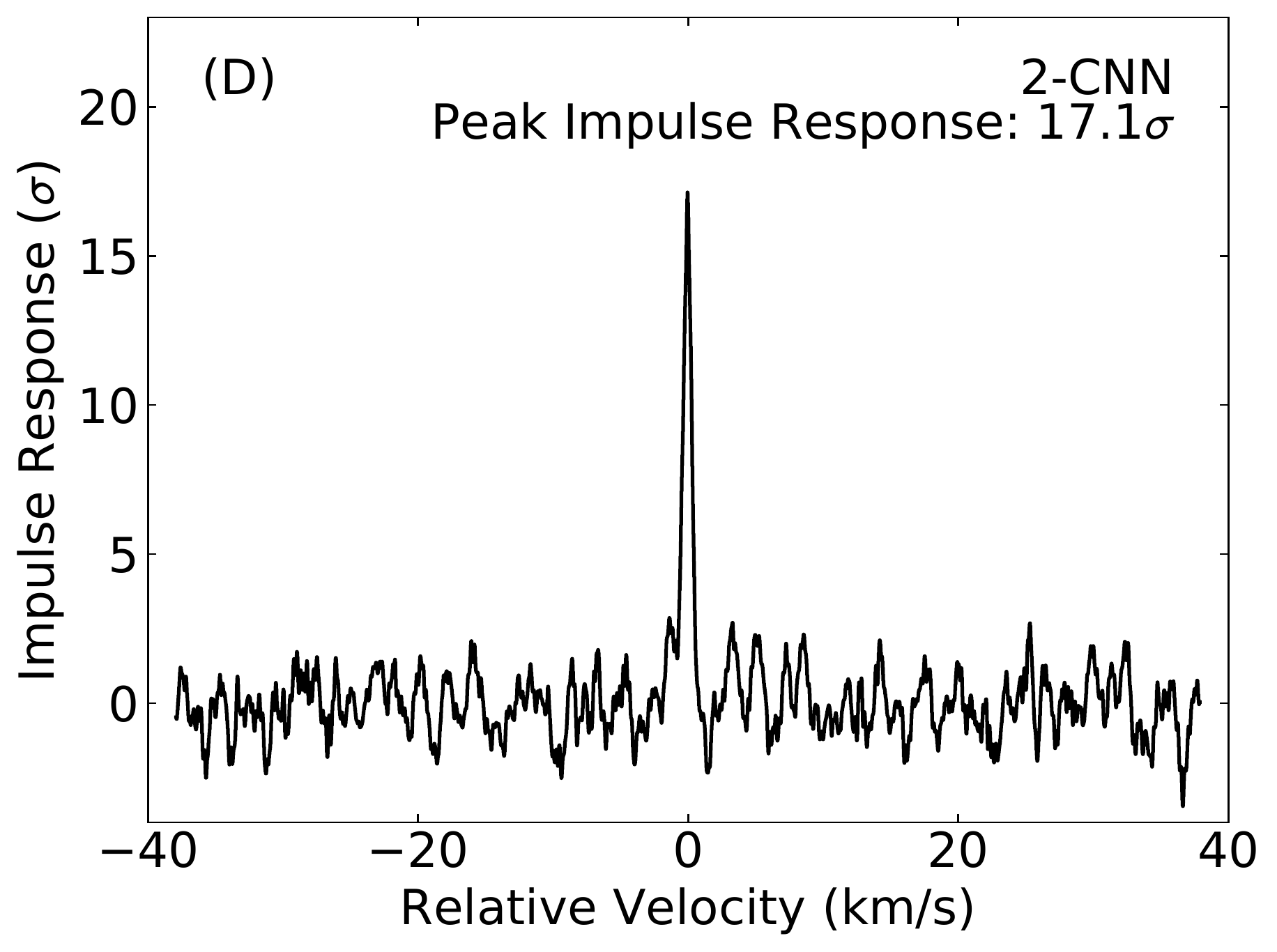}
    \caption{\textbf{Stacked Spectra and Impulse Responses for the Matched Filtering Analyses of 1-CNN and 2-CNN.} The stacked spectra of (A)  1- and (C) 2-CNN from the GOTHAM DR2 data in black, overlaid with the line profile in red from an MCMC analysis of the DR2 data.  The signal-to-noise ratio is shown on a per-channel basis. Impulse response functions of the stacked spectra of (B) 1-CNN and (C) 2-CNN using the simulated line profiles as matched filters. The peak of the impulse response functions provide a minimum significance for the detections of 13.5 and 17.1$\sigma$, respectively.}
    \label{cnn_stacks}
\end{figure*}

The details of this methodology, including analysis of the robustness, are presented elsewhere  (\citealt{Loomis:2021aa} and Appendix).  Briefly, we extract a small portion of the observations centered around the predicted frequency of each spectral line, discarding any windows with a spectral feature $>$5$\sigma$ to avoid interloping signals from other species. A signal-to-noise weighted average of the spectra was then calculated based on the expected intensity of the line (derived from the MCMC parameters) and the RMS noise of the observations, which we have verified contain no red-noise contamination (Fig.~\ref{fadein}).  The results are shown in  Fig.~\ref{cnn_stacks} (A and C) in units of the SNR.  To calculate the overall significance of any detections requires that we consider not only the peak SNR in the central channel, but the SNR of all channels with molecular signal.  To do this, the model spectra are also stacked using the same weights as used for the observations, and that stacked model is then used as a matched filter that is cross-correlated with the stacked observations.  The resulting impulse response spectrum provides a lower limit on the statistical significance (Fig.~\ref{cnn_stacks} B and D): 13.5$\sigma$ for 1-CNN and 17.1$\sigma$ for 2-CNN. We performed additional tests (see Appendix) to check the robustness of this methodology including jack-knife tests (Fig.~\ref{jacknives}) and checks for spurious detections (Figs. \ref{accuracy}, \ref{bn_errs}, \ref{injected_histos}, \ref{randos}, and \ref{rando_inj}). 

We conclude that both CNNs are detected in TMC-1.  The presence of these PAH molecules in the interstellar medium supports the hypothesis that PAHs are responsible for the unidentified infrared emission bands (UIRs) \citep{Tielens:2008fx}.   The UIR bands consist of numerous features at wavelengths characteristic of \ce{C-C} and \ce{C-H} stretching and bending motions of aromatic molecules consistent with PAHs. Although the infrared spectra of different PAH molecules are readily distinguishable in the laboratory at high spectral resolution \citep{Hudgins:1998gq}, the assignment of individual PAHs as carriers of the UIR features has not been possible because the differences in frequency are smaller than the width of the observed interstellar band profiles.  It is therefore likely that many different PAHs contribute to the UIR emission bands \citep{Tielens:2008fx}.  While the carriers of the UIRs must have substantial aromatic character, and PAHs are likely responsible for a sizable fraction of the overall emission, the specific structures and elemental compositions of the carriers remain a subject of debate \citep{Kwok:2011iv}. This has limited detailed analysis of sources that emit UIR bands \citep{Bauschlicher:2018ej}.  

We now consider the interstellar formation and destruction chemistry of 1-CNN and 2-CNN as individual molecules, rather than PAHs in aggregate.  Two scenarios have been proposed to explain the formation of PAHs: `top-down' and `bottom-up' formation chemistry.  In the top-down scenario, small interstellar carbon clusters, or carbon soot in the envelopes of evolved stars, are broken down by ultraviolet radiation to form a variety of PAHs \citep{Berne:2015fi}.  PAHs may also be formed on the surfaces of interstellar dust grains in the envelopes of evolved stars \citep{Martinez:2019ud}.  These PAHs then are distributed into the interstellar medium, including molecular clouds. In the bottom-up scenario, PAHs are built up \emph{in situ} in molecular clouds from smaller precursors through more common chemical evolutionary pathways such as gas-phase ion-molecule and neutral-neutral reactions and reactions occurring on grain surfaces \citep{Woods:2002gs,Jones:2011yc}.

Any population of PAHs inherited by TMC-1 from prior top-down formation must have survived in the diffuse ISM. However, PAHs comprised of less than $\sim$20--30 atoms cannot radiatively stabilize upon absorption of a UV photon, and are therefore destroyed in the diffuse ISM \citep{chabot_coulomb_220} The presence of small PAHs (the CNNs) in TMC-1 suggests at least some in situ, bottom-up formation.

Existing astrochemical reaction networks do not include detailed PAH formation and destruction chemistry \citep{Bettens:1996io,Bettens:1997iq}. We have extended a gas-grain chemical network \citep{shingledecker_cosmic-ray-driven_2018} to include reactions relevant to naphthalene, 1-CNN, and 2-CNN, along with reactions relevant to other detections from the GOTHAM survey \citep{Xue:2020aa,Loomis:2021aa,McCarthy:2020aa,McGuire:2020bb}. 

We included two major formation routes for naphthalene. The first is formation of \ce{C10H8} directly from the reaction of a phenyl radical with vinylacetylene: 
\begin{equation}
  \ce{C6H5 + CH2CHC2H -> C10H8 + H}.
  \label{napForm}
  \tag{R1}
\end{equation}
\noindent
This barrierless gas-phase reaction has been found to be viable under TMC-1-like conditions \citep{parker_low_2012}.
The second route involves a dihydronaphthalene (\ce{C10H10}) precursor that forms via the reaction between phenyl radical and 1,3-butadiene 
\begin{equation}
  \ce{C6H5 + CH2CHCHCH2 -> C10H10 + H}.
  \label{dialinForm}
  \tag{R2}
\end{equation}
\noindent
This reaction is also expected to occur in the ISM \citep{kaiser_pah_2012}. Successive abstraction of two hydrogen atoms from \ce{C10H10}, both of which have no energy barrier \citep{jensen_identification_2019}, yields \ce{C10H8}. For all species added to our network, we assume gas-phase depletion via both reaction with ions (with rate coefficients calculated using the Langevin formula) and adsorption onto grains. 

For 1- and 2-CNN, we added the formation routes
\begin{equation}
    \ce{C10H8 + CN -> 1-CNN + H}
    \label{cnnForm1}
    \tag{R3}
\end{equation}
\begin{equation}
    \ce{C10H8 + CN -> 2-CNN + H}
    \label{cnnForm2}
    \tag{R4}
\end{equation}
\noindent
assuming equal branching fractions. We assume reactions \eqref{cnnForm1} and \eqref{cnnForm2} occur on every collision, because analogous reactions between CN radicals and unsaturated hydrocarbons such as benzene are known to be both barrierless and exothermic \citep{Cooke:2020we}. Similar to naphthalene \citep{parker_low_2012}, we included an analogous process for 1-CNN involving the reaction of phenyl radicals (\ce{C6H5}) with cyanovinylacetylene (\ce{HC2CHCHCN}), which we assume occurs at the collisional rate. However, we find the contribution of this reaction to the 1-CNN abundance is small.

Our reaction network assumes that the chemistry of PAHs in molecular clouds such as TMC-1 is predominantly gas phase, because the rate of thermal desorption of CNN from dust grains at $\sim$10 K is negligible. Non-thermal desorption mechanisms might allow for the introduction of aromatic molecules into the gas from icy surfaces \citep{marchione_efficient_2016} but are not included in our calculations. 

\begin{figure}[t]
    \centering
    \includegraphics[width=0.49\textwidth]{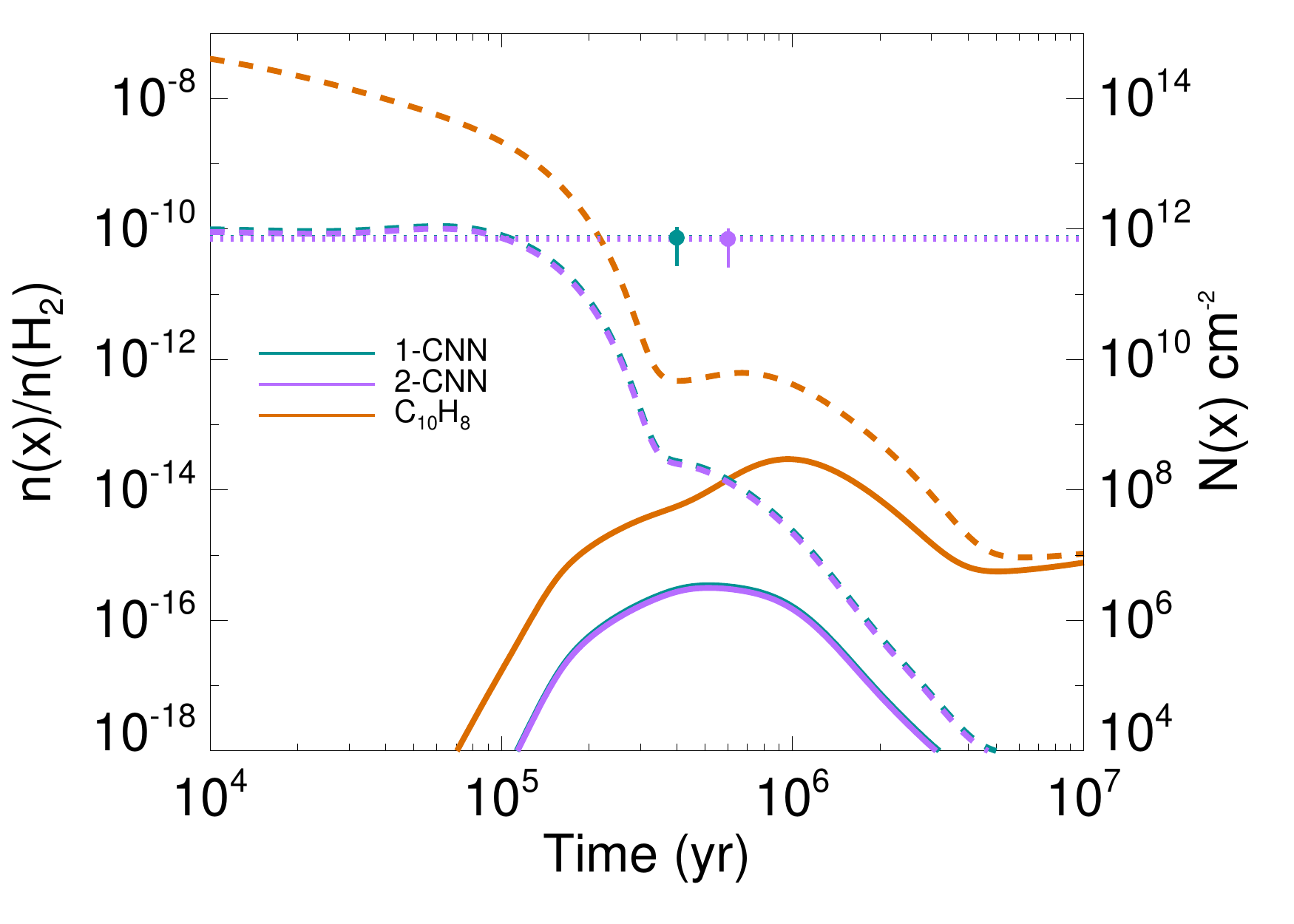}
    \caption{Calculated abundances of naphthalene (color), 1-CNN (color), and 2-CNN (color) under TMC-1 conditions in our fiducial model (solid lines) and a model which assumes an initial naphthalene abundance of n(\ce{C10H8})/n(\ce{H2})=$1.0\times10^{-7}$ (dashed lines), where n(\emph{x}) is the column density of molecule \emph{x}.  The right axis shows equivalent column densities assuming n(\ce{H2}) = $10^{22}$ cm$^{-2}$.  The circles are the values for these species as derived from our DR2 observations, with 1$\sigma$ error bars.  The dashed lines show these values extended over the range of the x-axis for display.}
    \label{fig:model}
\end{figure}

The results of our simulations are shown in Fig.~\ref{fig:model}.  In the fiducial model, the abundances of both 1- and 2-CNN are underpredicted by $\gtrsim6$ orders of magnitude. CN is abundant in the model, so naphthalene is the limiting reagent.  The large disparity between observations and predictions suggests: i) additional production pathways for naphthalene may be missing; ii) the efficiency of existing pathways is substantially underestimated; or iii) there is a non-negligible initial  abundance of naphthalene inherited from prior top-down chemistry. 

To test scenario iii), we performed a second calculation to ascertain how PAHs inherited from earlier stages of cloud evolution affect the abundances of 1- and 2-CNN under TMC-1 conditions.  For this model, we added an initial abundance of naphthalene (representing an inherited population) tuned to reproduce the observational data. We found agreement with the observations for a naphthalene abundance of $n(\ce{C10H8})/n(\ce{H2})=1.0\times10^{-7}$ at early times. This would correspond to roughly $1\%$ of the total carbon. However, the calculated abundances of 1- and 2-CNN again deviate from the observational values at times $>10^5$ years, the typical age for sources like TMC-1.

Our calculations neglect grain-surface formation pathways (see Appendix). Nevertheless, we regard the required initial abundance of naphthalene as unrealistically large, which in turn implies that a purely top-down formation pathway is also disfavoured. Conversely, the pathways we have considered for in situ bottom-up formation are also insufficient to reproduce observations. Other pathways must be required, such as ion-neutral reactions. Such processes are generally efficient under interstellar conditions and can lead to \ce{C10H9+} \citep{anicich_index_2003}, which might form naphthalene as a product of its dissociative recombination with electrons in a manner analogous to the benzene formation route from \ce{C6H7+} \citep{McEwan:1999ia}.

We conclude that two polycyclic aromatic hydrocarbons -- 1-cyanonaphthalene and 2-cyanonaphthalene -- are present in the molecular cloud  TMC-1.  We are unable to explain the derived abundances with either top-down or bottom-up scenarios, indicating that other formation routes may be required or existing routes may be more efficient than previously thought.

\section*{acknowledgements}

The National Radio Astronomy Observatory is a facility of the National Science Foundation operated under cooperative agreement by Associated Universities, Inc.  The Green Bank Observatory is a facility of the National Science Foundation operated under cooperative agreement by Associated Universities, Inc. \textbf{Funding:} B.A.M. was supported by NASA through Hubble Fellowship grant \#HST-HF2-51396 awarded by the Space Telescope Science Institute, which is operated by the Association of Universities for Research in Astronomy, Inc., for NASA, under contract NAS5-26555. A.M.B. acknowledges support from the Smithsonian Institution as a Submillimeter Array (SMA) Fellow. M.C.M and K.L.K.L. acknowledge financial support from NSF grants AST-1908576, AST-1615847, and NASA grant 80NSSC18K0396. C.N.S. thanks the Alexander von Humboldt Stiftung/Foundation for their generous support, as well as V. Wakelam for use of the \texttt{NAUTILUS} v1.1 code. I.R.C. acknowledges funding from the European Union’s Horizon 2020 research and innovation programme under the Marie Skłodowska-Curie grant agreement No 845165-MIRAGE. S.B.C. and M.A.C. were supported by the NASA Astrobiology Institute through the Goddard Center for Astrobiology. E.H. thanks the National Science Foundation for support through grant AST 1906489.   C.X. is a Grote Reber Fellow, and support for this work was provided by the NSF through the Grote Reber Fellowship Program administered by Associated Universities, Inc./National Radio Astronomy Observatory and the Virginia Space Grant Consortium.  \textbf{Author Contributions:} Conceptualization: B.A.M., A.M.B., A.J.R., S.K.; Methodology: B.A.M., A.M.B., K.L.K.L., R.A.L.; Software: R.A.L., K.L.K.L., B.A.M.; Data Collection: B.A.M., R.A.L., A.M.B., K.L.K.L., C.X., M.A.S., A.J.R., I.R.C.; Modeling: C.N.S., A.M.B., C.X., E.R.W., S.B.C., E.H.; Writing: B.A.M., A.M.B., R.A.L., C.N.S., K.L.K.L., M.C.M.; Review \& Editing: All authors. \textbf{Competing Interests:} We declare no competing interests. \textbf{Data and Materials Availability:}  GBT and VLA data are available in the NRAO archive (\url{https://archive.nrao.edu/archive/advquery.jsp}) under project codes GBT17A-164, GBT17A-434, GBT18A-333, GBT18B-007, GBT19A-047, and TCAL0003. Observational data windowed around these transitions, the full catalogs including spectroscopic properties of each transition, and the partition function used in the MCMC analysis are provided in the Harvard Dataverse repository \citep{GOTHAMDR2}.  Information on obtaining the \textsc{nautilus} code is available at \url{http://perso.astrophy.u-bordeaux.fr/~vwakelam/Nautilus.html}.  Input and output files for the astrochemical models are available in a second Harvard Dataverse Repository \citep{Models}.

%\bibliography{bibliography,shingledecker,dataverse,misc,science}

%\bibliographystyle{aasjournal}

\clearpage

\appendix

\renewcommand{\thefigure}{A\arabic{figure}}
\renewcommand{\thetable}{A\arabic{table}}
\renewcommand{\theequation}{A\arabic{equation}}
\setcounter{figure}{0}
\setcounter{table}{0}
\setcounter{equation}{0}

\section{Observations and Data Reduction}

The GOTHAM project, overall observing strategy, and detailed data reduction procedures are described in detail in Ref.~\citep{McGuire:2020bb} for the DR1 data.  Here, we provide an overview of DR1, as well as describing the additional observations for DR2.  

\subsection{Project Codes and Observing Procedure}

Details of the observing procedures are found elsewhere (\citealt{McGuire:2020bb}, their section 3), and only summarized here. The observations contained in DR2 were carried out between February 2018 and June 2020 on the Robert C. Byrd 100-m Green Bank Telescope in Green Bank, West Virginia under project codes GBT18A-333, GBT18B-007, and GBT19A-047. The target was at right ascension 04$^h$41$^m$42.50$^s$, declination +25$^{\circ}$41$^{\prime}$26.8$^{\prime\prime}$, a region in TMC-1 known as the ``cyanopolyyne peak."  PKS 0528+134 (J0530+1331) was the calibrator source for pointing and focus observations, which were performed at the beginning of each observing session, and subsequently every 1--2~hours, depending on the weather, and typically converged to $\lesssim$5$^{\prime\prime}$.  Observations were conducted using position switching (ON-OFF), in which the target was observed for 2\,min and the off position (1$^{\circ}$ off target) was then observed for 2\,min.  The off position was inspected and found to be clear of emission.  In addition we incorporated archival data from project GBT17A-164 and GBT17A-434.  The detailed observing strategy for these archival data is outlined in Ref.~\citep{McGuire:2018it}, but is largely identical to that used for the GOTHAM data.  The archival data were re-calibrated and re-reduced uniformly with the new GOTHAM observations, to ensure consistency.  

\subsection{Spectral Configurations}

\begin{figure}[tbh!]
    \centering
    \includegraphics[width=\textwidth]{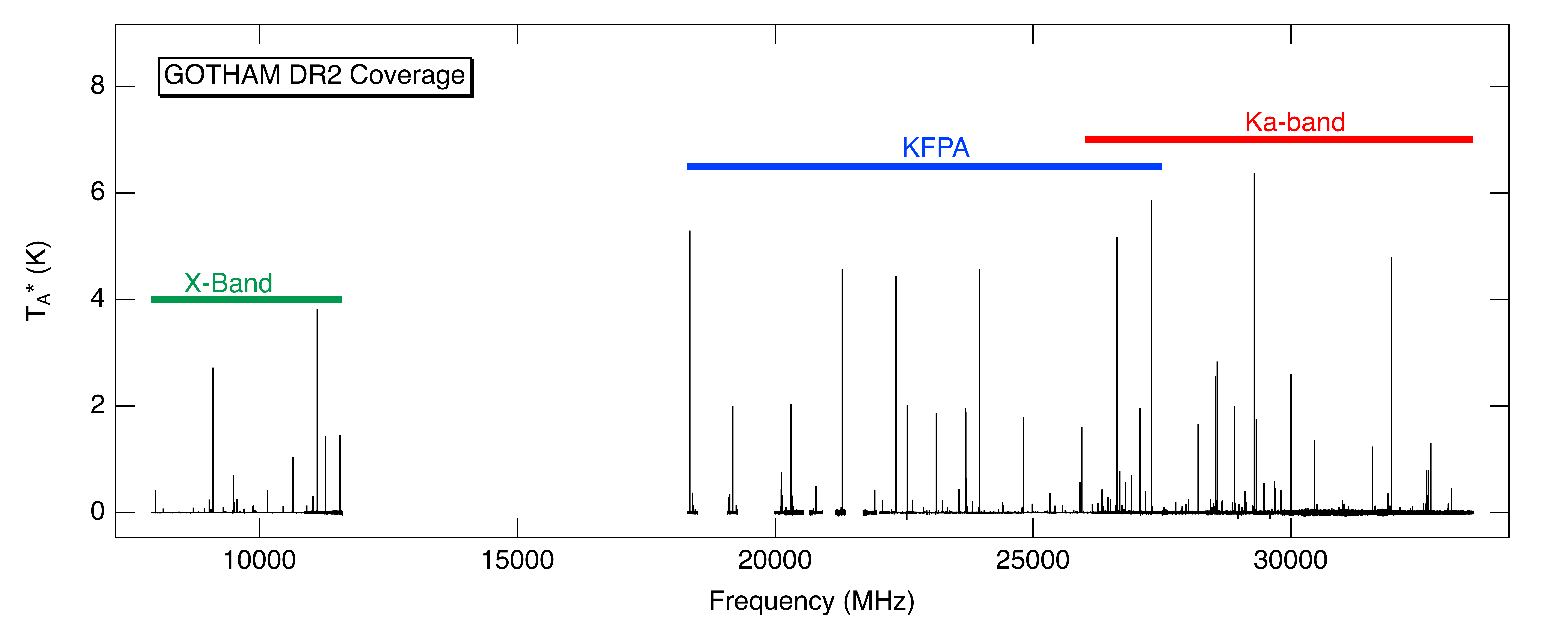}
    \caption{\textbf{Spectral Coverage of GOTHAM DR2.}  The DR2 coverage of the GOTHAM survey in black.  The receivers used in each frequency range are shown as colored bars above the spectra.}
    \label{gotham_coverage}
\end{figure}

Details of the spectral setups are found elsewhere (\citealt{McGuire:2020bb}, their section 3.1), and only summarized here.  The DR2 coverage of the GOTHAM survey is shown in Fig.~\ref{gotham_coverage}.

All spectra were obtained using the VErsatile GBT Astronomical Spectrometer (VEGAS) \citep{Roshi:2012he}.  The receivers used and the associated back-end parameters are summarized in Table~\ref{rx}.

\begin{table}[tbh!]
    \scriptsize
    \caption{\textbf{Summary of Receiver and Modes Used.}  Bandwidth and channels listed are on a per-spectral-window basis. Resolution and half-power beamwidth (HPBW) calculated at a central frequency of the receiver and will vary at the top and bottom ends of the range \citep{McGuire:2020bb}. }
    \centering
    \begin{tabular}{c c c c c c c c c}
    \toprule
    Band    &  Frequency    &    \# Windows &   Bandwidth   &   Channels    & \# Polarizations    &   \multicolumn{2}{c}{Resolution}  &   Half-Power Beam Width \\
            &  (GHz)        &               &   (MHz)       &               &                       &   (kHz)       &   (km\,s$^{-1}$)  &   ($^{\prime\prime}$)  \\
    \midrule
    X       &  8.0--11.6    &   8           &   187.5       &   131,072     &   2   &   1.4         &   0.05  &   80  \\
    K-Band  &  18.0--27.5   &   8           &   187.5       &   131,072     &   2   &   1.4         &   0.02  &     33  \\
    Ka-Band &  26.0--39.5   &   4           &   187.5       &   131,072     &   1   &   1.4         &   0.015 &     27    \\
    \bottomrule
    \end{tabular}
    
    \label{rx}
\end{table}

\subsection{Calibration}

Details of the calibration are found elsewhere (\citealt{McGuire:2020bb}, their section 3), and only summarized here. All three receivers were primarily calibrated using an internal noise diode, nominally providing an uncertainty of at best, $\sim$30\%.  The noise diode in the X-band receiver was calibrated in 2018 and referenced to Karl G. Jansky Very Large Array (VLA) flux density measurements \citep{gbtcal}, and is therefore assumed to be better than 30\%. We sought to improve both the absolute flux calibration of the KFPA and Ka-band measurements, and to ensure relative agreement between the two (and with X-band).  The pointing source, J0530+1331, is bright enough to use as a flux calibrator but has shown long-term variability (as well as short-term variability of order $\sim$20\%; \citep{Gorshkov:2016hy}).  We therefore obtained VLA flux density measurements of this source (project TCAL0003) and have used these to calibrate the KFPA data.  We then use the $J = 3 \rightarrow 2$ transition ($J$ is the principle rotational quantum number) of \ce{HCC^{13}CN} at 27,181~MHz that is detected by both receivers to bring the Ka-receiver measurements into the same calibration as the KFPA. Due to the short-term variability of J0530+1331 we assume the calibration is still accurate only to $\sim$20\%.

\subsection{Data Reduction}

\begin{figure}[htb!]
    \centering
    \includegraphics[height=2.5in]{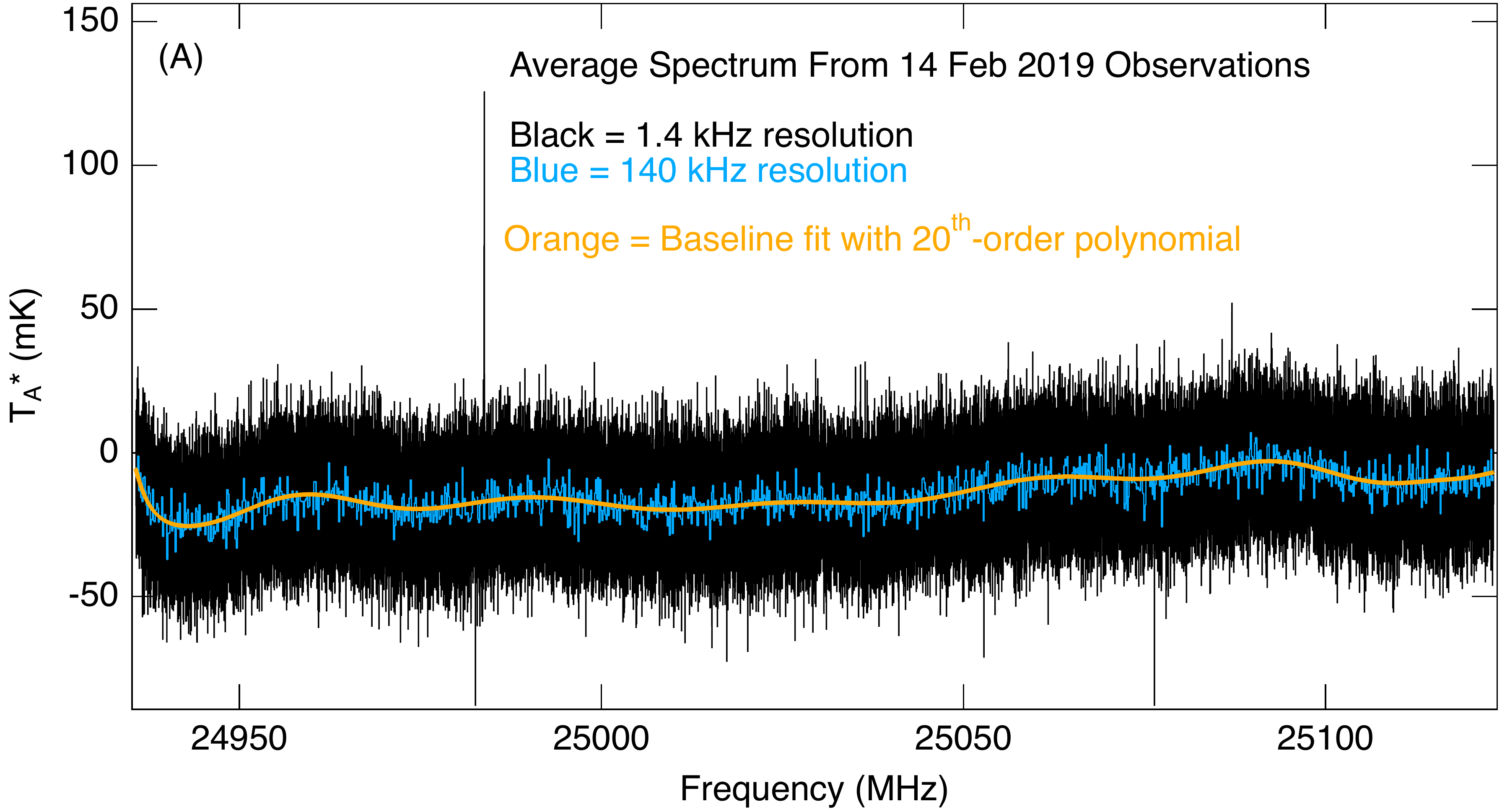}
    \includegraphics[height=2.5in]{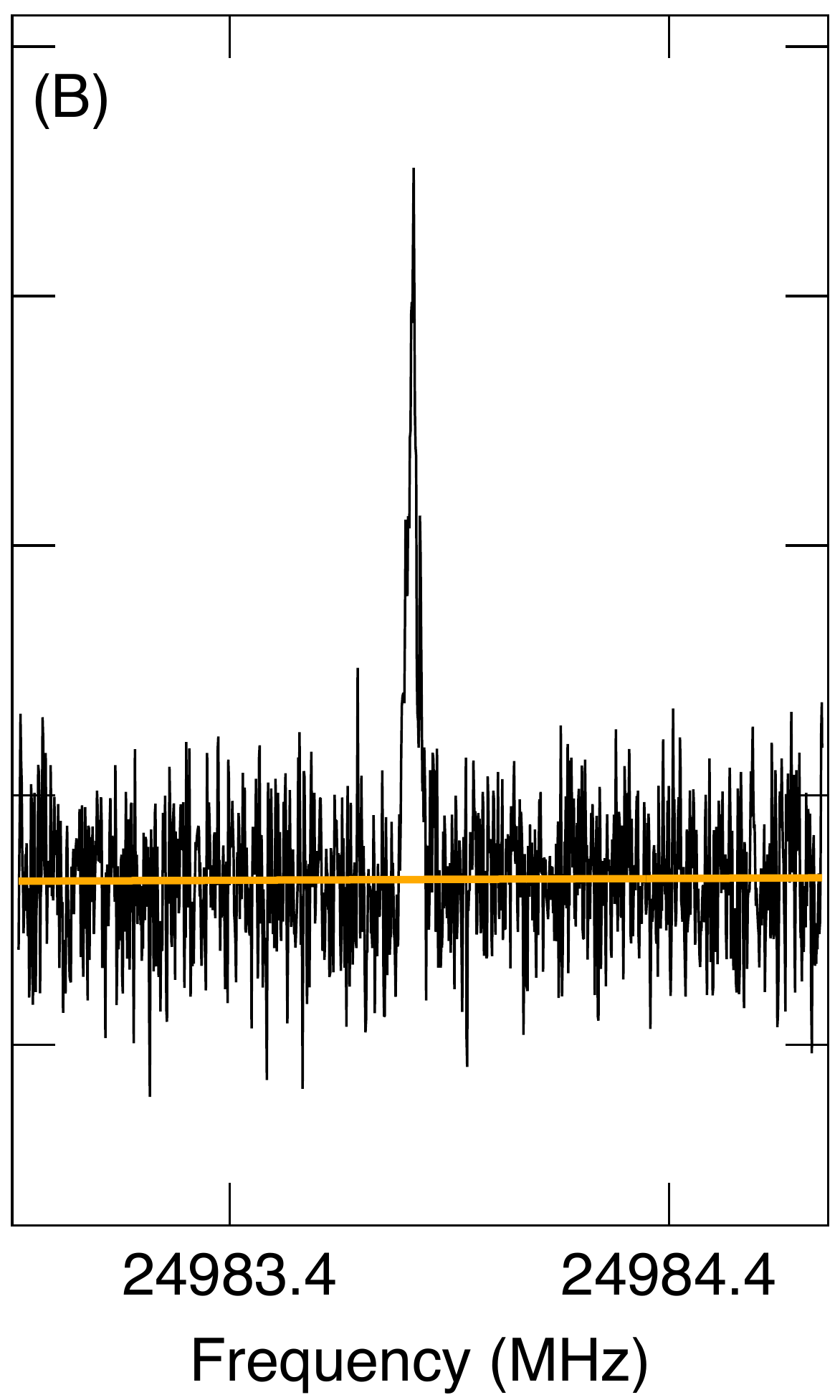}
    \caption{\textbf{Example of a 20$^{th}$-order baseline subtracted from the GOTHAM data.} (A) the averaged (full-resolution) spectrum acquired from observations on 14 Feb 2019 is shown in black.  In blue, the observations have been re-sampled to a resolution of 140 kHz (100$\times$ coarser) to show the ripple in the bandpass.  The baseline model obtained using \texttt{gbtidl} using a 20$^{th}$-order polynomial is overlaid in orange.  Lower-order polynomials were unable to remove the broad ripple, whereas higher-order polynomials were not needed.  (B) a portion of the full-resolution data is shown in black over the baseline model in orange.}
    \label{baselines}
\end{figure}

Details of the data reduction procedure are found elsewhere (\citealt{McGuire:2020bb}, their section 4), and only summarized here.  Initial data processing and calibration was performed using \textsc{gbtidl} \citep{gbtidl}. Each ON-OFF position switched scan pair was corrected for Doppler Tracking, calibrated to the internal noise diodes, and then placed on the atmosphere-corrected $T_A$* intensity scale \citep{Ulich:1976yt}.  When available (for X-band and the KFPA), both recorded polarizations were averaged together to improve the signal-to-noise ratio.  The spectra were then inspected and manually cleaned of radio-frequency interference (RFI) and artifacts. RFI below the noise level of the observations is discussed below. Baselines were removed with a polynomial fit of order appropriate to the baseline ripple observed, which typically ranged from 1--20 (see Fig.~\ref{baselines}).  As with any baseline fitting, some error in the absolute offset from $y=0$ is possible. The extremely narrow linewidths in TMC-1 relative to the broad baselines (bandpass:linewidth ratio of between 15000:1 and 59000:1 across the observations) that are being fit and removed would make the introduction of narrow features, which might mimic molecular signal, exceptionally unlikely even using substantially higher-order polynomials.  The primary sources of bandpass ripple are temperature-dependent variations in contributions from the receiver itself, the intermediate frequency system, and digital filters within the VEGAS spectrometer.  In all cases, the continuum model was inspected to ensure that the line profiles, which are much narrower than the baseline model variations, were not affected. 

A noise-weighted average was performed to arrive at the final spectrum.  When convolved to a uniform velocity resolution of 0.05~km~s$^{-1}$ across the spectrum (corresponding to the lowest-resolution data at X-band), the RMS noise level varies from $\sim$2--20~mK, dependent entirely upon the integration time at each frequency. We make use of the full 1.4~kHz resolution when possible.

\section{MCMC, Stacking, and Matched Filtering Analysis Results}

Both techniques were developed in the signal processing and radar communities \citep{Woodward:1953vn,North:1963kl}, but have been adopted by the astrophysics and astrostatistics communities over the last decade as a method for detecting weak spectral signals -- often below the local RMS noise -- and quantifying their significance in interstellar clouds \citep{Langston:2007bm}, protoplanetary disks \citep{Loomis:2020io}, comets \citep{Biver:2015fx}, and (exo)planetary atmospheres \citep{Birkby:2013ko}.

\subsection{Spectral Line Coverage and Parameters}

\begin{figure}[tbh!]
    \centering
    \includegraphics[width=\textwidth]{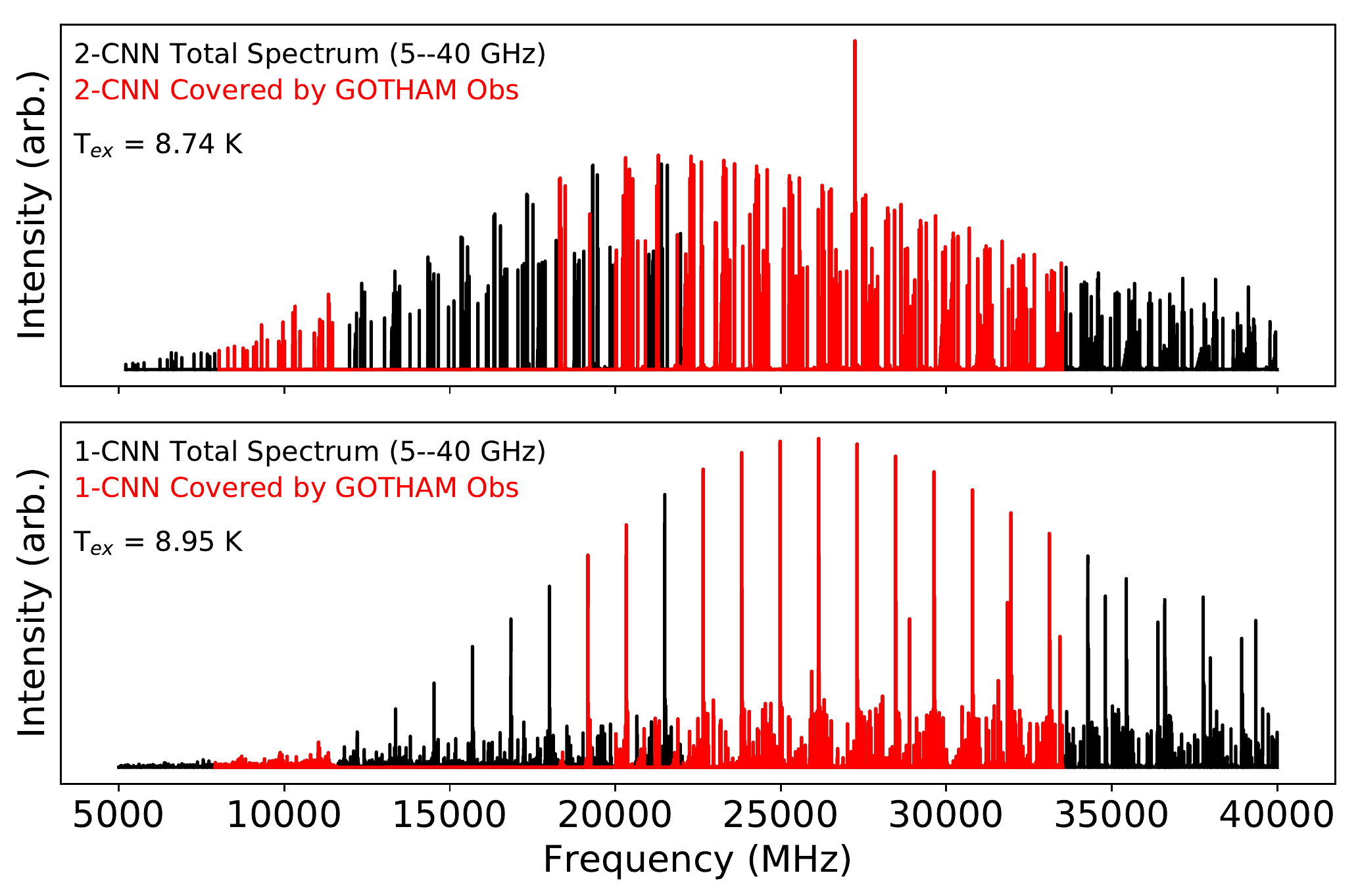}
    \caption{\textbf{Coverage of CNN spectra in GOTHAM DR2 data}.  Simulations of the spectra of 1-CNN (bottom) and 2-CNN (top) between 5--40 GHz at the derived excitation temperatures in the GOTHAM data, and accounting for beam dilution effects.  The lines which are covered by GOTHAM data are shown in red.  The line of 2-CNN near 27\,GHz is particularly intense due to a fortuitous overlap of individual transitions at that frequency.  Detection of that individual signal at 5$\sigma$ with the GBT would take $\sim$350--400 hours of dedicated integration time.}
    \label{cnns_coverage}
\end{figure}

The specific transition parameters for each species are obtained by generating spectral line catalogs directly from laboratory spectroscopy of benzonitrile \citep{McGuire:2018it} and the CNNs \citep{McNaughton:2018op}. Table~\ref{app:lines} shows the total number of transitions (including hyperfine components) of the molecules analyzed in this paper that were covered by the DR2 observations and were above our predicted flux threshold of 5\% of the peak intensity \citep{Loomis:2021aa}.  Also included are the number of transitions, if any, that were coincident with interfering transitions of other species, and the total number of lines used after excluding interlopers.  Observational data windowed around these transitions, the full catalogs including spectroscopic properties of each transition, and the partition function used in the MCMC analysis are provided in the Harvard Dataverse repository \citep{GOTHAMDR2}.  Fig.~\ref{cnns_coverage} visually demonstrates the coverage of the GOTHAM data in relation to the wider spectra of the CNNs.  

\begin{table*}[bht!]
    \centering
    \caption{Total number of transitions of a given species within the range of the GOTHAM data, number of interfering lines, and total number included in MCMC fit.}
    \begin{tabular}{l c c c}
    \toprule
                &   Transitions Covered &   Interfering Lines   &   Total Transitions   \\
    Molecule    &   By GOTHAM           &      In Data         &       Used in MCMC        \\
    \midrule
    1-CNN &   1508                &   6                   &   1502    \\
    2-CNN &   957                &   0                   &   957    \\
    benzonitrile &   255             &   0                   &   255     \\
    \bottomrule     
    \end{tabular}
    \label{app:lines}
\end{table*}

\subsubsection{Likelihood of Interfering Transitions}

We estimate the approximate number of interfering transitions we should expect to have excluded in this analysis based on the line density of the data.  The density of lines $>$5$\sigma$ above the local RMS noise in the GOTHAM data is $\sim$1 per $\sim$240\,km\,s$^{-1}$ \citep{McGuire:2020bb}.  Although the individual velocity components of transitions have linewidths of $\sim$0.1\,km\,s$^{-1}$, it is the aggregate linewidth of signals in our spectra ($\sim$0.3\,km\,s$^{-1}$) that determine chance coincidences.  Thus, we expect to find an interfering $>$5$\sigma$ transition within a full-width at half-maximum (FWHM) of our target lines approximately 1 out of every 800 lines.  As shown in Table~\ref{app:lines}, this is close to the rate at which rejected lines from the MCMC analysis ($\sim$1 every 500 lines).  A similar trend holds for the spectral stacking and matched filtering, which includes all lines (no initial intensity threshold applied), but as a consequence rejects a larger number that contain interfering emission. 

\subsection{Robustness Tests}

Analysis of the robustness of our analysis is described elsewhere both specifically in the context of this work \citep{Loomis:2021aa}, and in other spectroscopic applications (e.g. \citep{Loomis:2018bt,Cumming:2005dd,Abbott:2016ki,Ruffio:2017ev}).  Here, we provide a number of additional tests of the technique's robustness against false positives and spurious interfering signals hidden under the noise.

\subsubsection{Jack-knife Tests}

\begin{figure}[tbh!]
\centering
\includegraphics[width=0.40\textwidth]{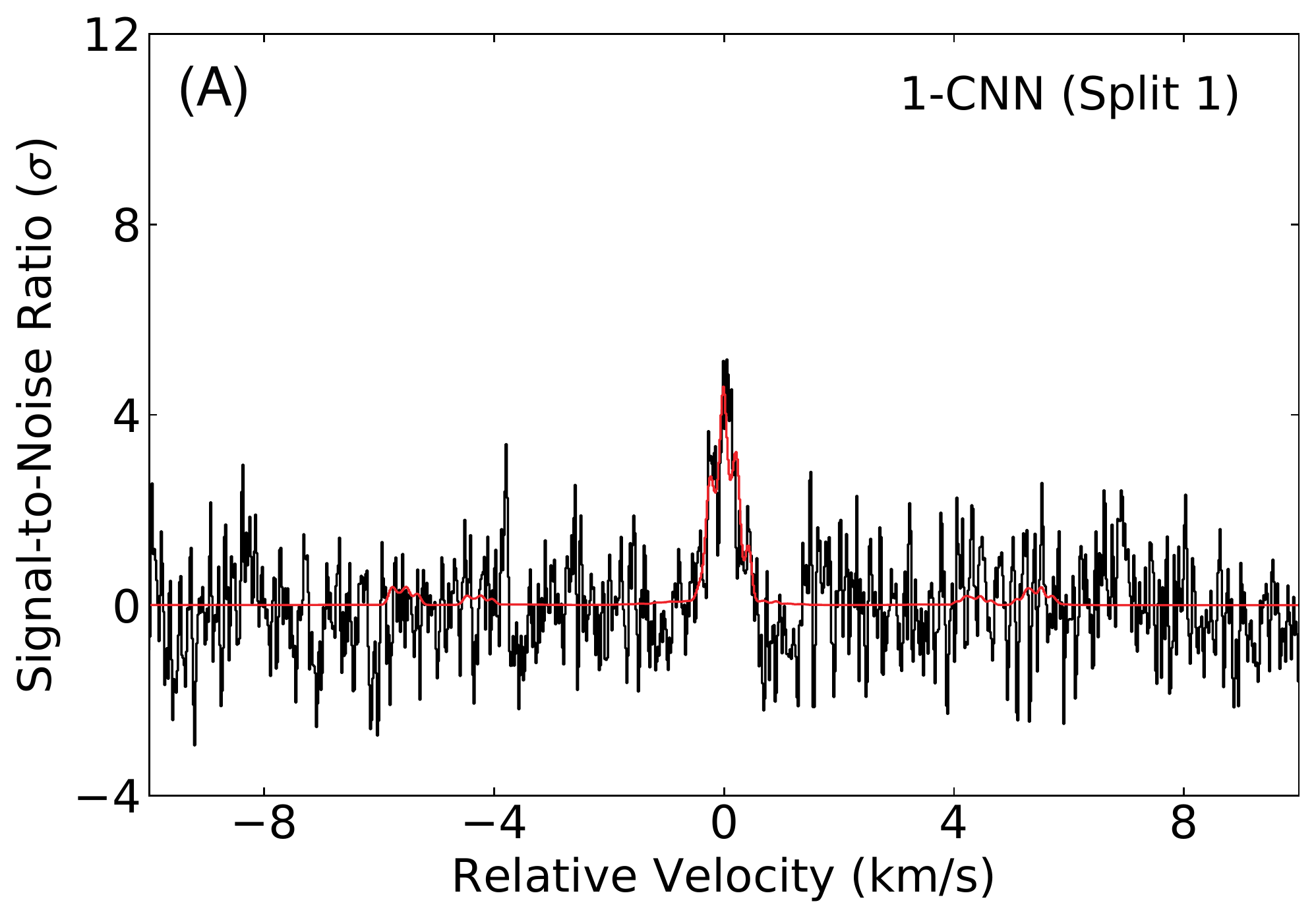}
\includegraphics[width=0.40\textwidth]{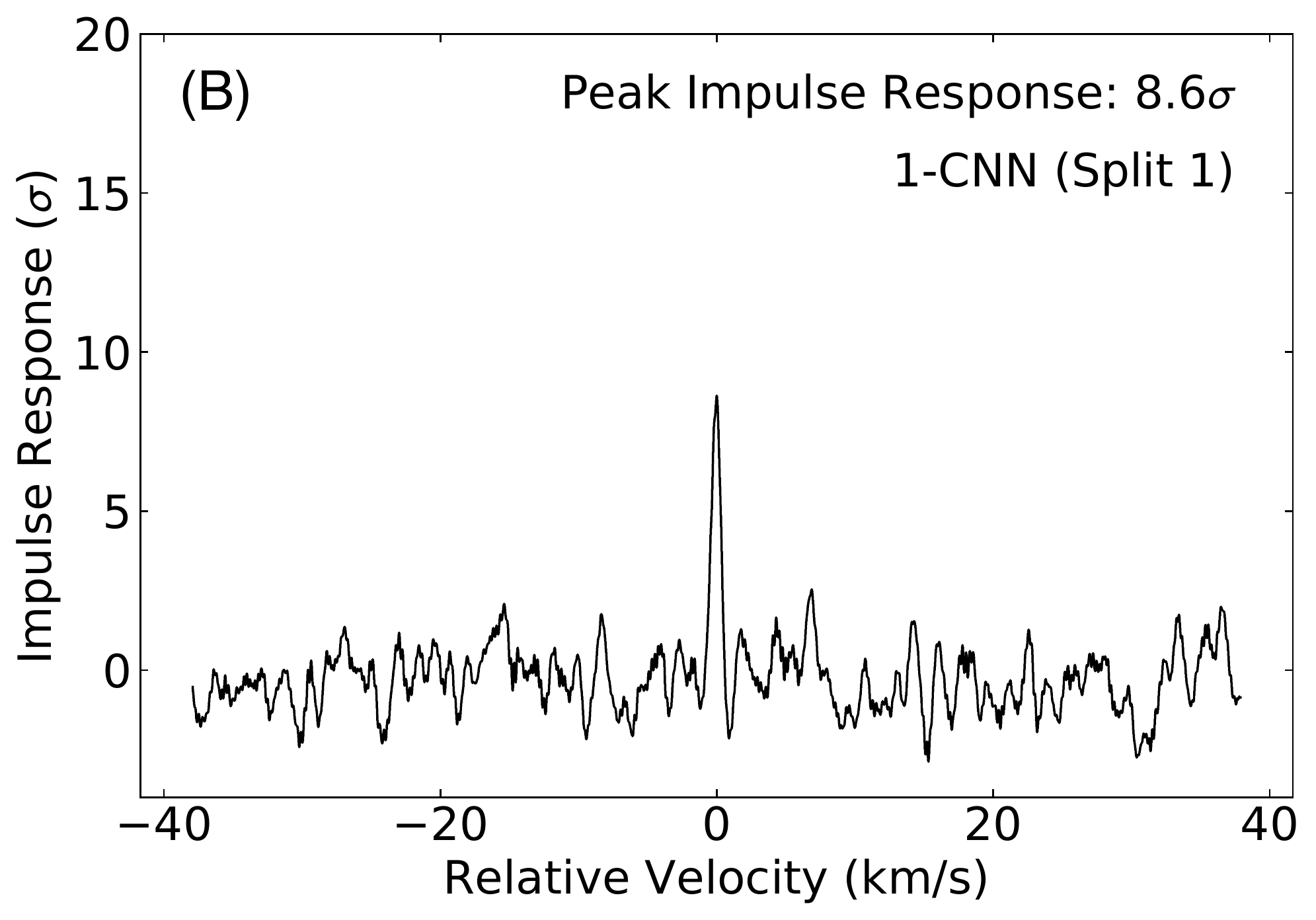}
\includegraphics[width=0.40\textwidth]{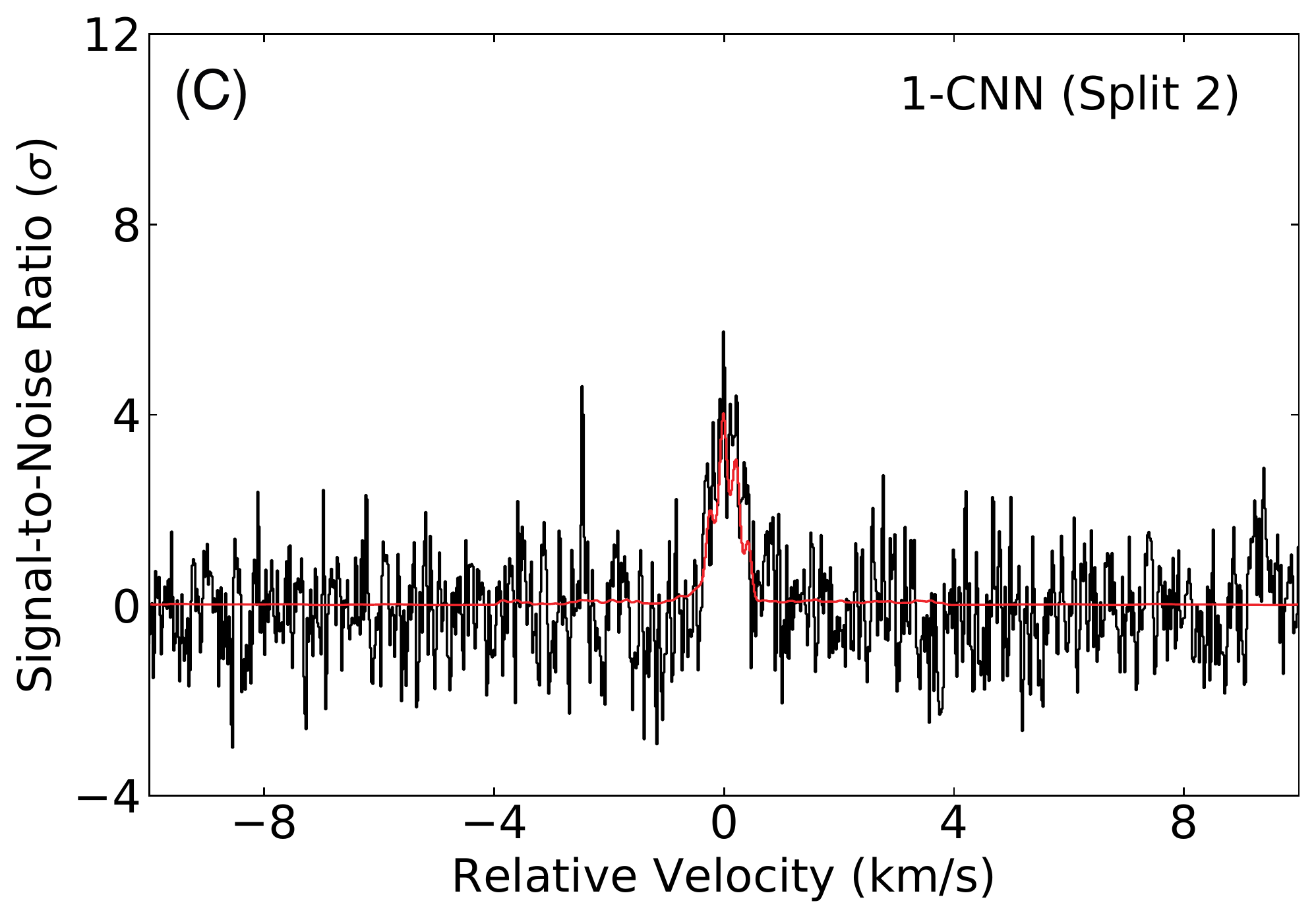}
\includegraphics[width=0.40\textwidth]{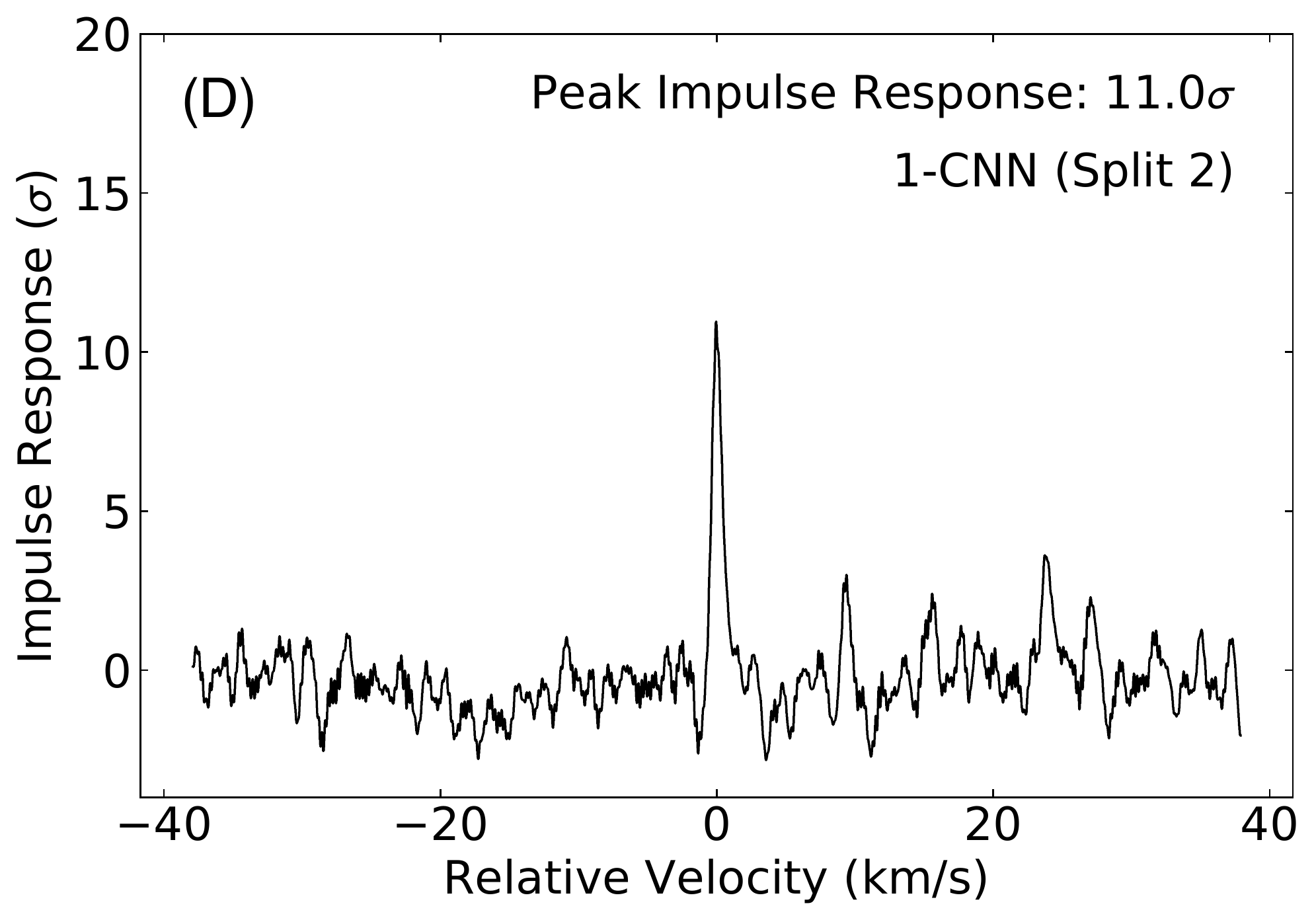}\\
\vspace{1em}
\includegraphics[width=0.40\textwidth]{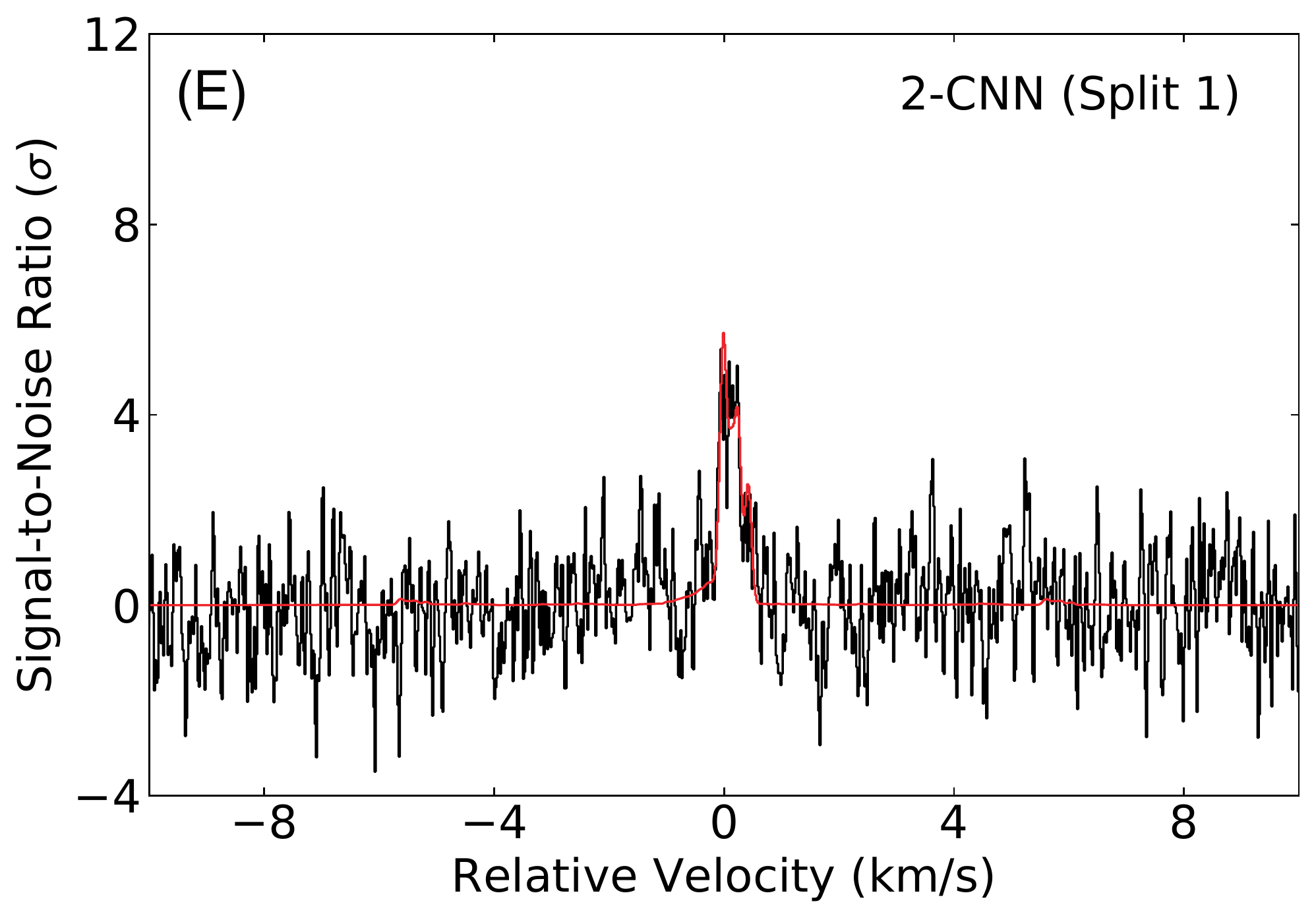}
\includegraphics[width=0.40\textwidth]{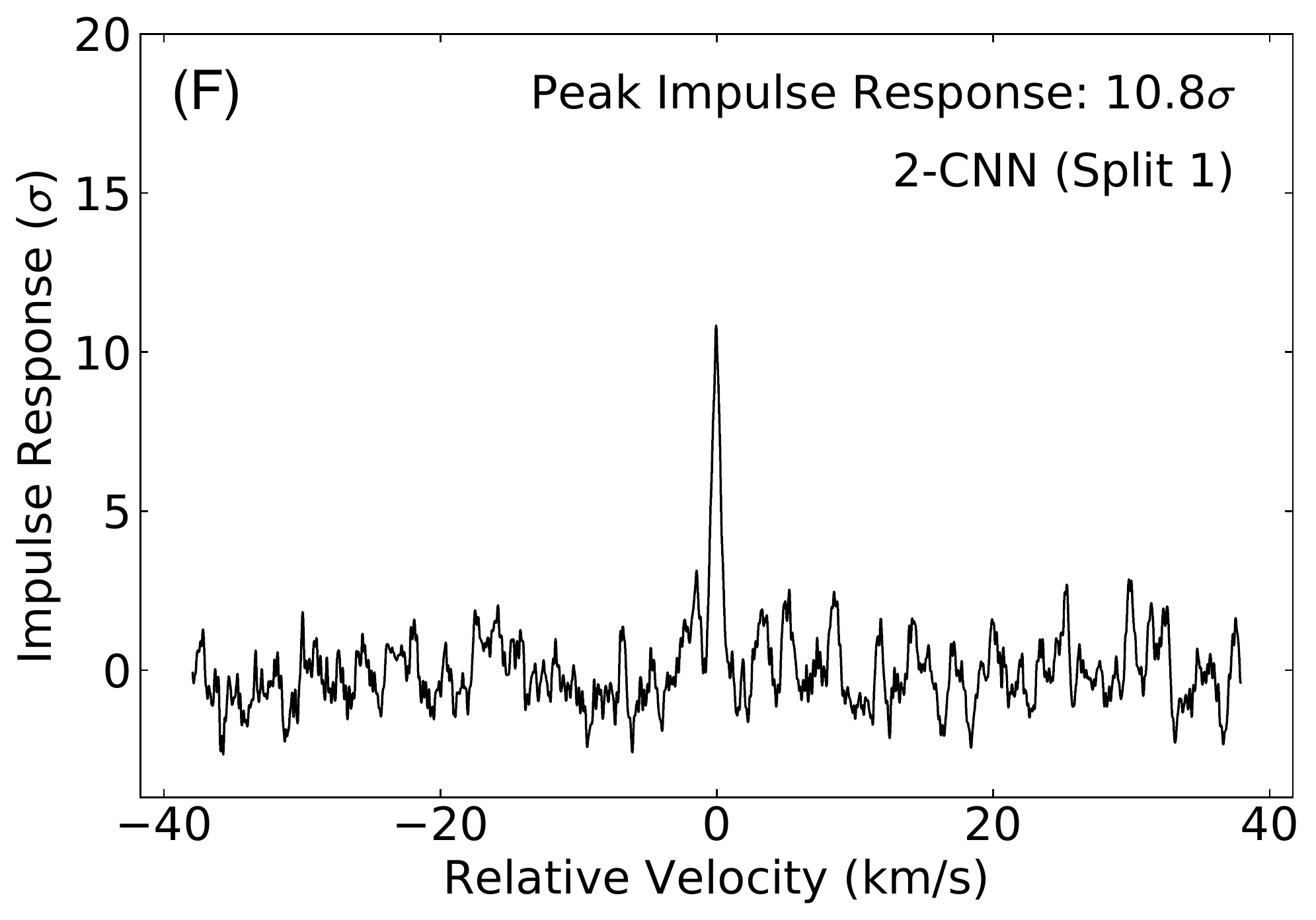}
\includegraphics[width=0.40\textwidth]{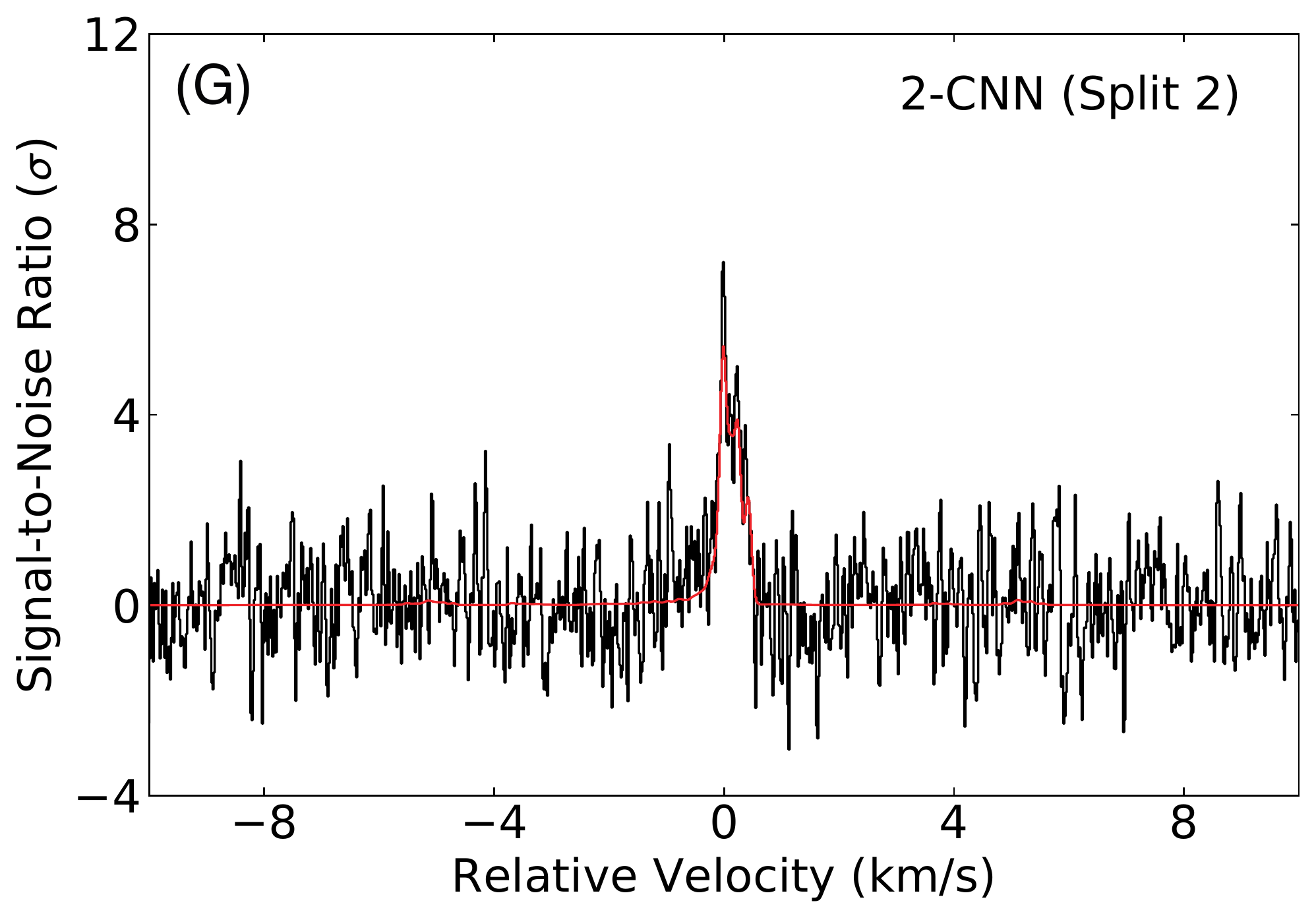}
\includegraphics[width=0.40\textwidth]{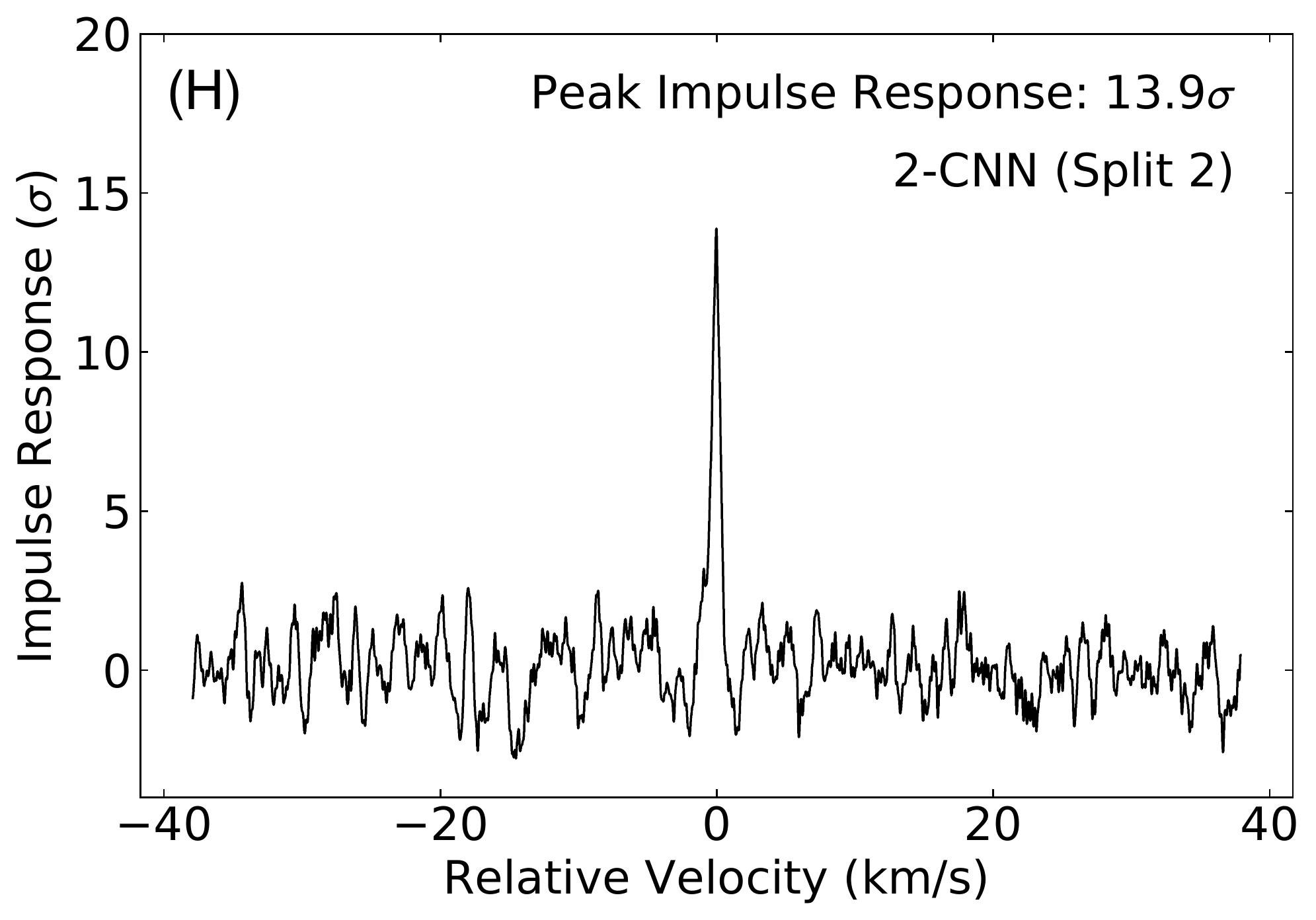}
\caption{\textbf{Results of jack knife tests for 1-CNN and 2-CNN}. Same as Fig.~4, but showing the jack knife results for 1-CNN and 2-CNN.}
\label{jacknives}
\end{figure}

We perform a jack-knife test of the data.  By splitting the catalog of lines to be stacked for a molecule into two parts and perform identical stacking and matched filtering analyses on each half.  If the detected signal is not from the molecule of interest, then we expect the two halves of the data to produce different results. If the signal is real, each half of the data is expected to have roughly $\frac{1}{\sqrt{2}} \times$ the filter response of the full dataset - it will not be exact, as the catalog halves are not selected to produce identical total intensity.  Instead, we select every other line (taken here as an observable line which may be a collection of hyperfine components).  Figures~\ref{jacknives} shows jack-knifing tests performed for the CNNs.  

Adding the responses of each molecule in quadrature yields values of 13.9$\sigma$ and 17.6$\sigma$ for 1-CNN and 2-CNN, respectively.  This is nearly identical to the full datasets for 1-CNN (13.5$\sigma$) and 2-CNN (17.1$\sigma$).  Because of the close density of lines, from both $K-$components of the transitions and hyperfine parameters, it is challenging to generate two completely independent catalogs for these molecules. As a result, some small amount of intensity bleed over from lines assigned to catalog A will nevertheless be close enough in frequency to lines in catalog B that they will blend and lend intensity.  The effect, however, is small.

\subsubsection{Accuracy of Rotational Constants}

We perform additional tests to determine whether it is likely to recover false-positive emission signal from the spectra using these techniques. False-positives become less likely to occur as the number of lines used in the stacking analysis increases.  This is because while undetected red-noise might potentially occur in one of the windows, that particular window is increasingly down-weighted in the ultimate average as the total number of lines used increases.  Only in the case where emission is present (but below the noise) in every window at precisely the predicted frequency used for extraction will a signal emerge in the average.

We test whether a molecule is only detectable if we know the frequencies of its transitions to high precision.  Deviations from the true frequency would lead to the emission being no longer coherently combined and instead noise is added to the accumulated data, thus reducing the signal-to-noise ratio.  We test this effect here using benzonitrile and the CNNs.  

\begin{figure}[tbh!]
    \centering
    \includegraphics[width=0.75\textwidth]{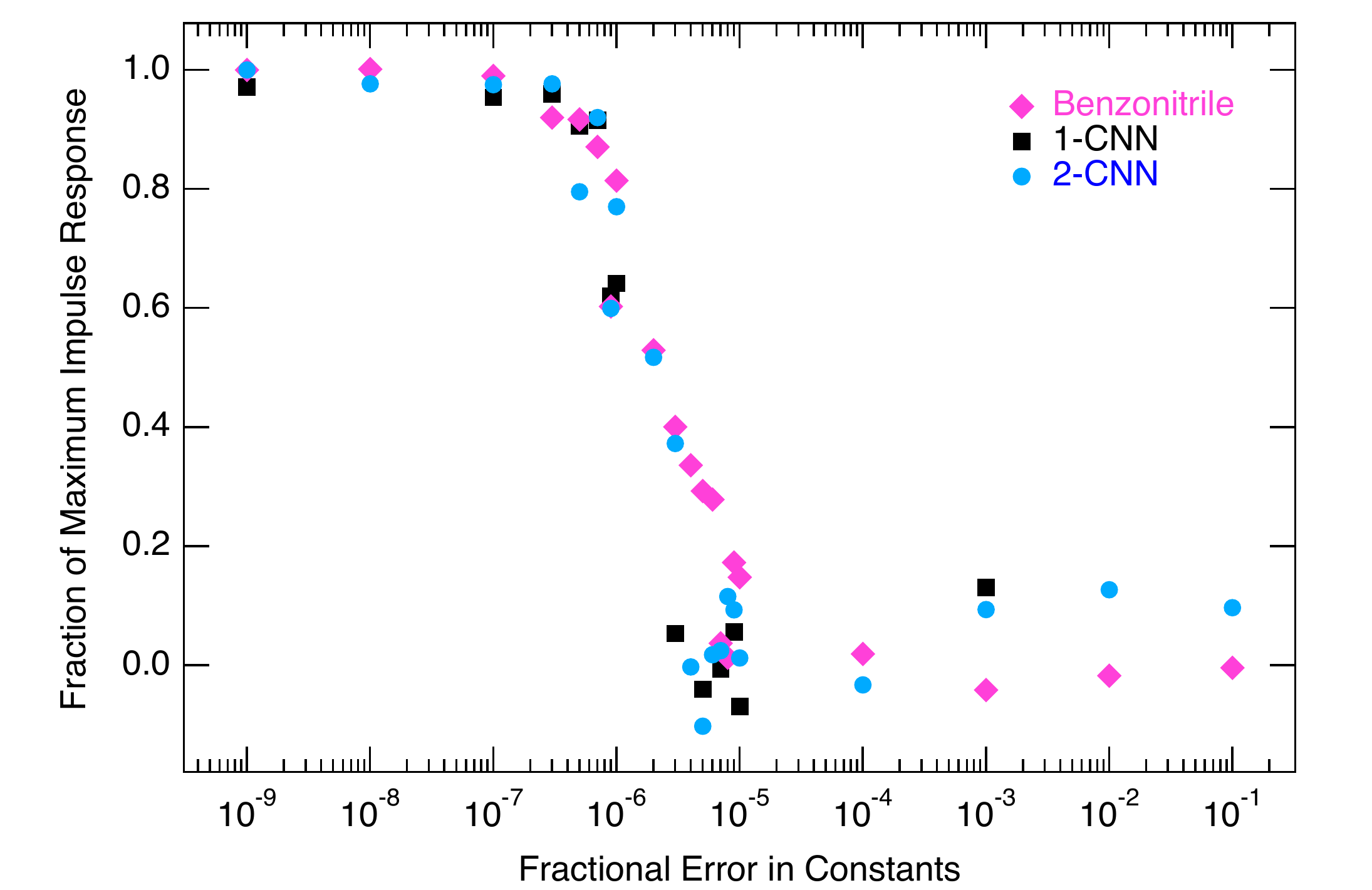}
    \caption{Fractional modification to rotational constants plotted versus normalized matched filter response for three species -- benzonitrile (fuscia), 1-CNN (black), and 2-CNN (light blue).}
    \label{accuracy}
\end{figure}

First, we introduce a random error of a given magnitude into the rotational constants used to predict the transition frequencies.  Then, we perform the stack and matched filtering analysis using the same parameters derived from the MCMC model fitting of the true catalog and determine how much of the impulse response is recovered.  This is shown in Figure~\ref{accuracy}.

The accuracy of the constants derived from laboratory measurements is $\sim$5$\times10^{-8}$ \citep{McNaughton:2018op}, and this is where we recover the maximum impulse response.  Below this the magnitude of the response drops rapidly.  An accuracy of $\sim$100\,ppb recovers most of the signal, and $\sim$10\,ppb recovers all of it.  Assuming a few hundred transitions are measured in the laboratory to obtain these constants, individual line frequencies need to be known to $\sim$1\,kHz accuracy for this technique to produce detectable signal in our data.  By design, this is approximately the resolution of the GBT observations (1.4\,kHz).  Thus, if the center of each extraction window is not precisely (within about 1\,kHz) of the correct frequency, the signal begins to broaden and incorporate extra noise.  We conclude that the rotational constants are known with sufficient accuracy and any molecule with different constants could not explain the signal we detect.

\subsubsection{Accuracy of Individual Line Frequencies}

\begin{figure}[tbh!]
    \centering
    \includegraphics[width=0.9\textwidth]{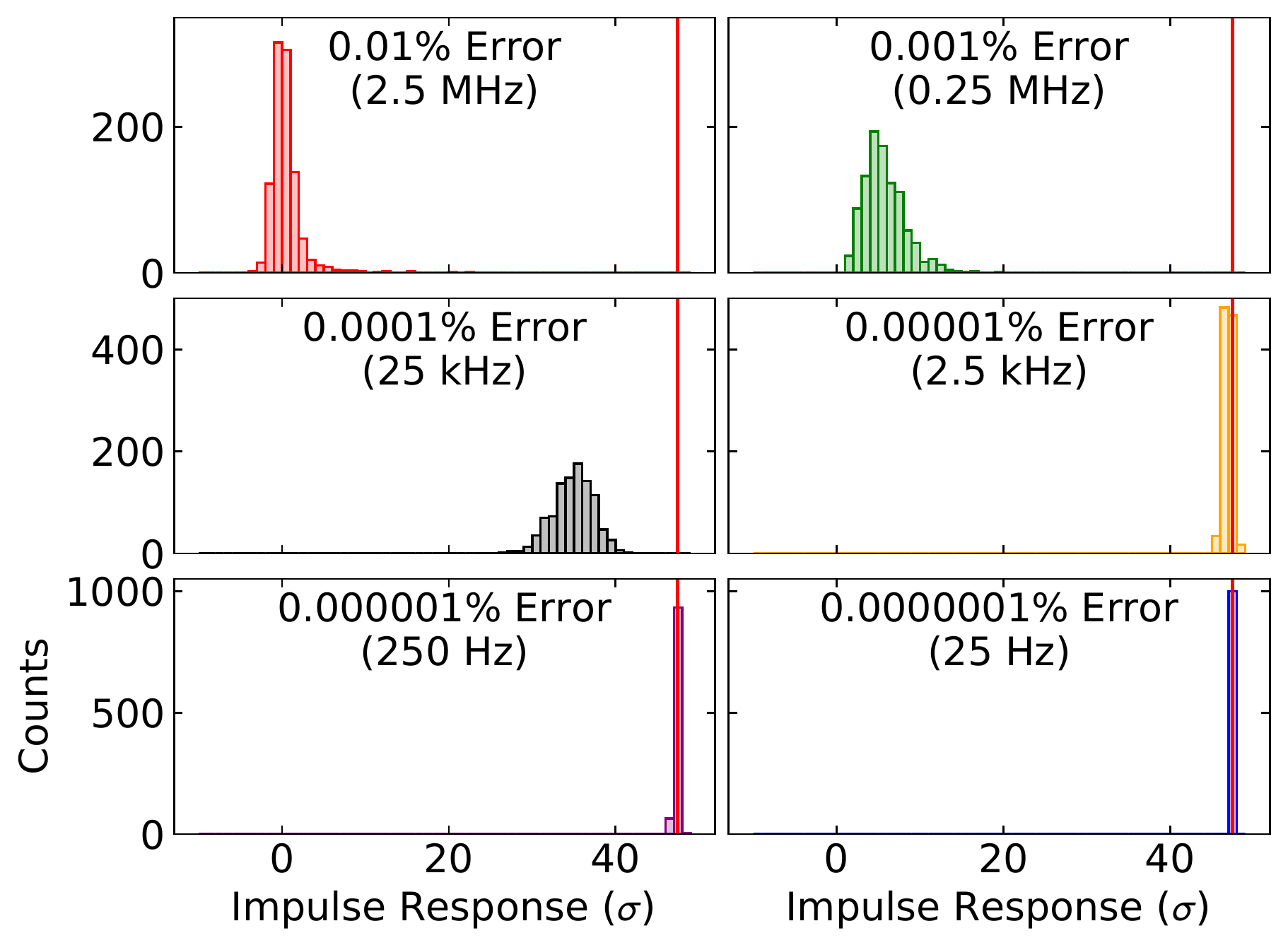}
    \caption{\textbf{Results of Introducing Errors into the Benzonitrile Frequencies}.  Histograms of the impulse response function obtained from 1000 stacks and matched filtering analyses using a benzonitrile catalog with errors of the given percentages applied to the line frequencies.  The red vertical line represents the impulse response recovered with the unaltered catalog.}
    \label{bn_errs}
\end{figure}

We performed two tests to determine the required accuracy of line frequencies.  First, we have taken the benzonitrile catalog and applied a random error to each line frequency drawn from a normal distribution with a given standard deviation.  Then, a spectral stack and matched filter analysis is performed and the resulting impulse response function is recorded.  This process was repeated 1000 times for errors from 0.01\% to 0.0000001\% in order of magnitude steps.  The results are shown in Fig.~\ref{bn_errs}.  The impulse response of the unaltered catalog is not recovered until an accuracy of 0.0001\% (corresponding to 2.5 kHz) or better is achieved.

\begin{figure}[tbh!]
    \centering
    \includegraphics[width=0.9\textwidth]{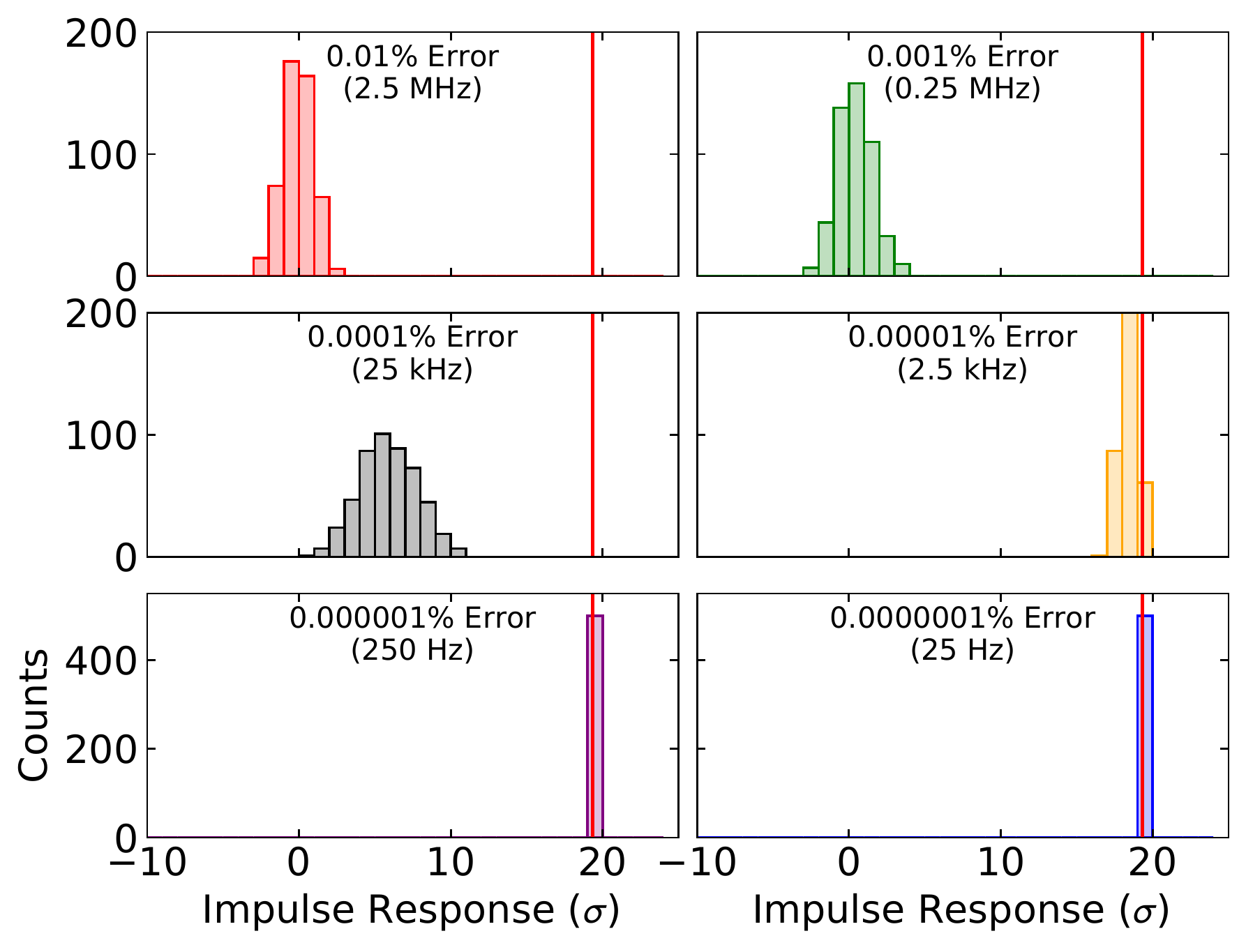}
    \caption{\textbf{Results of Introducing Errors into a Catalog For an Injected Signal}.  Histograms of the impulse response function obtained from 1000 stacks and matched filtering analyses using a fictitious molecule that has been injected into the GOTHAM data and with errors of the given percentages applied to its frequencies.  The red vertical line represents the impulse response recovered with the unaltered catalog.}
    \label{injected_histos}
\end{figure}

Second, because there is still inherent error in the underlying catalog of benzonitrile used as our fiducial model, we have generated an entirely fictitious molecule and injected that signal into our data.  In this way, we know the true maximum impulse response that can be obtained.  We then repeated the same analysis as for benzonitrile, using 500 trials, with the results shown in Fig.~\ref{injected_histos}.  This confirms that high accuracy transition frequencies are needed: the signal is not reliably recovered until 250 Hz accuracy is achieved.  

\subsubsection{Tests for Spurious Stacking Results}

\begin{figure}[tbh!]
    \centering
    \includegraphics[width=0.9\textwidth]{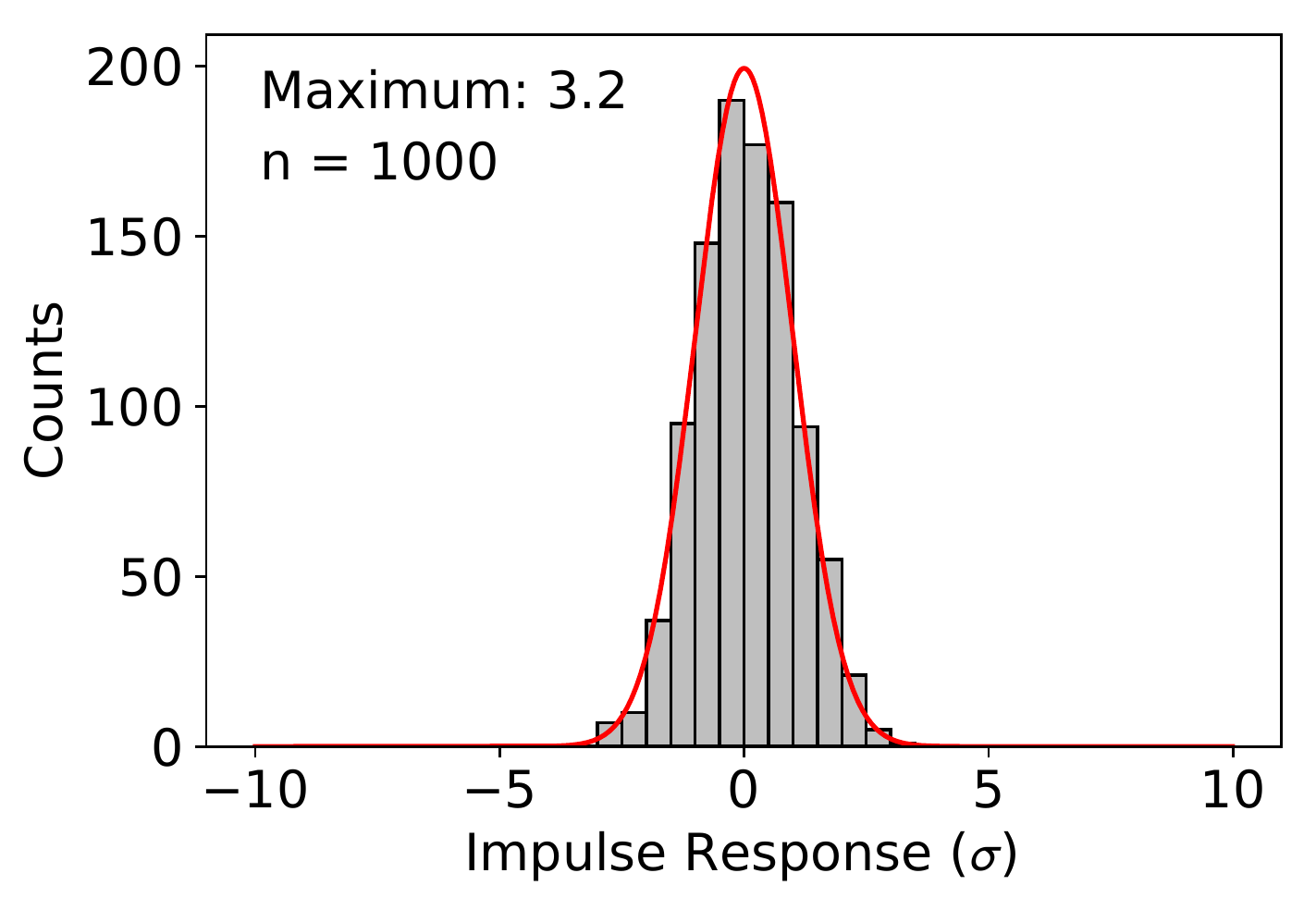}
    \caption{\textbf{Histogram of Impulse Response from Random Catalogs}.  Histogram of the impulse response function recovered from stacking and applying a matched filter to 1000 randomly generated catalog files in grey.  The maximum impulse response obtained was 3.2$\sigma$.  A simulation of the expected profile of the response assuming the noise being stacked was white is overlaid in red and matches the results.}
    \label{randos}
\end{figure}

The previous three analyses determine the required accuracy needed for rotational constants and transition frequencies to return a signal, and thus it is correspondingly extremely unlikely that a random set of frequencies could return a significant impulse response.  To further test this, we generated 1000 entirely random catalogs with random frequencies falling within the range of the GOTHAM data.  We then stacked these catalogs, performed a matched filtering analysis using the same catalogs, and recorded the impulse response function.  The results are shown in Fig.~\ref{randos}.  None of the random catalogs generated a significant ($>$5$\sigma$) response, and the distribution of responses obtained matched the expectations for averaging white noise.

\begin{figure}[tbh!]
    \centering
    \includegraphics[width=0.9\textwidth]{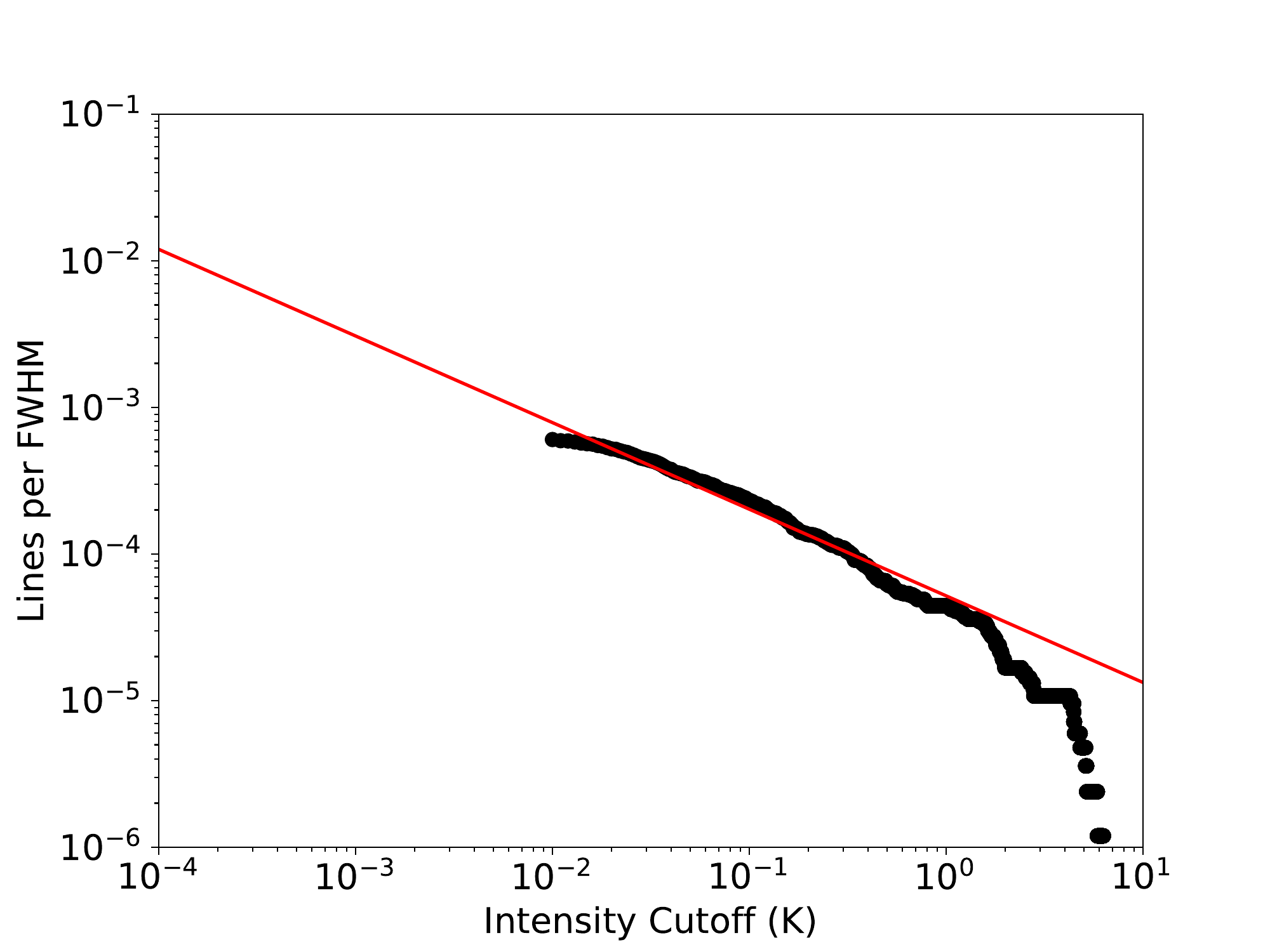}
    \caption{\textbf{Analysis of the Line Density in DR2}. A peak-fitting analysis of the DR2 data was performed to determine the number of lines below any given intensity threshold.  In black, that data is plotted in the form of the number of lines per average FWHM of a spectral feature ($\sim$20\,kHz) versus a given threshold intensity.  Overlaid on this plot in red is a power law model fitted to the data, assuming that the number of lines must always be positive or zero.  Both are presented on log-log scales.  The equation of best fit is $y=(5.1881\times10^{-5})x^{-0.59086}$ where $y$ is Lines per FWHM and $x$ is the Intensity Cutoff. There is no reason to assume that the actual number of lines will follow a power law, or any other functional form, when projecting to lower intensities.  However, the model appears to be adequate for the limited range of intensities we are able to measure.}
    \label{extrapolation}
\end{figure}

\begin{figure}[tbh!]
    \centering
    \includegraphics[width=0.9\textwidth]{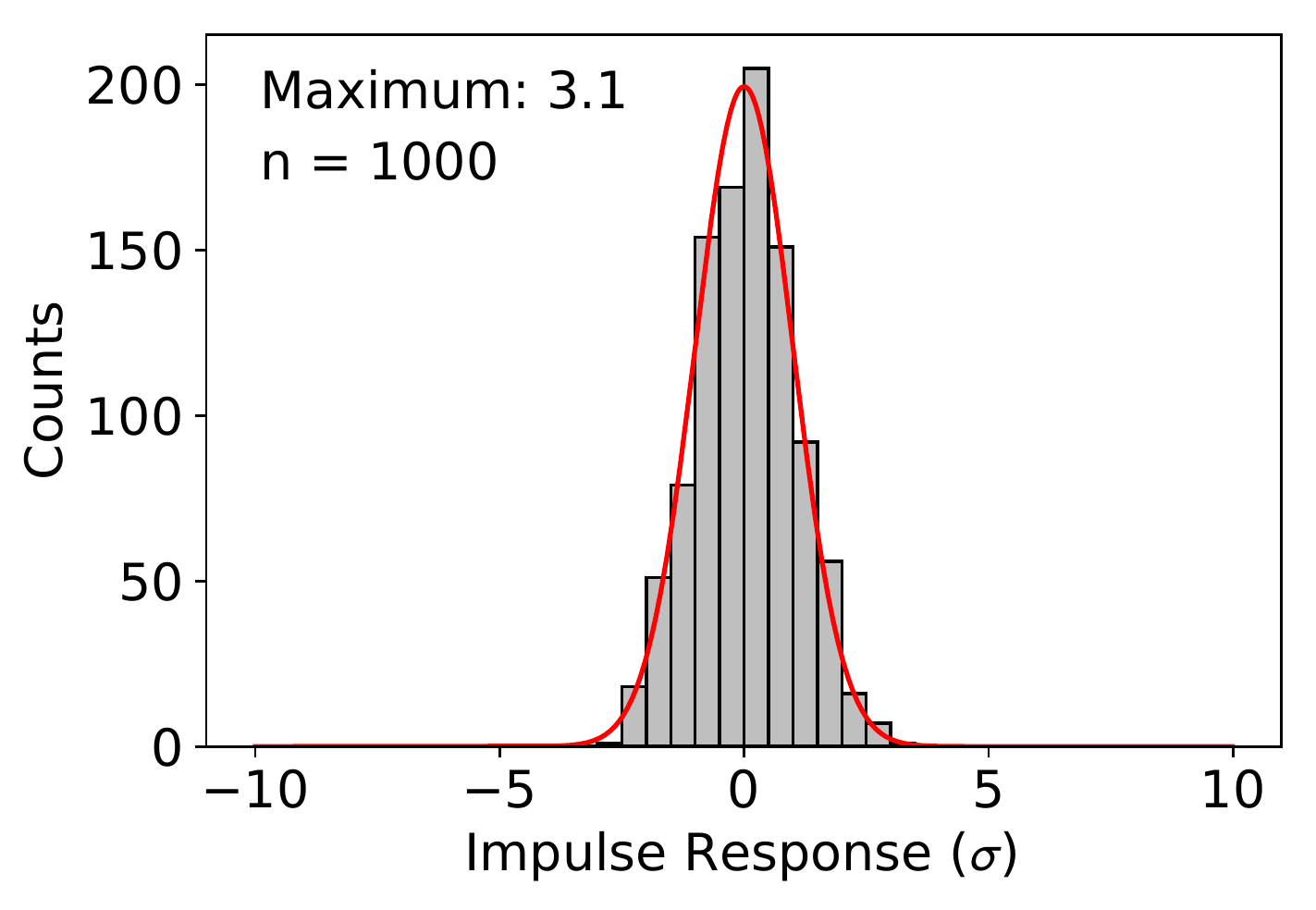}
    \caption{\textbf{Histogram of Impulse Response from Random Signals}.  Same as Fig.~\ref{randos}, but for 1000 randomly generated synthetic spectra in grey.  The maximum impulse response obtained was 3.1$\sigma$.}
    \label{rando_inj}
\end{figure}

We also generated entirely synthetic spectra with exactly known noise properties and signal components.  For each of 1000 trials, we first generate a 1.4\,kHz resolution spectrum from 8--33\,GHz with an RMS noise of 4\,mK (equivalent to our average RMS).  We then inject a random set of signals with intensity equal to 0.1$\sigma$ (0.4\,mK) with a linewidth of 0.3\,km\,s$^{-1}$ (equivalent to the overall line width of signals in TMC-1).  The number of signals (6600) was chosen by extrapolating the line density of the GOTHAM survey to the 0.1$\sigma$ level and then doubling that number to ensure a conservative estimate (Fig.~\ref{extrapolation}).  Using benzonitrile as a fiducial molecule, we then performed a spectral stack and recorded the impulse response function.  The results are shown in Fig.~\ref{rando_inj}.  None of the injected signal trials generated a significant ($>$5$\sigma$) response, and the distribution of responses again matches that expected for averaging white noise.

\subsubsection{Noise Properties of the Spectra}

\begin{figure}[tbh!]
    \centering
    \includegraphics[height=0.8\textheight]{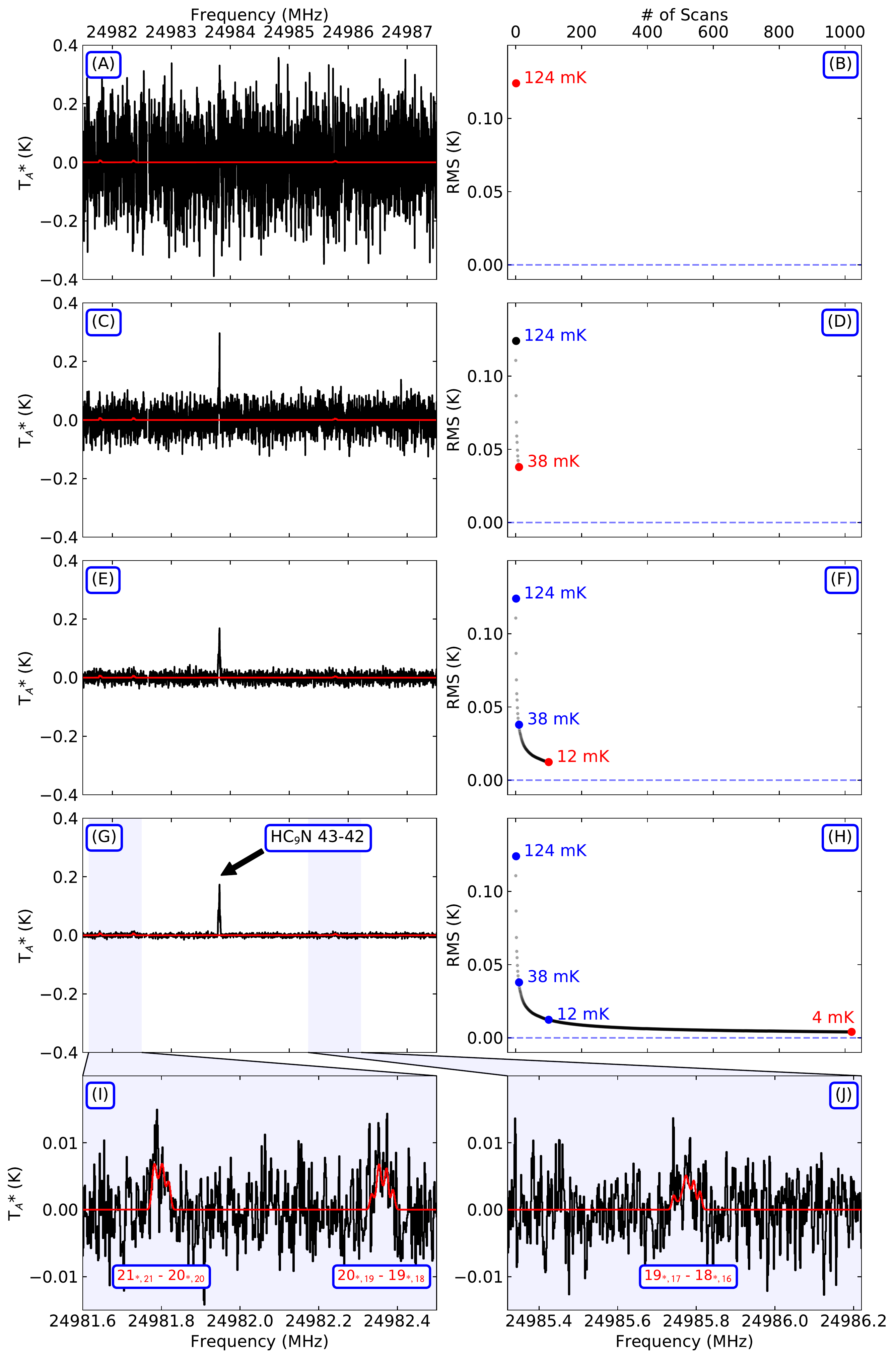}
    \caption{\textbf{Spectra of GOTHAM Data at Different Stages of Averaging.}  Scans of the 24.9\,GHz region where the 1-CNN lines are strongly detected.  (A) A single representative scan.  (C) An average of 10 scans.  (E) An average of 100 scans.  (G) The full data set (1020 scans) averaged, a line of \ce{HC9N} labeled, and zoomed outsets (I and J) showing detail on the detected 1-CNN lines (from Fig.~\ref{cnn_lines}C) in black and the model fit in red.  (B, D, F, H) These panels show the RMS in red for the corresponding spectrum at left, as well as the RMS for every individual step in the averaging in grey dots.}
    \label{fadein}
\end{figure}

We examined the signal averaging process using the spectral window containing the strongest 1-CNN detections.  We collected 1020 scans to average in this region, averaged them in series and recorded the RMS after each addition (i.e. 1 scan, 2 scans, 3 scans,~...~, 1020 scans).   Fig.~\ref{fadein} shows the results of this analysis, with snapshots at 1, 10, 100, and 1020 scans, the RMS at each of these averages, and every average in between.  The noise behaves as expected: the RMS noise follows a $\sqrt{\rm{Number of Scans}}$ decrease. 

\clearpage

\subsubsection{Number of Lines Required}

\begin{figure}[tbh!]
    \centering
    \includegraphics[width=\textwidth]{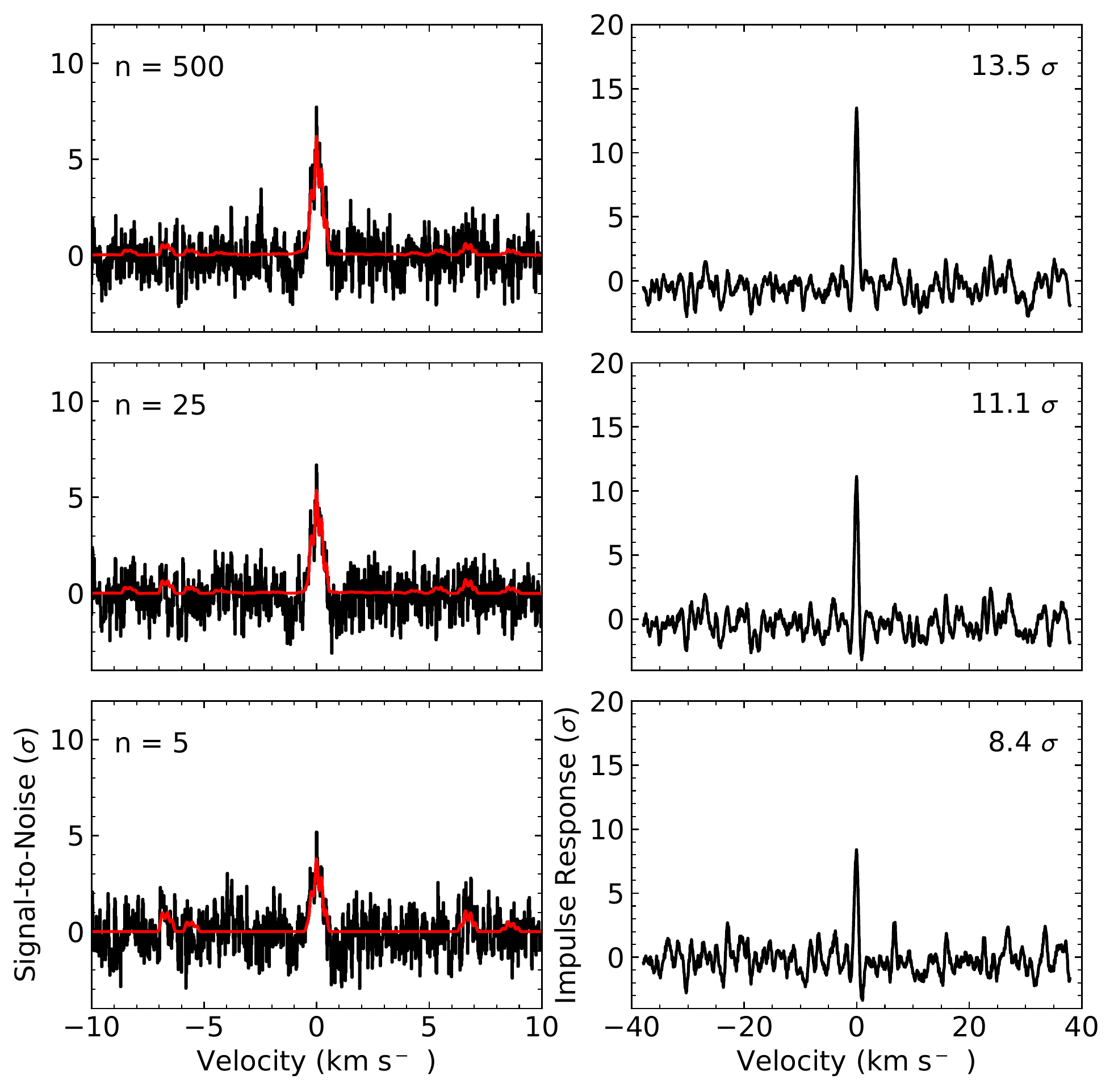}
    \caption{\textbf{Results of Stacking and Filtering Subsets of the Highest Signal-to-Noise Lines of 1-CNN.}  Same as for Fig.~\ref{cnn_stacks} but 5, 25, and 500 highest signal-to-noise lines of 1-CNN (left column) and the impulse response recorded (right column).}
    \label{1cnn_snr}
\end{figure}

\begin{figure}[tbh!]
    \centering
    \includegraphics[width=\textwidth]{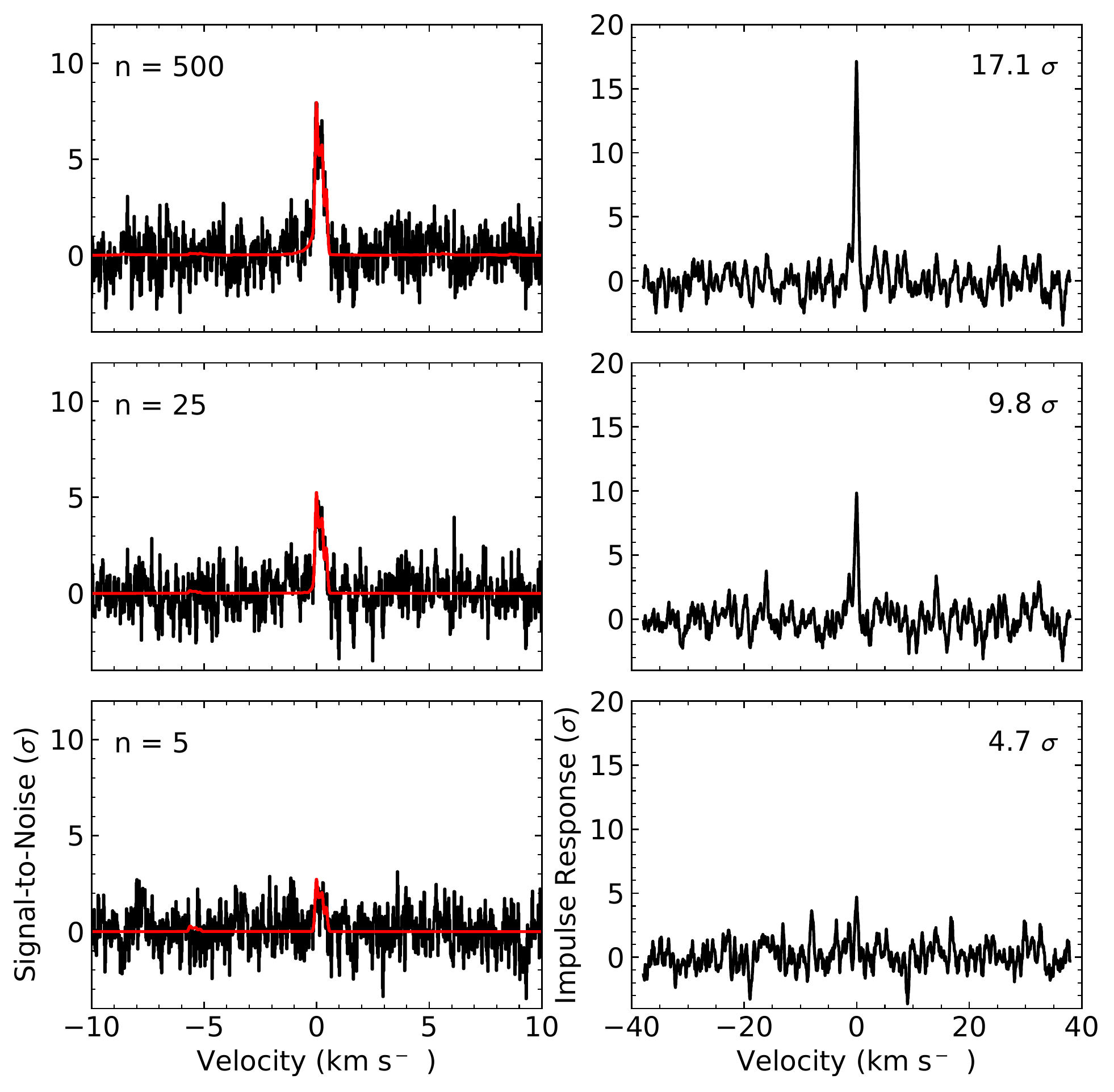}
    \caption{\textbf{Results of Stacking and Filtering Subsets of the Highest Signal-to-Noise Lines of 2-CNN.} Same as for Fig.~\ref{1cnn_snr} but for 2-CNN.}
    \label{2cnn_snr}
\end{figure}

Utilizing a large number of lines in the stacking and filtering analysis ensures that the maximum signal from a molecule is recovered.  However, that the bulk of the signal significance arises from a small subset of the highest signal-to-noise lines.  For 1-CNN and 2-CNN, a 5$\sigma$ impulse response is achieved after averaging only 5--10 of the highest signal-to-noise lines.  These are shown in Figs.~\ref{1cnn_snr} and \ref{2cnn_snr} for 1-CNN and 2-CNN, respectively, showing the results of including 5, 25, and 500 highest signal-to-noise lines as examples.  The results show that using only a fraction of the total number of lines, a substantial portion of the significance is recovered.

\clearpage

\subsubsection{Multiple hypothesis testing when matched filtering}

By applying only our best fit model from the MCMC analysis as our filter template for the matched filtering analysis, we are implicitly testing multiple hypotheses (i.e. each of the likelihood function calls from the MCMC process). Thus it is important to correct for this multiple hypothesis testing when considering the significance of the filter responses shown throughout this work (e.g., Fig.~\ref{cnn_stacks}). If our hypotheses were substantially independent, then the filter significance may be overstated without this correction, equivalent to overfitting. Conversely, if the hypotheses that were tested were substantially covariant and dependent, then the correction factor will be very small. This point is described in more detail in \citet{Loomis:2018bt}, \citet{Loomis:2021aa}, and references therein.

In our case, the hypotheses that have been tested during the MCMC process are very constrained and covariant. This is due to both the well understood quantum mechanical nature of the 1-CNN and 2-CNN spectral catalogues, as well as the well constrained source properties of TMC-1. Because the observed components of the TMC-1 cloud are relatively quiescent (compared to a more active region such as a star-forming complex), the physical parameters ($T_{ex}$, $\Delta V$, $v_{lsr}$ and relative $N_T$ between velocity components) for most detected species should be quite similar.  Because of this highly constrained nature, our hypotheses during the MCMC process are extremely similar, and the correction factor is essentially unity. Calculating this factor explicitly would require integrating the full Bayesian evidence, which is not computationally tractable, and thus we assume that it is unity.

\begin{figure}[hbt!]
    \centering
    \includegraphics[width=0.49\textwidth]{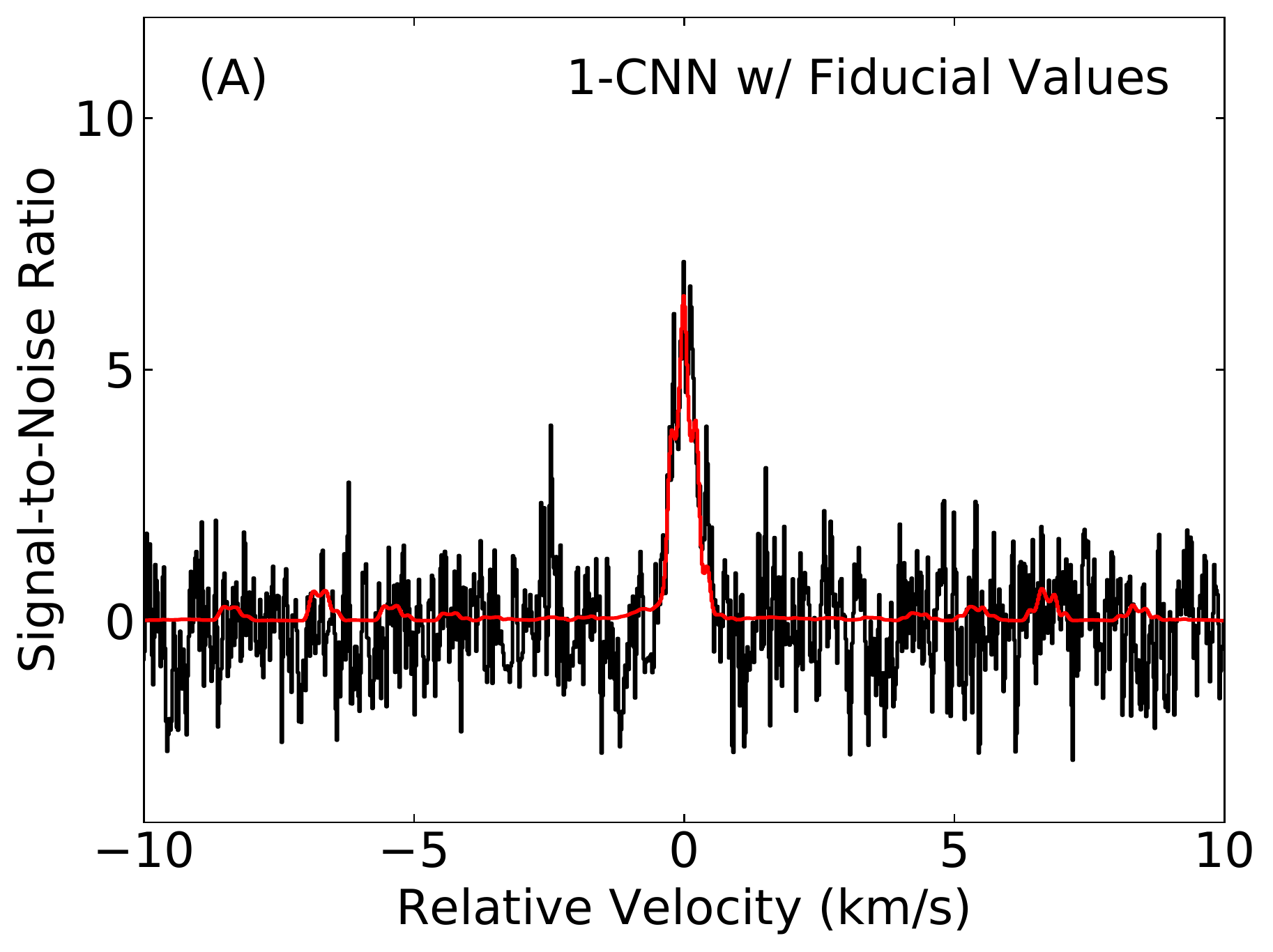}
    \includegraphics[width=0.49\textwidth]{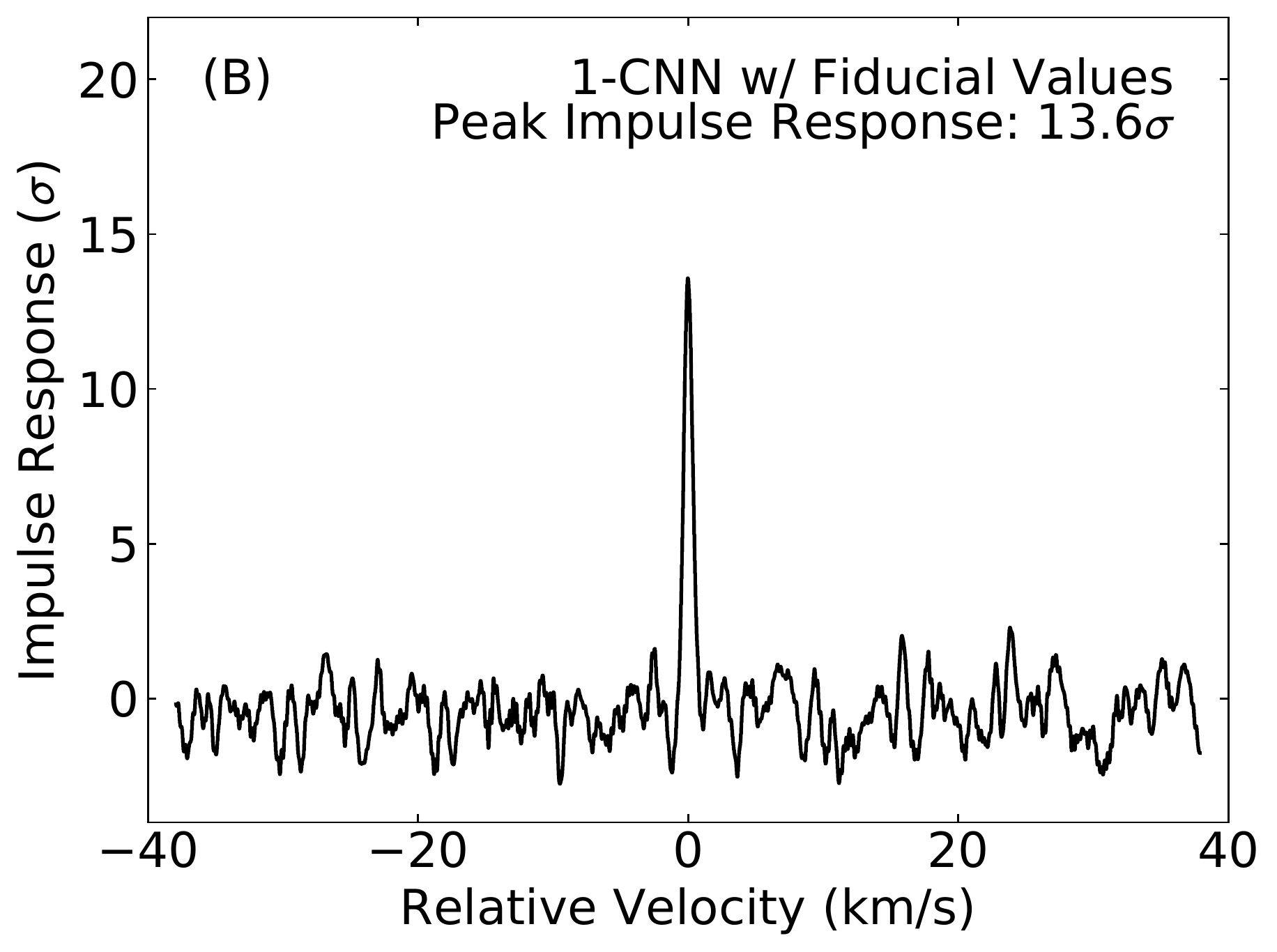}
    \includegraphics[width=0.49\textwidth]{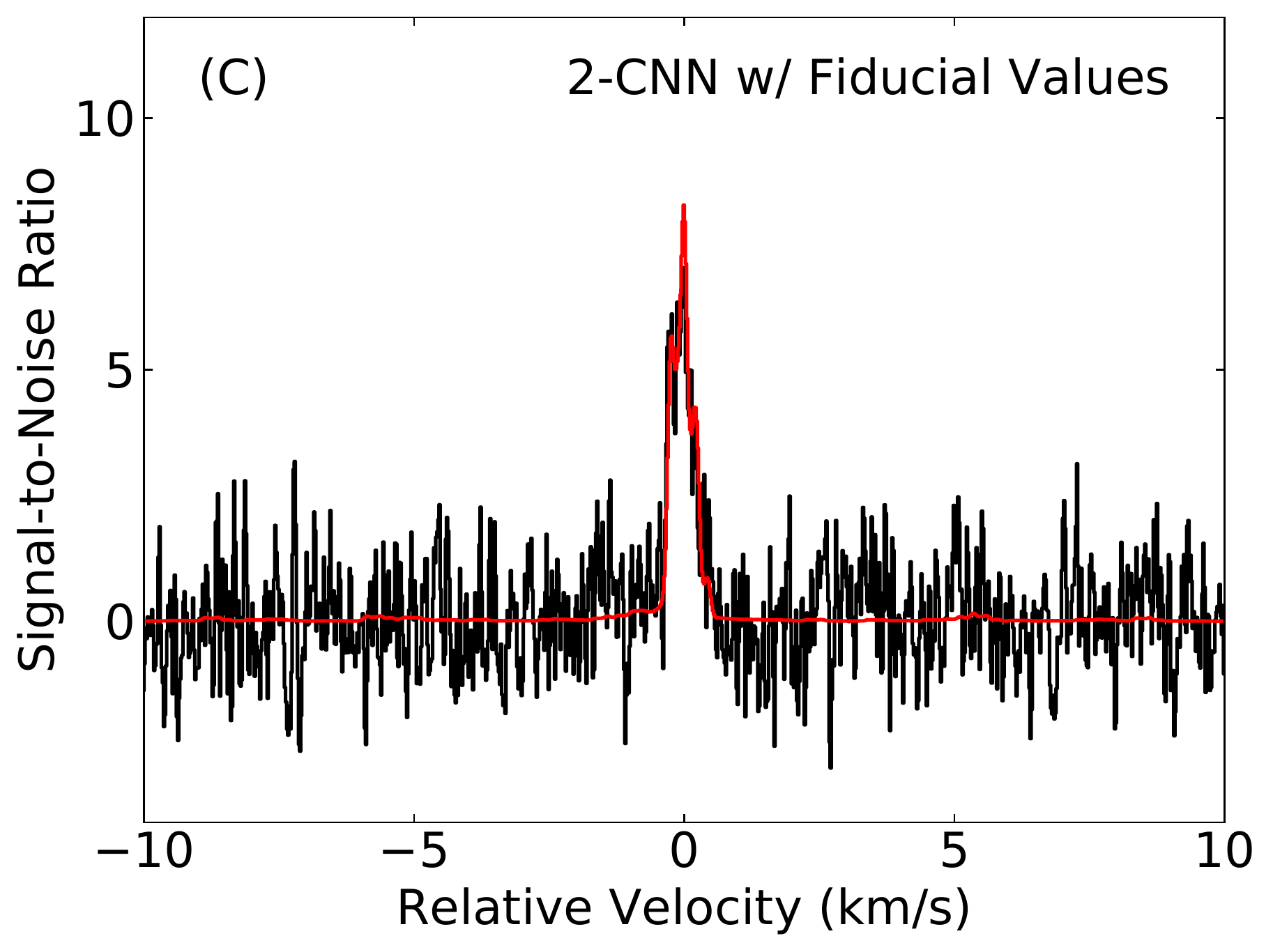}
    \includegraphics[width=0.49\textwidth]{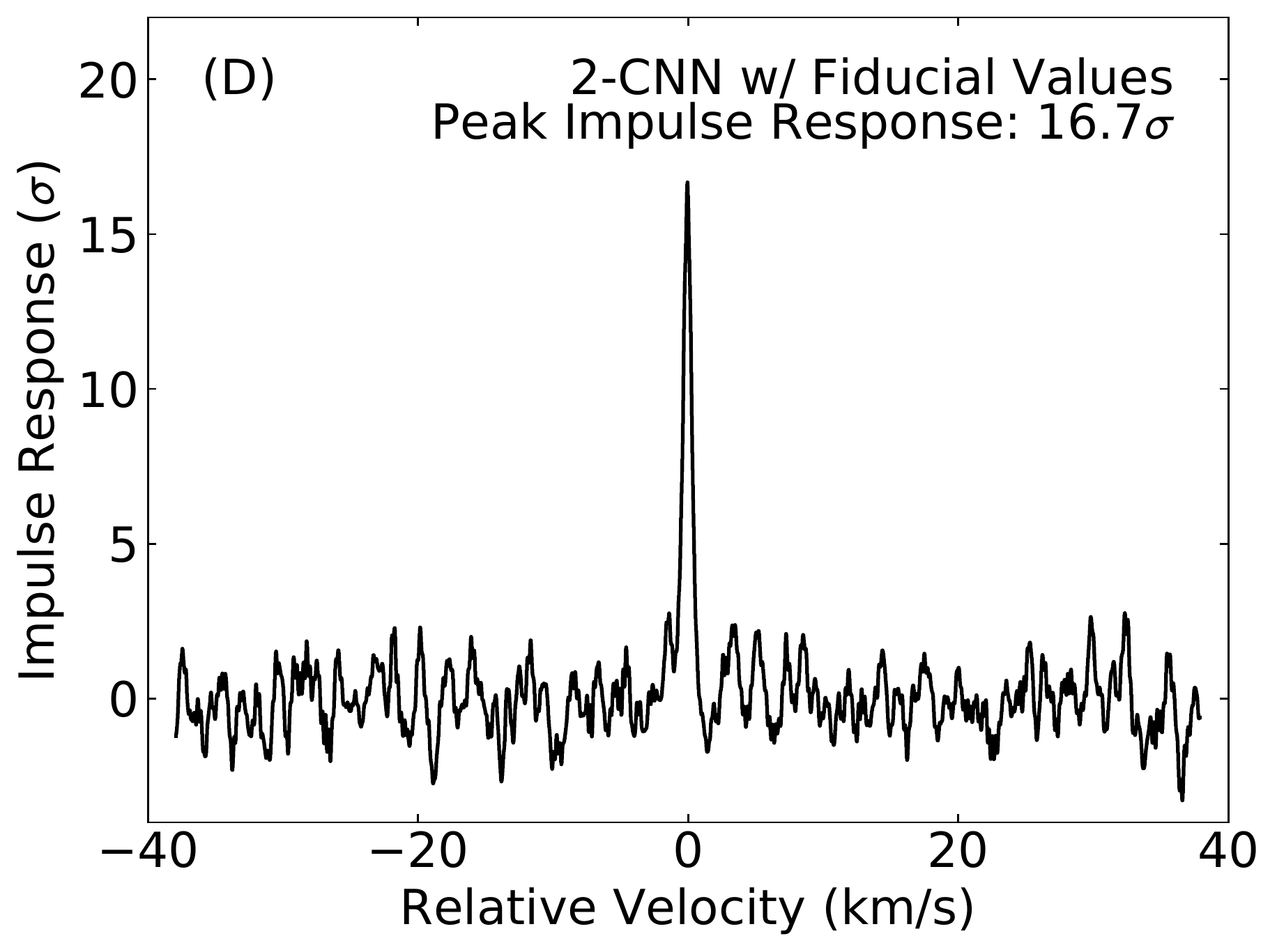}
    \caption{\textbf{Stacked Spectra and Impulse Responses for the Matched Filtering Analyses of 1-CNN and 2-CNN Using a Fiducial Model.} Same as Fig.~\ref{cnn_stacks}, except using the physical parameters listed in Table~\ref{fiducial}.}
    \label{fiducial_plot}
\end{figure}

To illustrate the robustness of this assumption, we show that most of the significance of a matched filter response for any species is recovered using an arbitrary fiducial set of model parameters. We create such a set of fiducial parameters (Table~\ref{fiducial}) starting from those of benzonitrile, which is strongly detected and representative.  These were then used to simulate the spectra of 1-CNN and 2-CNN, after which we performed a spectral stack (Fig.~\ref{fiducial_plot} A \& C) and matched filter (Fig.~\ref{fiducial_plot} B \& D) as before, recovering nearly identical impulse response functions to those derived when using the parameters for 1-CNN and 2-CNN obtained from our MCMC analysis for these species.  We therefore conclude that our assumption of a multiple hypothesis correction factor of unity is valid, and the impulse response spectra shown in this work are not overstated or overfit.

\begin{table}[htb!]
    \centering
    \caption{\textbf{Fiducial Parameter Set.} These parameters were chosen starting from the strongly detected benzonitrile and generally reproduce those of most molecules found in GOTHAM.  As discussed elsewhere, the source sizes are the least well constrained.}
    \begin{tabular}{c c c c c c}
    \toprule
    \multirow{2}{*}{Component}&	$v_{lsr}$					&	Size					&	\multicolumn{1}{c}{$N_T$}					&	$T_{ex}$							&	$\Delta V$		\\
    			&	(km s$^{-1}$)				&	($^{\prime\prime}$)		&	\multicolumn{1}{c}{(Relative)}		&	(K)								&	(km s$^{-1}$)	\\
    \midrule
    C1	&	5.575   &   75      &   1.00    &   9   &   0.125\\
    C2	&   5.767   &   100     &   2.50    &   9   &   0.125\\
    C3	&	5.892   &   200     &   1.50    &   9   &   0.125\\
    C4	&	6.018   &   200     &   2.25    &   9   &   0.125\\
    \bottomrule
    \end{tabular}
    \label{fiducial}
\end{table}

\subsection{MCMC priors from benzonitrile}

Our MCMC analysis of the cyanonaphthalenes requires a choice of prior likelihoods; ideally, these priors match physical and chemical characteristics of the target species.  We opted to use benzonitrile, which is strongly (39$\sigma$; \citep{Loomis:2021aa}) detected in the data and is likely to be chemically similar to the cyanonaphthalenes. We performed the MCMC analysis with the same four component model on benzonitrile using the DR2 data with uninformative priors on each parameter.  With the exception of the four velocity components, we applied no other constraints on parameter values save for physical constraints (e.g. positive values for $T_{ex}$,  \citealt{Loomis:2021aa}).

\begin{figure}[tbh!]
\centering
\includegraphics[width=\textwidth]{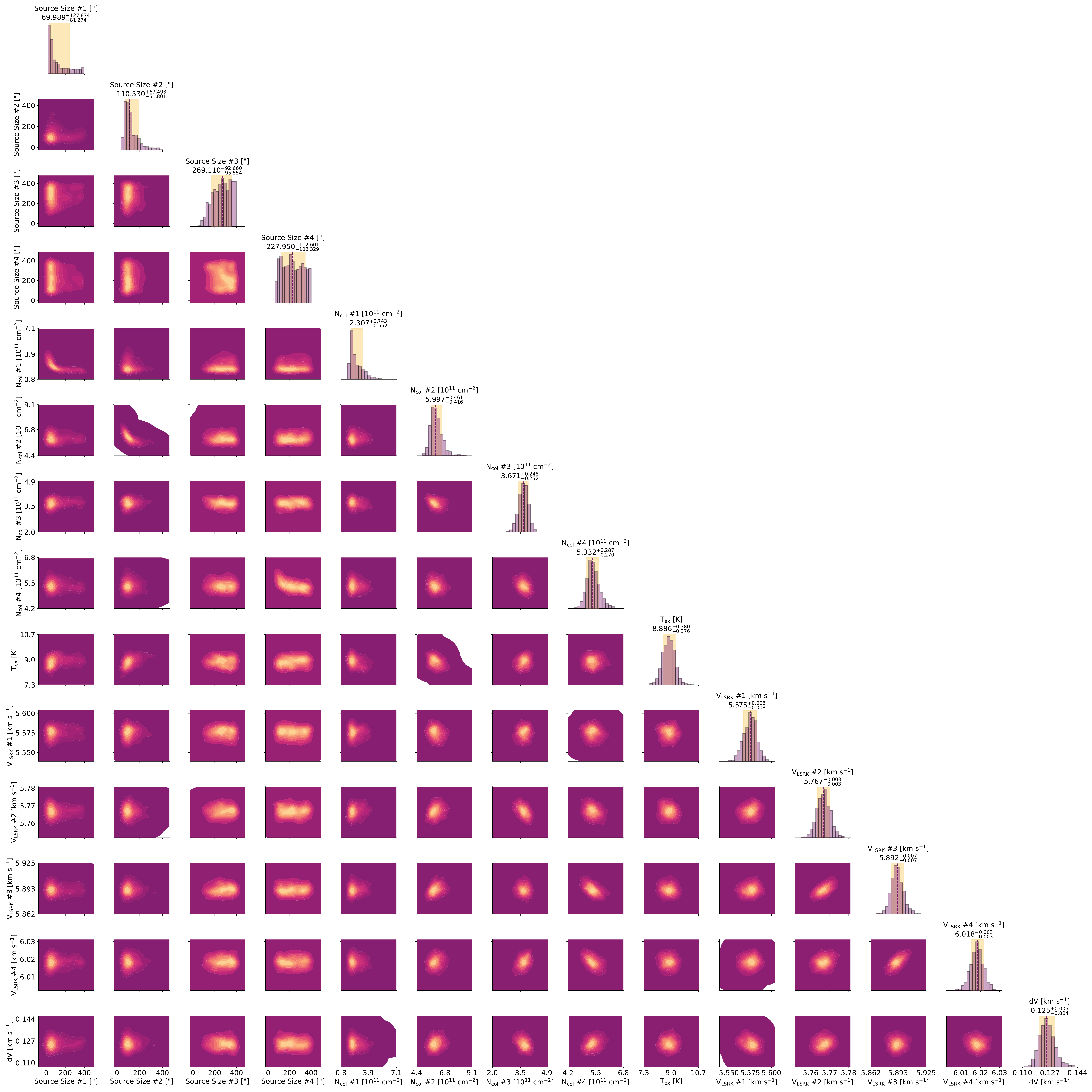}
\caption{\textbf{Corner Plot of MCMC Parameters for Benzonitrile.} Off-diagonal contours show how model parameters are covariant, while diagonal histograms represent the full space of parameter values considered in the coverged posterior (i.e. marginalized posterior) distributions for the benzonitrile MCMC fit with DR2 data. In the diagonal traces, the shaded regions comprise the 16$^{th}$ and 84$^{th}$ confidence intervals, with their numeric values annotated; these correspond to $\pm$1 sigma for a Gaussian posterior distribution.}
\label{bn_prior_triangle}
\end{figure}

A corner plot of the posterior distributions and covariances is shown in Fig.~\ref{bn_prior_triangle} following 1000 samples with 200 walkers after burn-in. The four source size parameters appear to be largely unconstrained in our model. The remaining parameters converge to mostly Gaussian distributions, with the largest covariances observed between source sizes and column densities. Subsequently, the parameters are assumed to be independent Gaussians, with their nominal mean and variance values summarized in Table \ref{bn_results}, which we adopt as priors for the CNN models.

\begin{table*}[hbt!]
\centering
\caption{\textbf{Best-fitting parameters from the MCMC analysis of benzonitrile.} The quoted uncertainties represent the 16$^{th}$ and 84$^{th}$ percentile ($1\sigma$ for a Gaussian distribution). Column density values are highly covariant with the derived source sizes.  The marginalized uncertainties on the column densities are therefore dominated by the largely unconstrained nature of the source sizes, and not by the signal-to-noise of the observations.  See Fig.~\ref{bn_prior_triangle} for a covariance plot, and Loomis et al.\citep{Loomis:2021aa} for a detailed explanation of the methods used to constrain these quantities and derive the uncertainties. Uncertainties derived by adding the uncertainties of the individual components in quadrature.}
\begin{tabular}{c c c c c c}
\toprule
\multirow{2}{*}{Component}&	$v_{lsr}$					&	Size					&	\multicolumn{1}{c}{$N_T$}					&	$T_{ex}$							&	$\Delta V$		\\
			&	(km s$^{-1}$)				&	($^{\prime\prime}$)		&	\multicolumn{1}{c}{(10$^{11}$ cm$^{-2}$)}		&	(K)								&	(km s$^{-1}$)	\\
\midrule
\hspace{0.1em}\vspace{-0.5em}\\
C1	&	$5.575^{+0.008}_{-0.008}$	&	$70^{+175}_{-35}$	&	$2.31^{+0.93}_{-0.36}$	&	\multirow{6}{*}{$8.9^{+0.4}_{-0.4}$}	&	\multirow{6}{*}{$0.125^{+0.005}_{-0.004}$}\\
\hspace{0.1em}\vspace{-0.5em}\\
C2	&	$5.767^{+0.003}_{-0.003}$	&	$111^{+83}_{-36}$	&	$6.00^{+0.51}_{-0.37}$	&	&	\\
\hspace{0.1em}\vspace{-0.5em}\\
C3	&	$5.892^{+0.007}_{-0.006}$	&	$269^{+90}_{-99}$	&	$3.67^{+0.24}_{-0.26}$	&	&	\\
\hspace{0.1em}\vspace{-0.5em}\\
C4	&	$6.018^{+0.003}_{-0.003}$	&	$228^{+117}_{-104}$	&	$5.33^{+0.30}_{-0.25}$	&	&	\\
\hspace{0.1em}\vspace{-0.5em}\\
\midrule
$N_T$ (Total)	&	 \multicolumn{5}{c}{$1.73^{+0.85}_{-1.00}\times 10^{12}$~cm$^{-2}$}\\
\bottomrule
\end{tabular}

\label{bn_results}

\end{table*}

\subsection{Best-fitting parameters from the MCMC analysis of the cyanonaphthalenes}

\begin{figure*}[tbh!]
\centering
\includegraphics[width=\textwidth]{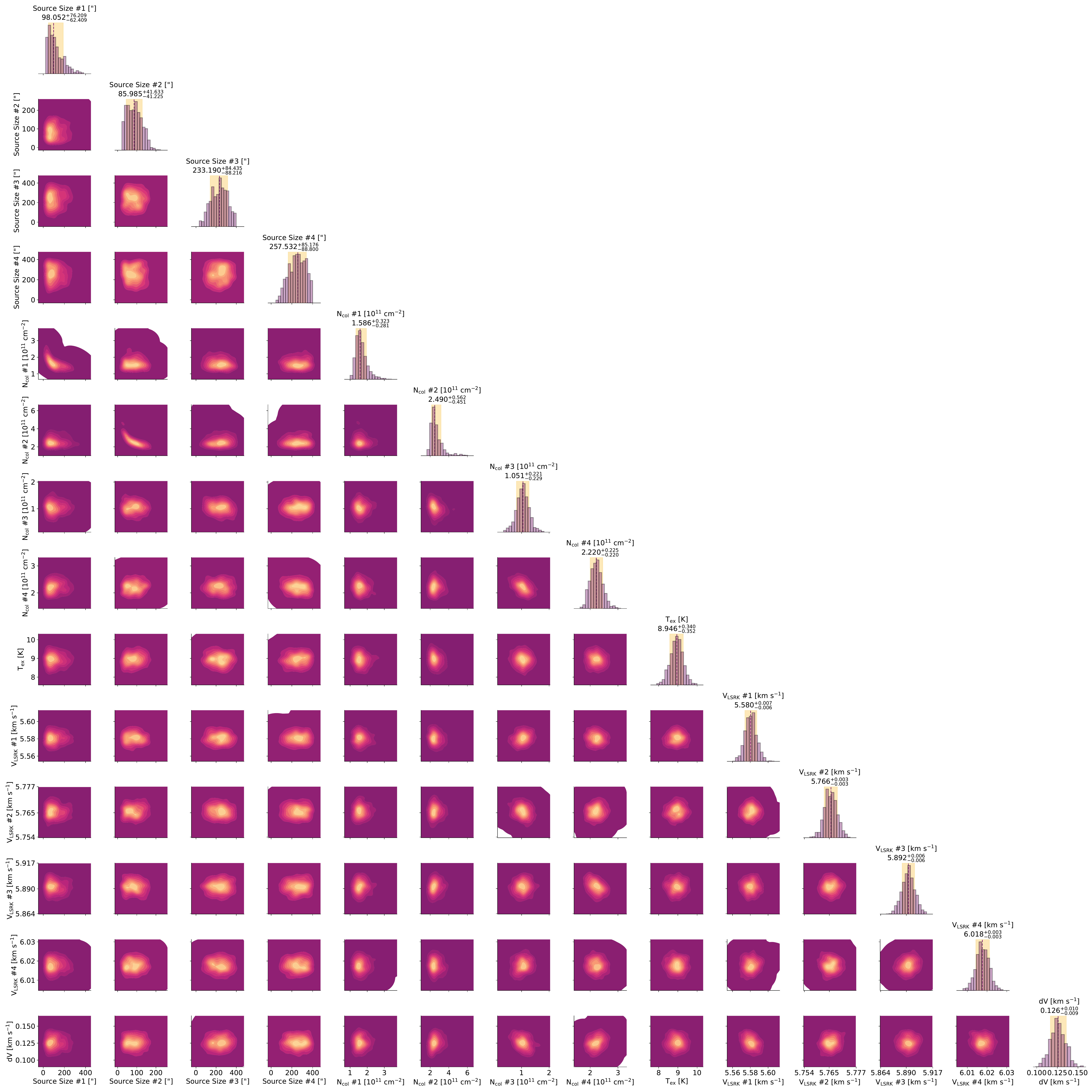}
\caption{\textbf{Corner Plot of MCMC Parameters for 1-CNN.} Same as for Fig.~\ref{bn_prior_triangle} but for 1-CNN.}
\label{1-CNN_triangle}
\end{figure*}

\begin{figure*}[tbh!]
\centering
\includegraphics[width=\textwidth]{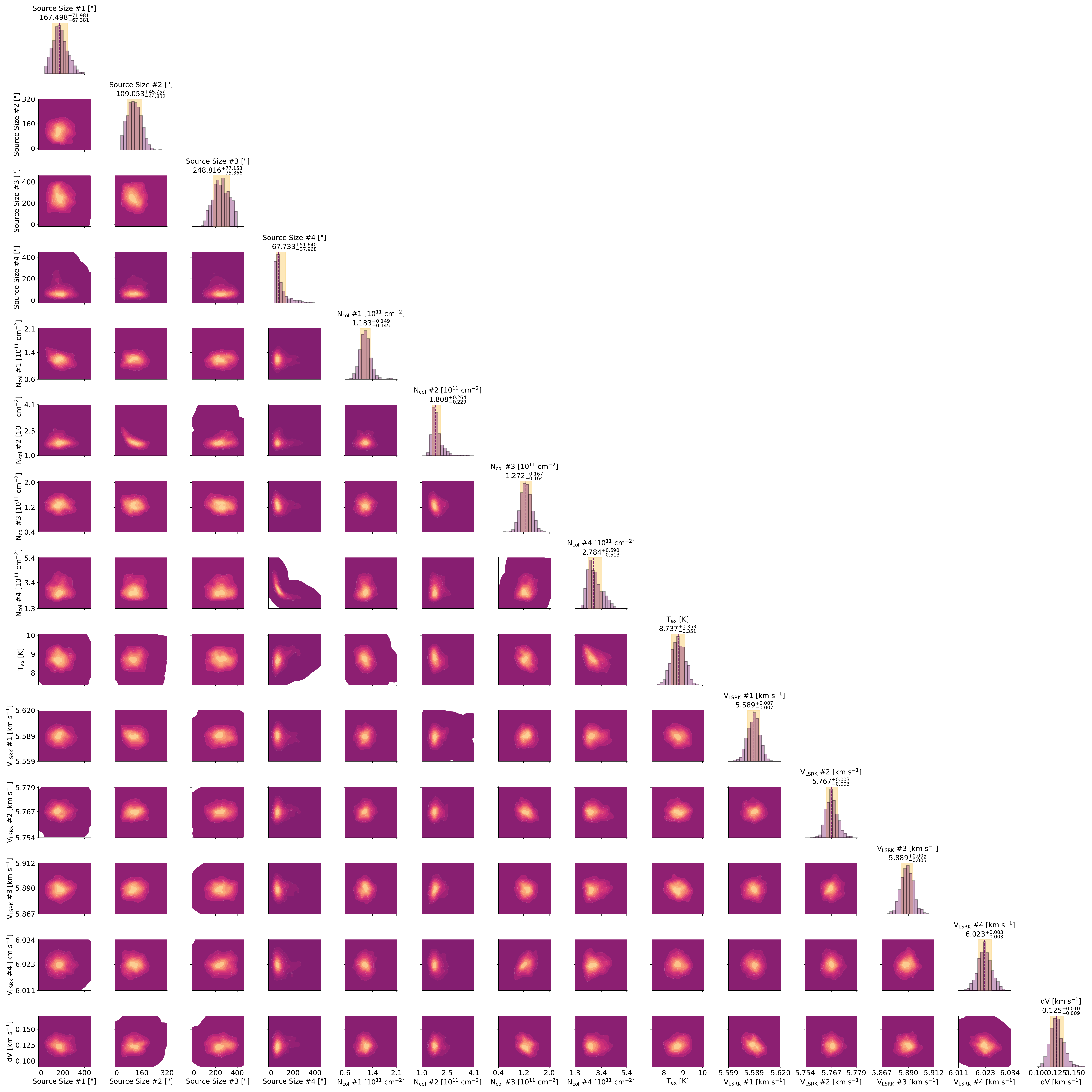}
\caption{\textbf{Corner Plot of MCMC Parameters for 2-CNN.} Same as for Fig.~\ref{bn_prior_triangle} but for 2-CNN.}
\label{2-CNN_triangle}
\end{figure*}

The best-fitting parameters from the MCMC analysis of 1-CNN and 2-CNN are given in Tables~\ref{1-CNN_results} and \ref{2-CNN_results}, respectively.  The stacked spectra and matched filter results are shown in Fig.~\ref{cnn_stacks}. Corner plots of the parameter covariances for the 1-CNN and 2-CNN fits are shown in Figs.~\ref{1-CNN_triangle} and~\ref{2-CNN_triangle}, respectively.

\begin{table*}[hbt!]
\centering
\caption{Same as Table~\ref{bn_results}, but for 1-CNN}
\begin{tabular}{c c c c c c}
\toprule
\multirow{2}{*}{Component}&	$v_{lsr}$					&	Size					&	\multicolumn{1}{c}{$N_T$}					&	$T_{ex}$							&	$\Delta V$		\\
			&	(km s$^{-1}$)				&	($^{\prime\prime}$)		&	\multicolumn{1}{c}{(10$^{11}$ cm$^{-2}$)}		&	(K)								&	(km s$^{-1}$)	\\
\midrule
\hspace{0.1em}\vspace{-0.5em}\\
C1	&	$5.580^{+0.007}_{-0.006}$	&	$98^{+90}_{-49}$	&	$1.59^{+0.36}_{-0.24}$	&	\multirow{6}{*}{$8.9^{+0.3}_{-0.4}$}	&	\multirow{6}{*}{$0.126^{+0.010}_{-0.009}$}\\
\hspace{0.1em}\vspace{-0.5em}\\
C2	&	$5.766^{+0.003}_{-0.003}$	&	$86^{+142}_{-41}$	&	$2.49^{+0.67}_{-0.34}$	&	&	\\
\hspace{0.1em}\vspace{-0.5em}\\
C3	&	$5.892^{+0.006}_{-0.006}$	&	$233^{+81}_{-92}$	&	$1.05^{+0.21}_{-0.24}$	&	&	\\
\hspace{0.1em}\vspace{-0.5em}\\
C4	&	$6.018^{+0.004}_{-0.003}$	&	$258^{+82}_{-92}$	&	$2.22^{+0.23}_{-0.22}$	&	&	\\
\hspace{0.1em}\vspace{-0.5em}\\
\midrule
$N_T$ (Total)	&	 \multicolumn{5}{c}{$7.35^{+3.33}_{-4.63}\times 10^{11}$~cm$^{-2}$}\\
\bottomrule
\end{tabular}

\label{1-CNN_results}

\end{table*}

\begin{table*}
\centering
\caption{Same as Table~\ref{bn_results}, but for 2-CNN}
\begin{tabular}{c c c c c c}
\toprule
\multirow{2}{*}{Component}&	$v_{lsr}$					&	Size					&	\multicolumn{1}{c}{$N_T$}					&	$T_{ex}$							&	$\Delta V$		\\
			&	(km s$^{-1}$)				&	($^{\prime\prime}$)		&	\multicolumn{1}{c}{(10$^{11}$ cm$^{-2}$)}		&	(K)								&	(km s$^{-1}$)	\\
\midrule
\hspace{0.1em}\vspace{-0.5em}\\
C1	&	$5.589^{+0.007}_{-0.007}$	&	$168^{+77}_{-63}$	&	$1.18^{+0.15}_{-0.14}$	&	\multirow{6}{*}{$8.7^{+0.4}_{-0.4}$}	&	\multirow{6}{*}{$0.125^{+0.010}_{-0.009}$}\\
\hspace{0.1em}\vspace{-0.5em}\\
C2	&	$5.767^{+0.003}_{-0.003}$	&	$109^{+47}_{-44}$	&	$1.81^{+0.30}_{-0.19}$	&		&	\\
\hspace{0.1em}\vspace{-0.5em}\\
C3	&	$5.889^{+0.005}_{-0.005}$	&	$249^{+79}_{-74}$	&	$1.27^{+0.17}_{-0.16}$	&		&	\\
\hspace{0.1em}\vspace{-0.5em}\\
C4	&	$6.023^{+0.003}_{-0.003}$	&	$68^{+65}_{-24}$	&	$2.78^{+0.67}_{-0.44}$	&		&	\\
\hspace{0.1em}\vspace{-0.5em}\\
\midrule
$N_T$ (Total)	&	 \multicolumn{5}{c}{$7.05^{+3.23}_{-4.50}\times 10^{11}$~cm$^{-2}$}\\
\bottomrule
\end{tabular}

\label{2-CNN_results}

\end{table*}

\clearpage

\section{Astrochemical Models}

\subsection{Initial Conditions}

Our reaction network simulations adopt similar physical conditions to previous chemical models of TMC-1: gas and dust temperatures of 10\,K, gas densities of $2\times10^4$ cm$^{-3}$, a cosmic-ray ionization rate of $1.3\times10^{-17}$\,s$^{-1}$ \citep{Loomis:2016js,McGuire:2018bz}. Initial elemental abundances were taken from \citep{hincelin_oxygen_2011} with the exception of atomic oxygen, where we utilize O/H $\approx1.5\times10^{-4}$, resulting in a slightly carbon rich C/O $\approx1.1$ \citep{Loomis:2021aa,burkhardt:2021aa}.

\subsection{Rate Coefficients}

A rate coefficient of $k_\mathrm{R1} =  2.5\times10^{-10}$ cm$^{3}$ s$^{-1}$ has been adopted for Reaction \ref{napForm}, a value typical for processes that occur with a single-collision \citep{parker_low_2012}. Similar calculations for Reaction \ref{dialinForm} of $k_\mathrm{R2} =  3\times10^{-10}$ cm$^{3}$ s$^{-1}$ were found to be also occur around the gas-phase collision rate \citep{kaiser_pah_2012}.

The assumption that Reactions \ref{cnnForm1} and \ref{cnnForm2} are barrierless and exothermic are supported by the difference between the ionization energy of the closed-shell reactant and the electron affinity of the radical, which is less than 8.75 eV. This criterion has been proposed \citep{smith_temperature-dependence_2006} for a neutral-neutral reaction to occur at approximately the collisional rate at low temperatures. Thus, Reactions \eqref{cnnForm1} and \eqref{cnnForm2} are assumed to have rate coefficients of $k_\mathrm{R3} = k_\mathrm{R4} = 1.5\times10^{-10}$ cm$^{3}$ s$^{-1}$.  Rate coefficients for destruction via ions were estimated assuming a Su-Chesnavich capture model to account for long-range Coulombic attractions \citep{woon_quantum_2009}.

\subsection{Relative Abundances of 1-CNN and 2-CNN}

Using the composite thermochemical method, G3//B3LYP, we estimate the relative energy difference between the two isomers to be on the order of 2 kJ mol$^{-1}$, with 1-CNN more stable than 2-CNN \citep{baboul_gaussian-3_1999}. Similar to the observed values (1-CNN/2-CNN ratio $\approx1.04\pm0.78$), the models also predict a slightly higher abundance for 1-CNN relative to 2-CNN (1-CNN/2-CNN\,$\approx$\,1.098) because the larger permanent dipole of 2-CNN results in slightly more efficient destruction rates with ions \citep{shingledecker_isomers_2020}. If we assume that the abundance difference between the two is real, then, based on the relative dipole principle \citep{shingledecker_isomers_2020}, the agreement between the model and observations suggests that the chemistry of 1- and 2-CNN is similar, except for their differing rates of ion-neutral reactions.

\section{The Formation of Naphthalene and its Precursors}

The major precursors for naphthalene are phenyl, vinylacetylene, and 1,3-butadiene. In our network, phenyl is formed mainly via the photodissociation of benzene, while the major pathway for benzene formation is a series of ion-neutral reactions and dissociative recombination processes \citep{McEwan:1999ia}, which culminate in

\begin{equation}
    \ce{C6H7+ + e- -> C6H6 + H.}
\end{equation}

\noindent
The dominant formation routes for vinylacetylene are the gas-phase reaction

\begin{equation}
    \ce{CH + CH3CCH -> CH2CHC2H + H}
\end{equation}

\noindent
as well as the following grain-surface H-addition reaction

\begin{equation}
    \ce{H + C4H3 -> CH2CHC2H}
\end{equation}

\noindent
where some fraction of the resulting vinylacetylene is introduced into the gas via reactive desorption.

\noindent
The dominant gas-phase formation routes for 1,3-butadiene are

\begin{equation}
    \ce{CH + CH3CHCH2 -> H + CH2CHCHCH2}
\end{equation}

\begin{equation}
    \ce{CCH CH3CHCH2 -> CH + CH2CHCHCH2.}
\end{equation}

At model times of greater than $\sim10^5$ yr, the fractional abundances of \ce{C6H5} and \ce{CH3CHC2H} are $\sim10^{-10}$, while \ce{CH2CHCHCH2} is even less abundant at $\sim10^{-13}$. Thus, though reactions (R1) and (R2) occur barrierlessly, the low abundances of the reactants leads to an overall slow rate of naphthalene formation \citep{burkhardt:2021aa}.

Given that our model underpredicts the abundances of 1- and 2-CNN -- and by implication also of naphthalene  -- it is likely, that our chemical network is missing dominant formation route(s) for naphthalene precursors. Another possibility, not mutually exclusive with missing precursors, is that ion-neutral/dissociative combination routes also lead to the formation of naphthalene, similar to the dominant benzene formation route \citep{McEwan:1999ia}.

\section{Carbon-chain Chemistry}

\begin{figure*}[tbh!]
\centering
\includegraphics[width=\textwidth]{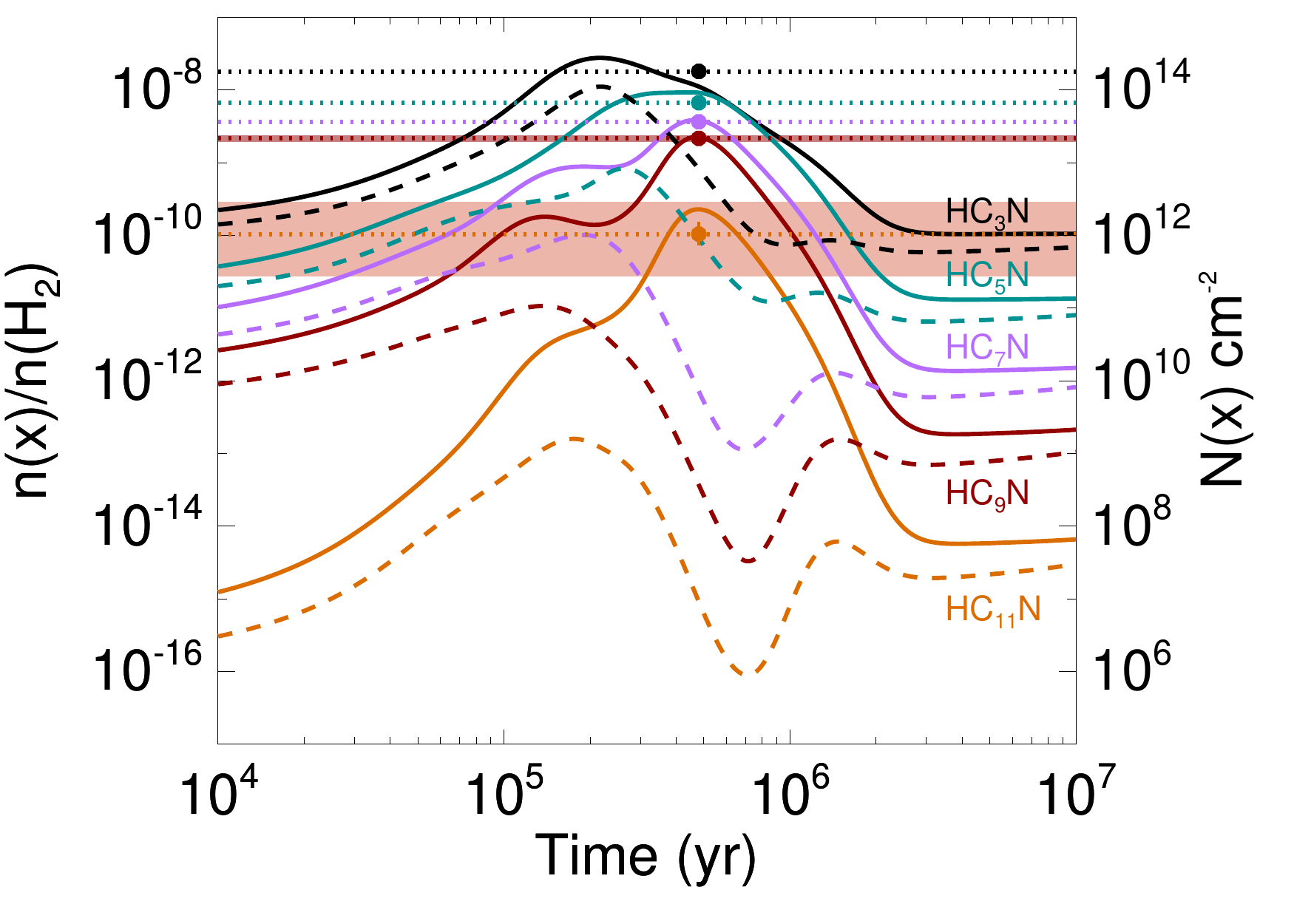}
\caption{\textbf{Reaction network results for cyanopolyynes.}  Same as for Fig.~\ref{fig:model} but for the cyanopolyynes \ce{HC3N}, \ce{HC5N}, \ce{HC7N}, \ce{HC9N}, and \ce{HC11N}. Solid lines represent model results using C/O=1.0974, while dashed lines give results for C/O=0.7 for comparison. Observational abundances and errors from \citep{Loomis:2021aa} are given by horizontal bars (too small to display for \ce{HC7N}, \ce{HC5N}, \ce{HC3N}), and abundances at the model time of best agreement are represented by filled circles.}
\label{cyanopolyynes}
\end{figure*}

We compare our  modeling results for 1- and 2-CNN (Fig.~\ref{fig:model}) with those for carbon-chain species, specifically the cyanopolyynes (Fig.~\ref{cyanopolyynes}). From Fig.~\ref{cyanopolyynes}, it  is clear that our simulations agree with the observed abundances for these species. The model time of closest agreement with fitted co-spatial column densities, represented by the filled circles, occurs at $\sim10^5$\,yr, a reasonable age for a cold core such as TMC-1 \citep{hincelin_oxygen_2011}. 

Figure~\ref{cyanopolyynes} illustrates the sensitivity of carbon-chain abundances to the assumed C/O value. As depicted there, a C/O = 0.7 (represented by the dashed lines) results in much poorer agreement than does C/O = 1.0974 (represented by the solid lines). The comparison of different cyanopolyyne abundances as a function of the C/O ratio with derived values \citep{Loomis:2021aa} is the basis for the empirical C/O=1.0974 used here, i.e.~it was chosen to reproduce the derived column densities of this family of interstellar molecules. We conclude that our chemical reaction network is adequate for the cyanopolyynes, but incomplete for the CNNs.

\section{Destruction of PAHs}

\begin{figure*}[tbh!]
\centering
\includegraphics[width=\textwidth]{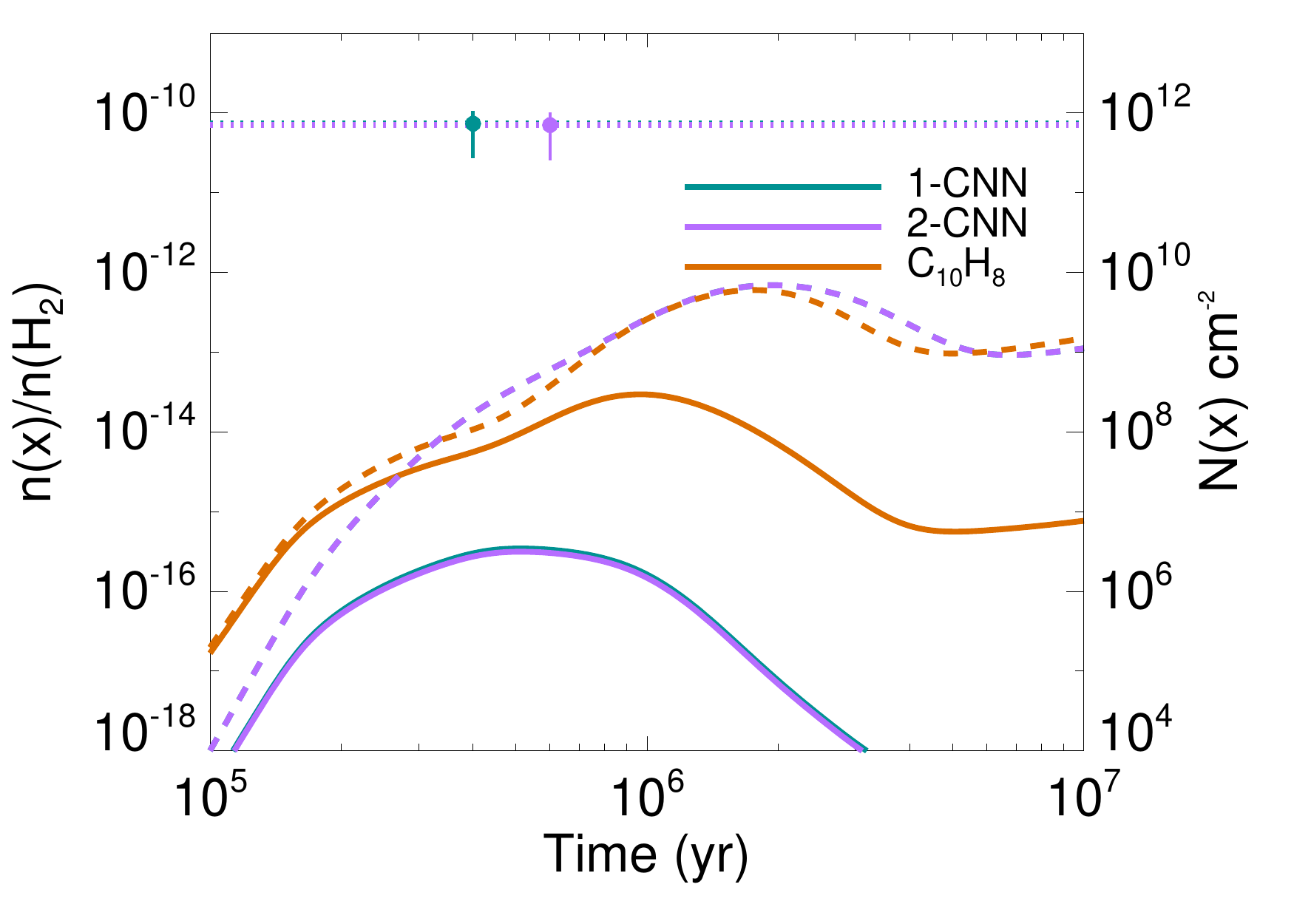}
\caption{\textbf{Reaction network results for CNNs when considering ion-neutral destruction pathways.} Same as for Fig.~\ref{fig:model} but 1- and 2-CNN with (solid lines) and without (dashed line) ion-neutral destruction pathways with ions for cyclic, aromatic species. The abundance of 1-CNN is identical to that of 2-CNN in the latter simulation because their destruction by ions is neglected.}
\label{pah_noion}
\end{figure*}

In our chemical network, naphthalene, 1-CNN, and 2-CNN are each destroyed via reaction with ions (\ce{H3+}, \ce{H3O+}, \ce{C+}, \ce{H+}, and \ce{He+}), as well as depletion onto grains, albeit to a lesser extent. To avoid introducing ``chemical loops,'' we have taken a cautious approach in adding these destruction pathways to our PAH network. We assume that reaction with an ion leads to dissociation into smaller linear fragments (e.g. \ce{C2H4}) rather than to cyclic cations (e.g. \ce{C10H9+}). It is possible that species like protonated naphthalene are produced in such processes, which could reform \ce{C10H8} via dissociative recombination. Thus, our model may overestimate the destruction rates of these small PAHs, contributing to the low predicted abundances for 1-CNN  and 2-CNN. As a test of the ion-neutral destruction reactions on our calculated PAH abundances, we have run an additional cold core model where those destruction routes with all benzene-, naphthalene-, and cyclopentadiene-related molecules have been disabled. The results of this simulation are given in Fig.~\ref{pah_noion}, which shows that ion-neutral destruction pathways do play a substantial role on the overall abundance of 1-CNN and 2-CNN, though their absence still does not bring the calculated and observed abundances into agreement. 1- and 2-CNN have different rates of destruction with ions \citep{shingledecker_isomers_2020} because of differences in their permanent dipole moments, resulting in a $\sim$10\% difference in peak abundances. With those reactions removed, the chemistry of 1- and 2-CNN become identical, and thus, their abundance curves in Fig. \ref{pah_noion} lie on top of one another.

More generally, the behavior of PAH cations upon recombination with electrons (or upon reaction between neutral PAHs and ions) is likely to affect our models, and is sensitive to molecular size. Previous studies found that PAHs with more than around 20--30 carbon atoms can efficiently undergo radiative stabilization \citep{montillaud_evolution_2013,rapacioli_formation_2006}, and thus are not prone to dissociation, unlike smaller PAHs such as naphthalene. Thus, this phenomenon may dominate in the inventory of PAHs inherited from earlier, more diffuse stages of cloud evolution. Because radiative stabilization processes are not efficient for smaller PAHs, they may be quickly depleted by the time densities reach $\sim10^4$ cm$^{-3}$ due to ionization by the less-attenuated UV photons of the Interstellar Radiation Field and subsequent dissociative recombination in the earlier, less UV-shielded environments.

\end{document}